\documentclass[numberedappendix]{emulateapj}

\pdfoutput=1
\usepackage{apjfonts}
\usepackage{amssymb}
\usepackage{amsmath}
\usepackage{subfigure}
\usepackage{multirow}

\shorttitle{}
\shortauthors{Mao et al.}

\begin{document}
\title{Magnetic Field Structure of the Large Magellanic Cloud from Faraday Rotation Measures of Diffuse Polarized Emission}
\author{S. A. Mao,\altaffilmark{1,2,3,4} 
N. M. McClure-Griffiths,\altaffilmark{2}
B. M. Gaensler,\altaffilmark{5}
M. Haverkorn,\altaffilmark{6,7}
R. Beck,\altaffilmark{8}
D. McConnell,\altaffilmark{2}
M. Wolleben, \altaffilmark{9}
S. Stanimirovi\'{c}, \altaffilmark{4}
J. M. Dickey,\altaffilmark{10}
L. Staveley-Smith\altaffilmark{11}
}

\altaffiltext{1}{Harvard-Smithsonian Center for Astrophysics, Cambridge, MA 02138; mao@astro.wisc.edu}
\altaffiltext{2}{Australia Telescope National Facility, CSIRO Astronomy \& Space Science, Epping, NSW 1710, Australia}
\altaffiltext{3}{Jansky Fellow, National Radio Astronomy Observatory, P.O. Box O, Socorro, NM 87801}
\altaffiltext{4}{Department of Astronomy, University of Wisconsin, Madison, WI 53706}
\altaffiltext{5}{Sydney Institute for Astronomy, School of Physics, The University of Sydney, NSW 2006, Australia}
\altaffiltext{6}{Department of Astrophysics, Radboud University, P.O. Box 9010, 6500 GL Nijmegen, The Netherlands}
\altaffiltext{7}{Leiden Observatory, Leiden University, P.O. Box 9513, 2300 RA Leiden, The Netherlands}
\altaffiltext{8}{Max-Planck-Institut f\"{u}r Radioastronomie, 53121 Bonn, Germany}
\altaffiltext{9}{Square Kilometre Array South Africa, The Park, Pinelands, 7405, South Africa}
\altaffiltext{10}{Physics Department, University of Tasmania, Hobart, TAS 7001, Australia}
\altaffiltext{11}{International Centre for Radio Astronomy Research(ICRAR), The University of Western Australia, Crawley, WA 6009, Australia}

\begin{abstract}
We present a study of the magnetic field of the Large Magellanic
Cloud (LMC), carried out using diffuse polarized synchrotron emission data at 1.4 GHz acquired at the Parkes Radio Telescope and the Australia Telescope Compact Array. 
The observed diffuse polarized emission is likely to originate above the LMC disk on the near side of the galaxy.
Consistent negative rotation measures (RMs) derived from the diffuse emission indicate that the line-of-sight magnetic field in the LMC's near-side halo is directed coherently away from us.
In combination with RMs of extragalactic sources that lie behind the galaxy, we show that the LMC's large scale magnetic field is likely to be of quadrupolar geometry, consistent with the prediction of dynamo theory.
On smaller scales, we identify two brightly polarized filaments southeast of the LMC, associated with neutral hydrogen arms. The filaments' magnetic field potentially aligns with the direction towards the Small Magellanic Cloud. We suggest that tidal interactions between the Small and the Large Magellanic Clouds in the past 10$^9$ years is likely to have shaped the magnetic field in these filaments.
\end{abstract}


\section{Introduction}
\label{section:introduction}
Magnetic fields are a key ingredient in the interstellar medium (ISM) of galaxies -- they accelerate and confine cosmic rays, trigger star  formation and exert pressure to balance gas against gravity. Therefore, understanding the structure and origin of magnetic fields in galaxies is crucial for a complete picture of galaxy evolution. 

Normal spiral galaxies possess spiral-like coherent magnetic fields on galactic scales. These fields are thought to be generated by the standard $\alpha$-$\omega$ dynamo, which relies on small scale turbulence ($\alpha$) and differential rotation ($\omega$) of the galactic disk to amplify and order magnetic fields on a time scale of a few Gyrs \citep[see for example,][]{beck1996}. However, standard dynamo has difficulties explaining the presence of coherent magnetic fields in slowly rotating dwarf irregular galaxies, such as NGC 4449 \citep{chyzy2000} and galaxies currently undergoing strong gravitational interactions, such as the Large and the Small Magellanic Clouds \citep{gaensler2005,mao2008}. For the Magellanic Clouds, the gravitational interaction time scale is short compared to the $e$-folding time of magnetic field growth predicted by dynamo theory, and thus could prevent the build up of coherent galactic scale magnetic fields. It has been suggested that a more efficient form of dynamo, driven by cosmic rays produced in star formation episodes, can amplify magnetic fields on time scales of $\sim$ 200 Myrs \citep{parker1992,hanasz2004}. The Large Magellanic Cloud (LMC), the closest external galaxy of irregular type at a distance of 50 kpc, is the ideal test bed to examine if coherent magnetic fields exist and if dynamos operate in such galaxies.

Characterizing the strength and geometry of large scale magnetic fields in dwarf irregular galaxies such as the LMC is also important in evaluating the role they play in magnetizing the surrounding intergalactic medium:  magnetic pressure can become dynamically important  and can aid to expel magnetized outflow from the galactic disk \citep[e.g.,][]{chyzy2000,kepley2010}. Finally, ram pressure effects experienced by the LMC as it travels through the Milky Way's hot halo and the LMC's on-going tidal interaction with the SMC both may have shaped the galaxy's large scale neutral hydrogen morphology  \citep[e.g.,][]{nidever2008,besla2010}. Because of its low density, the diffuse synchrotron emission of the LMC may be strongly affected by these effects as well. Hence, the diffuse polarized emission of the LMC may be the closest analog of the observed polarized emission of Virgo cluster spirals under ram pressure effects \citep[see for example,][]{wezgowiec2007}. Tracing the magnetic field in the ram pressure and tidally affected LMC allows one to investigate how well magnetic field lines follow large scale gas flows. 

The earliest studies of magnetic fields in the Large Magellanic Cloud were conducted using optical polarization measurements towards LMC stars \citep{johnson1959,wolstencroft1962,visvanathan1966,mathewson1970}. These data suggested a  possible spiral field geometry emerging near the star-forming 30 Doradus region. Diffuse synchrotron emission from the LMC has been studied at 1.4, 2.45, 4.75 and 8.4 GHz by  \cite{haynes1991}, \cite{klein1993} and \cite{gaensler2005a}. Based on the 2.45 GHz polarization data of the LMC, \cite{klein1993} found a trailing spiral magnetic field pattern in the LMC without correcting for Faraday rotation effects.  The authors also found large RMs of opposite sign in two polarized filaments southeast of the galaxy. \cite{gaensler2005} carried out a Faraday rotation measure (RM) study of extragalactic polarized sources (EGSs) behind the LMC to probe its large-scale coherent magnetic field. The authors found that the EGS RMs exhibit a sinusoidal variation with an amplitude of 50 rad$^{-2}$. This result is consistent with a 1 $\mu$G coherent azimuthal magnetic field in the disk of the LMC. The existence of $\mu$G coherent large-scale field in an irregular galaxy like the LMC suggests that instead the \cite{parker1992} type cosmic-ray driven dynamo could be in operation. Moreover, a $|$RM$|$ of 50 rad$^{-2}$ through the LMC disk \citep{gaensler2005} casts doubt on the validity of the  \cite{klein1993}  findings as polarization vectors can be rotated by as much as  40$^\circ$ at  2.45 GHz by a rotation measure of 50 rad$^{-2}$. The demonstration of an azimuthal field in the LMC disk in \cite{gaensler2005} is based solely on the line-of-sight component of the large scale field. If the Milky Way foreground polarized emission in the direction of the LMC can be accounted for, the plane-of-the-sky magnetic field component, as probed by polarized synchrotron emission, can be measured to completely characterize the large scale magnetic field in the LMC. 

Although diffuse polarized emission has been detected in the interferometer data used by \cite{gaensler2005a}, those observations lack the shortest spacing flux in Stokes Q and U as well as in total intensity. Extracting magnetic field information from diffuse polarized emission that lacks zero-spacing information can lead to incorrect interpretations of the magnetic field properties \citep{gaensler2001,haverkorn2004}. Single dish data can be used to fill in the missing short spacing flux. Unfortunately, the existing single dish polarization data at 1.4 GHz cover less than half of the LMC due to a faulty polarimeter \citep{klein1993}. Moreover, the single frequency polarization survey of  \cite{klein1993} cannot facilitate reliable rotation measure studies due to the lack of multiple closely-spaced frequency channels. Therefore, we have mapped the LMC at 1.4 GHz using the Parkes radio telescope with closely spaced frequency channels. We have combined these data with the ATCA data of \cite{gaensler2005a,gaensler2005} to fill in the zero spacing flux and to enable reliable rotation measure studies of diffuse emission.

In this paper, we present a study of polarization in the LMC on multiple spatial scales: we combine archival interferometric data \citep{kim1998,gaensler2005a,gaensler2005} with newly acquired single dish Parkes data to characterize the LMC's 3-D magnetic field. In Section~\ref{section:observations}, we describe observations and data reductions for both the single dish and interferometric data sets. We describe the large scale and small scale total intensity and polarization properties of the LMC in Section~\ref{subsection:thermal_nonthermal}. In Section~\ref{section:rm_determination}, we present the Faraday rotation measure of the diffuse emission and that of  the extragalactic background sources. The LMC's global magnetic field symmetry is discussed in Section~\ref{section:global_bfield}. We examine the validity of magnetic field equipartition in Section~\ref{section:eq_B_cr}. In Section~\ref{section:filaments}, we determine the magnetic field structure of the brightly polarized filaments in the southeastern part of the LMC. Finally, we discuss mechanisms that could produce the observed diffuse synchrotron polarization on large and small scales in Section~\ref{section:origin_large_scale} and Section~ \ref{section:origin_filaments}, respectively.

\section{Observations and Data Reduction}
\label{section:observations}
\subsection{Single dish Observations at 1.4 GHz}
Single dish observations of the LMC were conducted at the Parkes Radio Telescope over the period 2006 November 20 to November 27 using the H-OH receiver and the pulsar digital filterbank (PDFB1) with a total bandwidth of 256 MHz, which consisted of 2049 0.125-MHz-wide channels, centered on 1384 MHz. A quarter wave-plate was inserted in the circular waveguide between the feedhorn and the ortho-mode transducer to enable the measurement of right- and left-hand circular polarizations. Observing with circular rather than the native linear polarized feeds puts correlated errors into Stokes V, rather than into Stokes Q and U. All the observations were conducted between sunset and sunrise to minimize solar interference and ionospheric Faraday rotation.

The source PKS B1934-638, whose flux at 1384 MHz was assumed to be 14.94 Jy \citep{reynolds1994} was observed at the beginning of each night, and was used as the absolute flux and bandpass calibrator. The source 3C 138, whose polarization properties (fractional polarization of 8.4 \%  and a polarization angle of -16.5$^\circ$ at 1384 MHz) were determined at the ATCA, was used for polarization calibration. For every observing session, we conducted 10 scans across 3C 138 at a range of parallactic angles by rotating the feed from  -45$^\circ$ to +45$^\circ$ between individual scans. This allowed us to simultaneously solve for the polarimetric response as well as the Stokes parameters of 3C 138. To cross-check the calibration solution derived from 3C 138, another polarization calibrator; PKS 0637-75, was observed as well.

With the Parkes Radio Telescope, an area of 13$^\circ$$\times$14$^\circ$, centered on RA (J2000) = 5h20m, DEC (J2000) = $-$68d44m (the field covered was identical to the \citealt{kim1998} ATCA observations) was scanned along each of the RA and the DEC directions at a rate of 3.5$^\circ$/min with a 1 second sampling interval. Adjacent scans were spaced by 7 arcmin in order to achieve Nyquist sampling of the 14.4-arcmin beam at 1.4 GHz. Two full coverages of the area scanning in right ascension and three scanning in declination were obtained. The total observing time was approximately 70 hours including overheads and calibration.  In order to estimate and subsequently remove the contribution of ground emission to the observed Stokes parameters, we carried out zenith scans at the beginning of each observing run  between elevation angles 32$^\circ$ and 82$^\circ$ at various azimuth angles.

Calibration of the data was carried out using the Parkes Continuum Polarimetry Software\footnotemark[1]\footnotetext[1]{http://svn.atnf.csiro.au/trac/parkespol} developed by McConnell \& McClure-Griffiths. After bandpass calibration, channels affected by radio frequency interference (RFI) and edge channels affected by the poor performance of the quarter-wave plate were flagged and the 0.125MHz channels were rebinned into 8-MHz wide channels centered at 1384 MHz to match the frequency configuration of the ATCA observations (see Section~\ref{subsection:atca_obs}). The polarization calibration is based on the method outlined by \cite{johnston2002} -- instrumental polarization terms in the Mueller matrix and the properties of the polarized calibrator were solved simultaneously by observing the polarized calibrator multiple times at different feed angles.  Ground contribution to each Stokes parameter at each frequency channel was estimated by fitting a 3rd order polynomial to the average zenith scan profile as a function of elevation angle. The shape of the ground emission at various frequencies was stable over the observation while the offset was weakly time and azimuth dependent. A linear baseline was subtracted from each scan assuming that the emission at the edge of the scan was zero. This removes the absolute largest-scale structures in the maps. Elevation-dependent ground emission contribution was then removed.

We used an IDL map making routine based on that written to process the Parkes Galactic Meridian Survey (PGMS) data \citep{carretti2010}. First, Stokes I, Q, U channel maps of single direction scans were created. Then maps in orthogonal directions were combined in Fourier space using the algorithm described by \cite{emerson1988}. Raster noise was reduced by down-weighting regions in Fourier plane that corresponded to scales comparable to the scan width. The final noise in the channel maps was dominated by calibration errors and imperfect subtraction of ground emission (which can fluctuate as much as 2.5 times the theoretical noise; \citealt{carretti2010}) rather than by the instrumental sensitivity.

\begin{figure}
\centering
\subfigure{
\epsscale{0.42}
\plotone{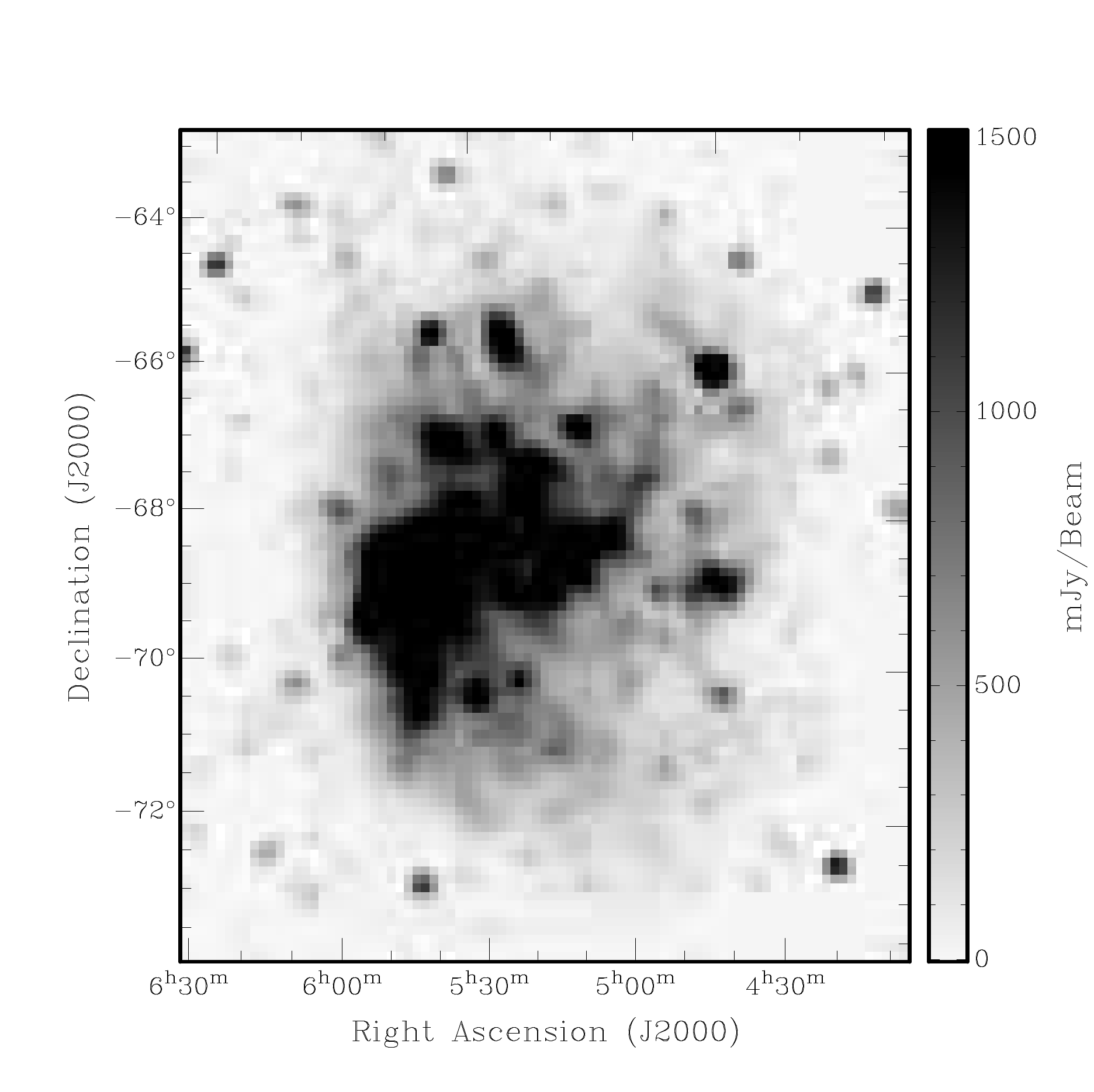}
}
\hspace{.1in}
\subfigure{
\epsscale{0.42}
\plotone{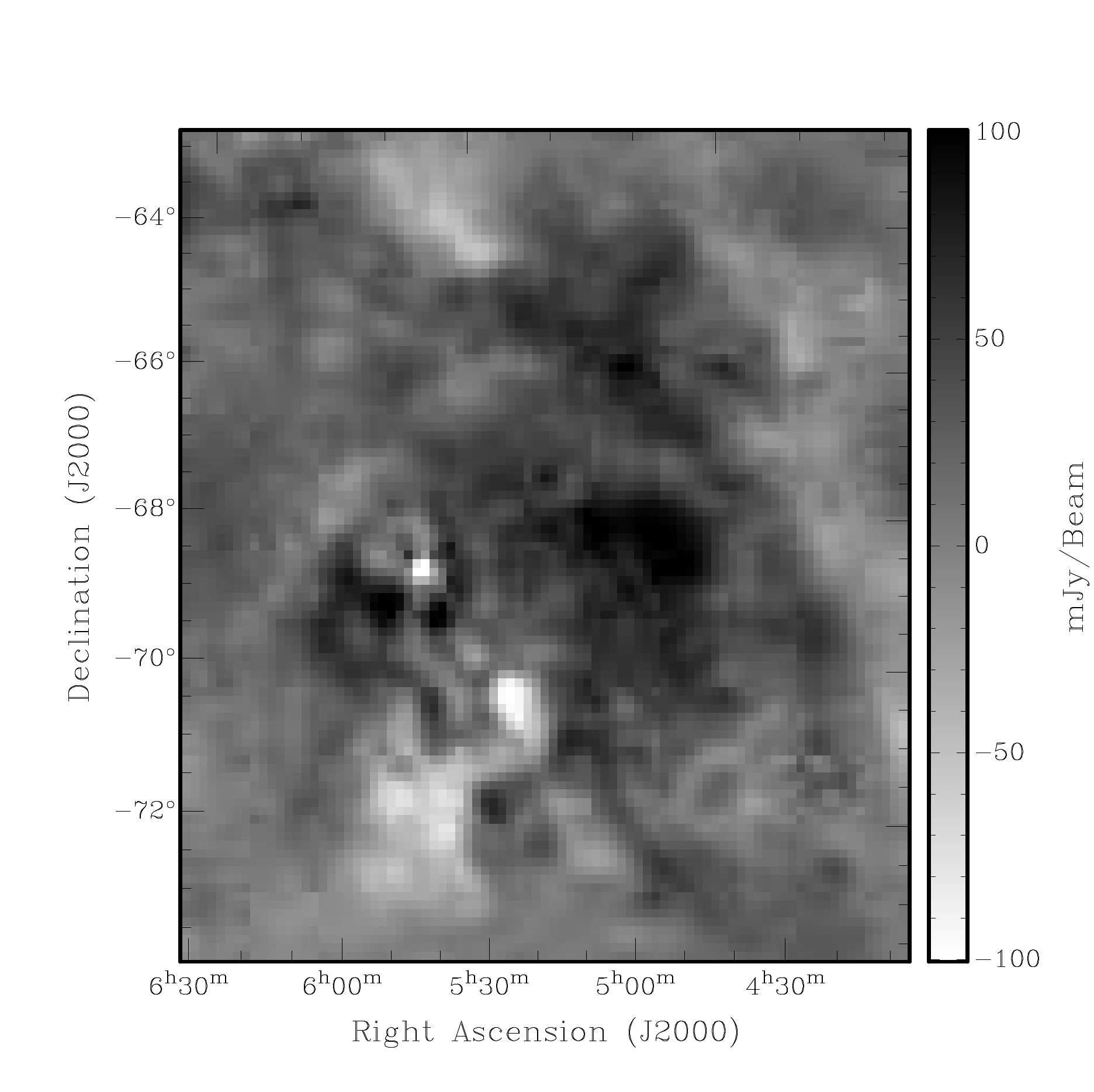}
}
\vspace{.1in}
\subfigure{
\epsscale{0.42}
\plotone{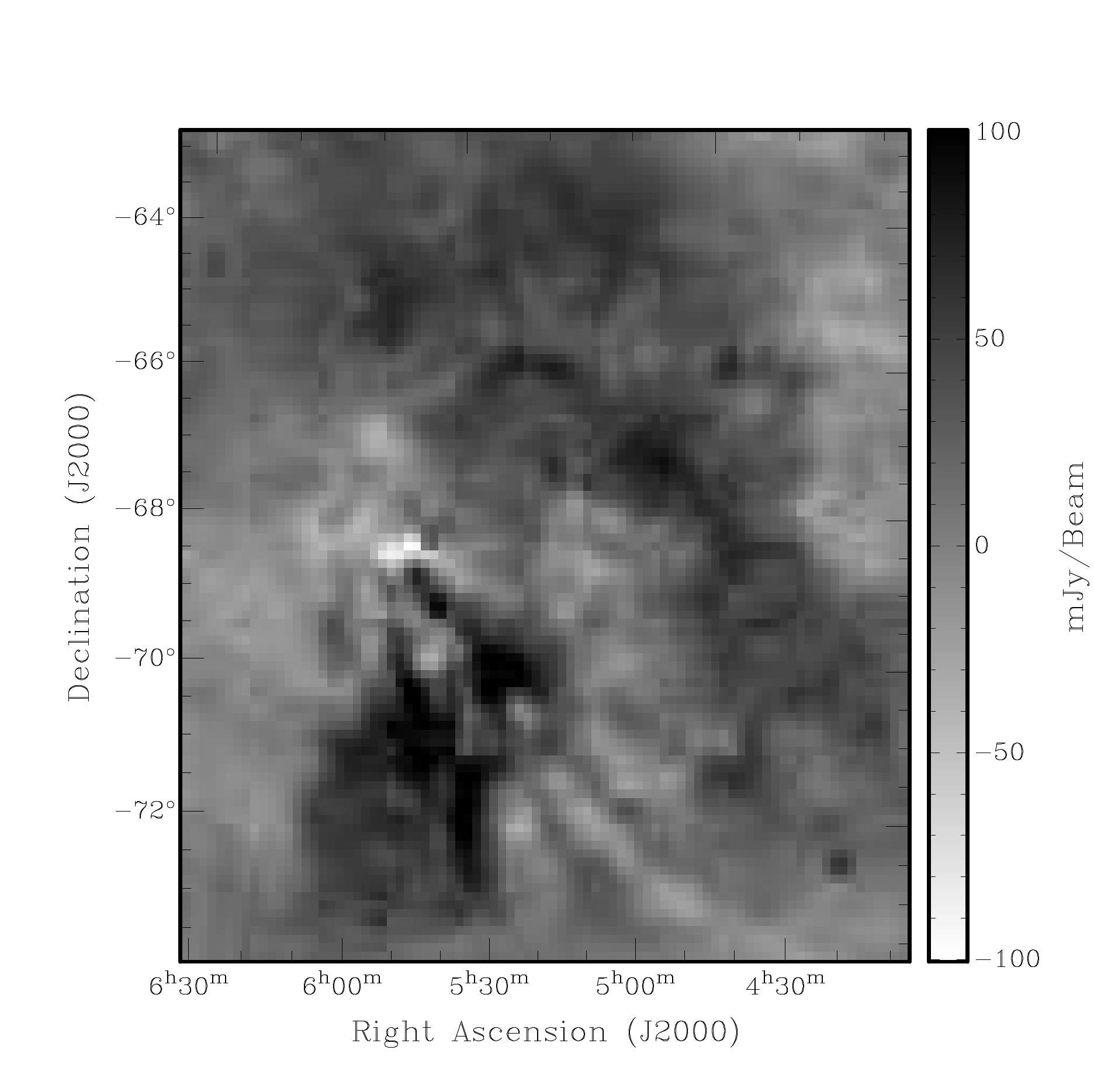}
}
\hspace{.1in}
\subfigure{
\epsscale{0.42}
\plotone{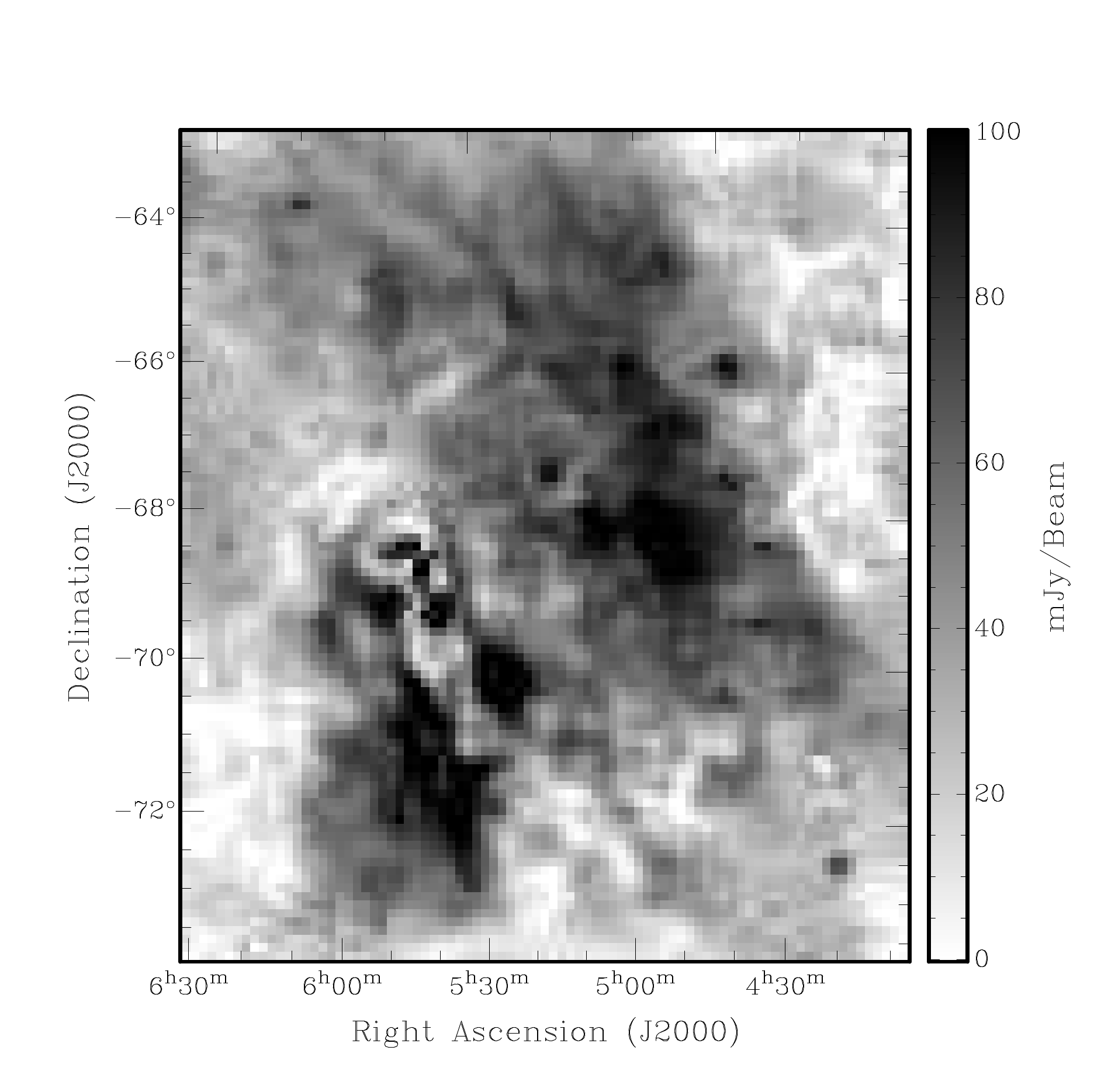}

}

\caption{Stokes I (upper left panel), Q (upper right panel), U (lower left panel) and de-biased linearly polarized intensity (lower right panel) maps  of the LMC at 1328 MHz from Parkes single dish observations.}
\label{fig:ParkesIQUV}
\end{figure}

The sensitivity in each final Stokes I channel map was $\sim$ 0.025 Jy/beam and that in Stokes Q and U channel maps was  $\sim$ 0.01Jy/beam at a resolution of $\sim$14'. Figure~\ref{fig:ParkesIQUV} shows  the Stokes I, Q, U maps and the corresponding de-biased polarized intensity image\footnotemark[2]\footnotetext[2]{The polarized intensity is de-biased to first order by subtracting the noise in individual Stokes Q or U channel maps from the polarized intensity in quadrature \citep{simmons1985}.} of the 8-MHz wide channel centered at 1328 MHz. It is clear that polarized emission is detected across the entire field of view. We note that even though the map-making routine has down-weighted raster noise, there still exist stripe-like artifacts in the channel maps with a maximum polarized intensity of $\sim$ 0.045 Jy/beam, which is evident for example in lower left hand corner of panels in Figure~\ref{fig:ParkesIQUV} . As a result, we only consider emission with polarized intensity larger than 0.045 Jy/beam to be physical.

\subsection{Interferometric Observations at 1.4 GHz}
\label{subsection:atca_obs}
\begin{figure}
\centering
\epsscale{0.8}
\plotone{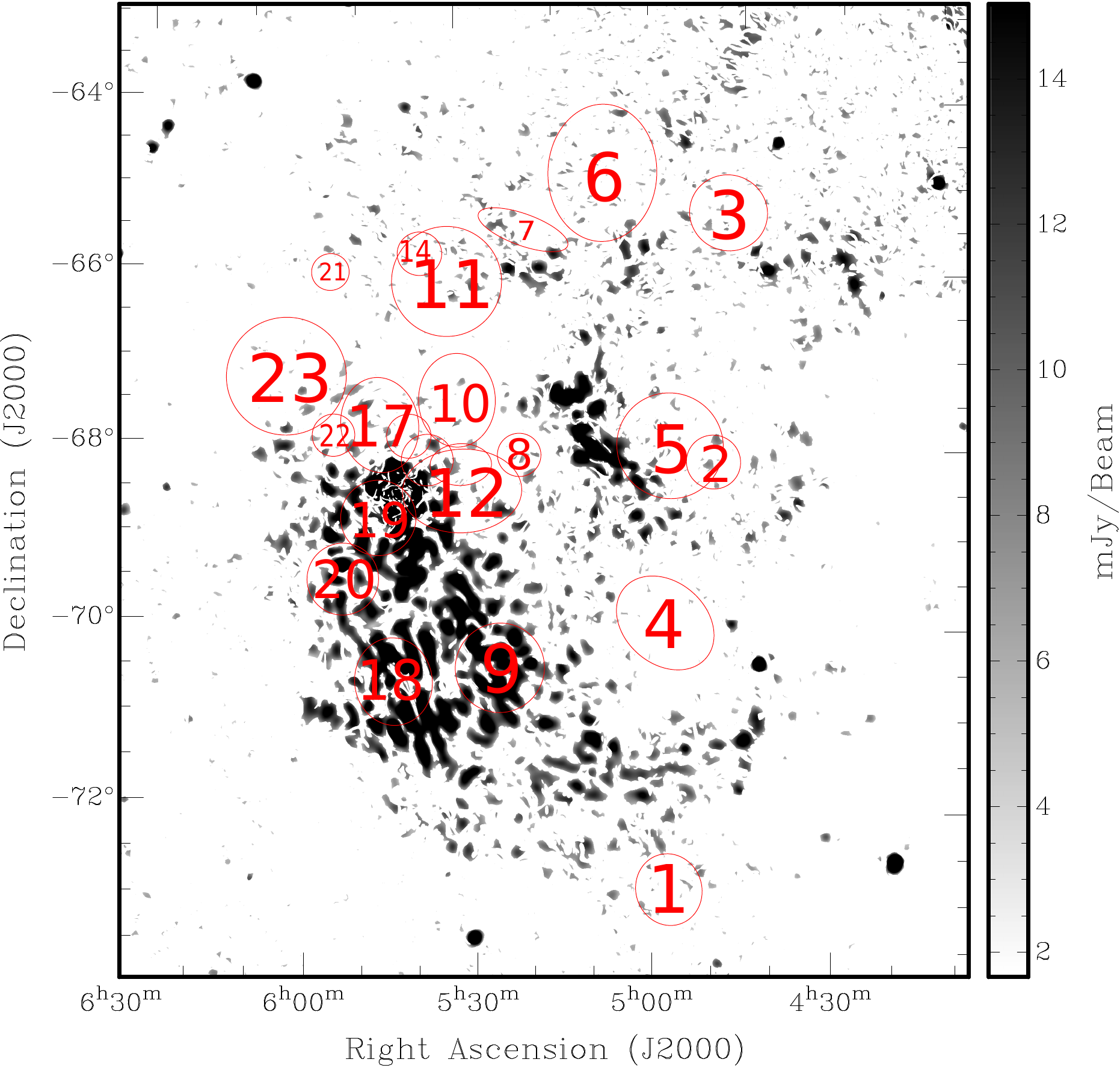}
\caption{A de-biased linearly polarized intensity map of the LMC at 1.4 GHz, made from ATCA mosaic observations. The map has been smoothed to a resolution of $6'$. The locations, dimensions and catalog numbers of supergiant shells defined by \cite{kim1999} are super-imposed on the image.}
\label{fig:atca_pi}

\end{figure}

Continuum polarimetric data for the LMC at 1.4 GHz were recorded as part of the LMC Australia Telescope Compact Array (ATCA) HI survey. The details of the observations were described by \cite{kim1998}. To summarize, the LMC was observed in mosaic mode consisting of 1300 pointings over the period 1994 October to 1996 February. Full polarization was recorded in 32 adjacent frequency channels each of bandwidth 4 MHz centered on 1384 MHz. The standard primary calibrator PKS B1934--638 was used to determine the absolute flux density and polarization leakages. A secondary calibrator (PKS B0407--68 or PKS B0454--810) was observed every 30 minutes to calibrate the time-dependent antenna gains. The MIRIAD package was used for data reduction \citep{sault2003}. Data were first flagged and calibrated. Flagging and rebinning the 32 4 MHz-wide channels resulted in 13 8-MHz wide channels. For each frequency channel, mosaiced images were formed in Stokes Q and U with a pixel size of 13 $\times$ 13 sq. arcsec. Two sets of channel maps were made for different purposes:  for rotation measure analysis of extragalactic point sources, we discarded visibilities from the shortest baselines ($<$ 100 m) so that diffuse polarized emission would not be imaged. For the purpose of analyzing diffuse polarized emission, channel maps were made without the longest baselines (without antenna 6) to increase surface-brightness sensitivity. In both cases, super-uniform weighting was used to suppress sidelobes. The images were then deconvolved using the task PMOSMEM, a maximum entropy algorithm that jointly deconvolves all Stokes parameters simultaneously \citep{sault1999}. Finally, the maps were restored using a $\sim$40$''$$\times$40$''$ Gaussian beam. After primary beam corrections, the sensitivity across mosaiced channel maps is not uniform, with a mean of roughly  0.7 mJy/beam. The final channel maps have spatial sensitivity from scales of 40$"$ to 30$'$. To ensure that no polarized emission suffer from bandwidth depolarization,  a de-biased linearly polarized intensity map was made at each frequency channel before averaging them together to make a single polarization map of the LMC. The de-biased polarized intensity map made without antenna 6, after smoothing to a resolution of 6 arcmin, is shown in Figure~\ref{fig:atca_pi}. The false-discovery rate algorithm SFIND developed by \cite{hopkins2002} was used to identify point sources in the de-biased polarized intensity map made without the shortest spacing. We reuse the same  324 polarized point sources identified by \citep{gaensler2005}. These sources have linear fractional polarizations between 0.3 and 50  \% and were confirmed to not correspond to catalogued pulsars or supernova remnants.

\subsection{Combining interferometric and single dish data}
The Parkes data were combined with the ATCA data to fill in the zero and short spacing flux absent in the interferometric observations. Even though mosaicing allows the recovery of flux on scales larger than that correspond to the shortest antenna spacing (it effectively reduces the shortest baseline by the diameter of the antenna), the largest scale flux still remains missing. Before adding in the short-spacing data, it is necessary to ensure that the Parkes and ATCA data are on the same flux density scales \citep[see][]{stanimirovic2002}. The calibration scale factor, defined as the ratio of flux densities of an unresolved compact source in the single-dish map to that in the interferometric map, can be found using several different methods. The most straightforward way is to compare directly fluxes of point sources far away from the diffuse emission. An alternative way is to compare the deconvolved Parkes (dividing the Fourier transform of the single dish data by the Fourier transform of the single-dish beam) and ATCA visibilities in the region of overlapping spatial frequencies. One can also compare the Fourier transform of the ATCA image convolved with the Parkes beam with the Fourier transform of the Parkes image in the region of overlapping spatial frequencies. All three methods yield similar calibration scale factors very close to 1. We choose to use the third method above, as implemented in the MIRIAD task IMMERGE. This task requires one to supply the beam information of the single dish data. We have fitted 2D gaussians to Stokes I point sources in the field of view to determine the effective beam size in each frequency channel. These values are listed in column 2 of Table ~\ref{table:cal_factor}. The calibration scale factor for each frequency channel is found by comparing the Stokes I Parkes and ATCA channel maps in the $uv$ annulus from 120 to 170 k$\lambda$, a region in the Fourier plane that is well sampled by both single dish and interferometric observations.  When determining the scale factor, we have masked out the 30 Doradus region where sidelobes, leakage and other artifacts in the ATCA total intensity channel map might result in systematics. The same scale factors are then used to combine the Stokes Q and U channel maps. Besides the channel maps, we have also performed similar single-dish and interferometric combination using the multi-frequency synthesis ATCA Stokes I map at 1.4 GHz, produced by a peeling algorithm to remove ring-like deconvolved artifacts \citep{hughes2007} and our Parkes Stokes I map to obtain a radio continuum map of the LMC sensitive to all scales. The calibration scale factor used for the  \cite{hughes2007} and our Stokes I combination procedure is determined to be 1.037.  The noise in the final combined Stokes Q and U channel maps is dominated by the ATCA data (with a sensitivity $\sim$ 0.7 mJy/ATCA beam). The ATCA and Parkes combined maps are sensitive to scales ranging from the resolution of $\sim$ 40$"$ up to the size of the entire observed field. The combined total and polarized intensity maps of the LMC smoothed to a resolution of 3' are shown in Figures~\ref{fig:sb_on_I} and ~\ref{fig:combined_pi_3am}, respectively.

\section{Results}

\label{subsection:thermal_nonthermal}
We first discuss the morphology of the combined ATCA and Parkes total intensity map shown in Figure~\ref{fig:sb_on_I}. Radio continuum emission of galaxies at centimeter wavelengths consists of both free-free emission (with a flat spectral index $\beta$\footnotemark[3]\footnotetext[3]{Throughout the chatper, we define the spectral index as $S\propto\nu^\beta$} of -0.1) and synchrotron emission with a steeper spectral index, typically $\sim$ $-$0.8 \citep{condon1992}. As \cite{hughes2007} has pointed out, the morphological similarities between the 1.4 GHz continuum map and the Southern H-Alpha Sky Survey Atlas (SHASSA) H$\alpha$ map of the LMC \citep{gaustad2001} suggests a large thermal fraction at  1.4 GHz. 

Ideally, one could construct a free-free radio template of the LMC using the SHASSA H$\alpha$ map. However, an accurate separation of the thermal and non-thermal emission from the LMC requires the knowledge of line-of-sight dust distribution in the LMC to correct H$\alpha$ for extinction as well as information on the electron temperatures in HII regions and in the diffuse medium. Since these quantities are poorly known, we choose to adopt an overall non-thermal flux of 14.5 Jy at 10 GHz and a non-thermal spectral index of -0.7 \citep{israel2010}. This corresponds to mean non-thermal fractions of 0.39, 0.58 and 0.63 at  4.75 GHz, 2.45 GHz and 1.4 GHz respectively.

\subsection{Total Intensity Morphology of the LMC at 1.4 GHz} 

\begin{figure}
\centering
\plotone{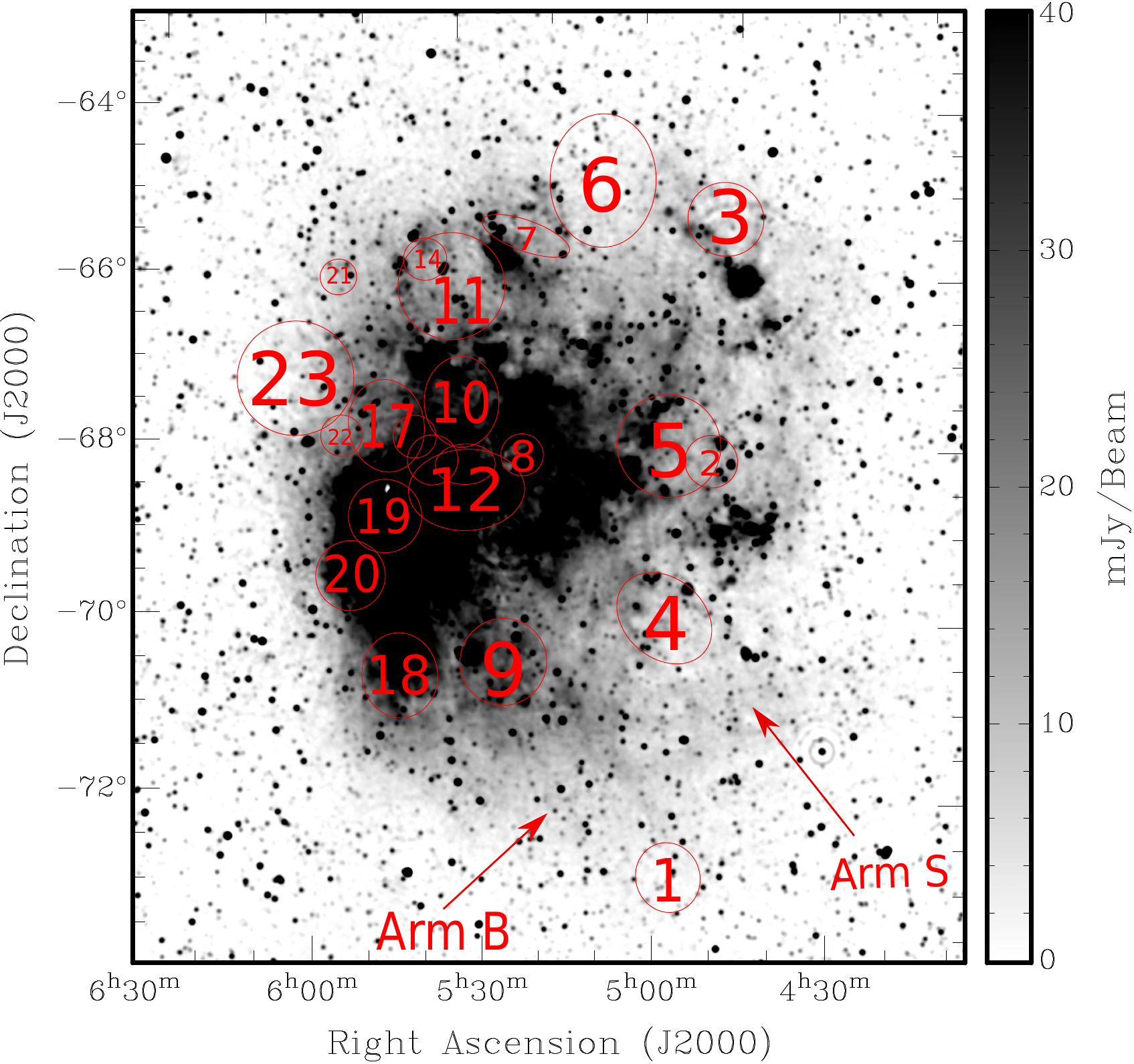}
\caption{Total intensity map of the LMC from the combined Parkes and ATCA data at 1.4 GHz, smoothed to a resolution of 3'. The locations of supergiant shells defined by \cite{kim1999} are overlaid. The locations of HI Arm B and Arm S as defined by \cite{staveleysmith2003} are overplotted.}
\label{fig:sb_on_I}
\end{figure}

\begin{figure}
\centering
\plotone{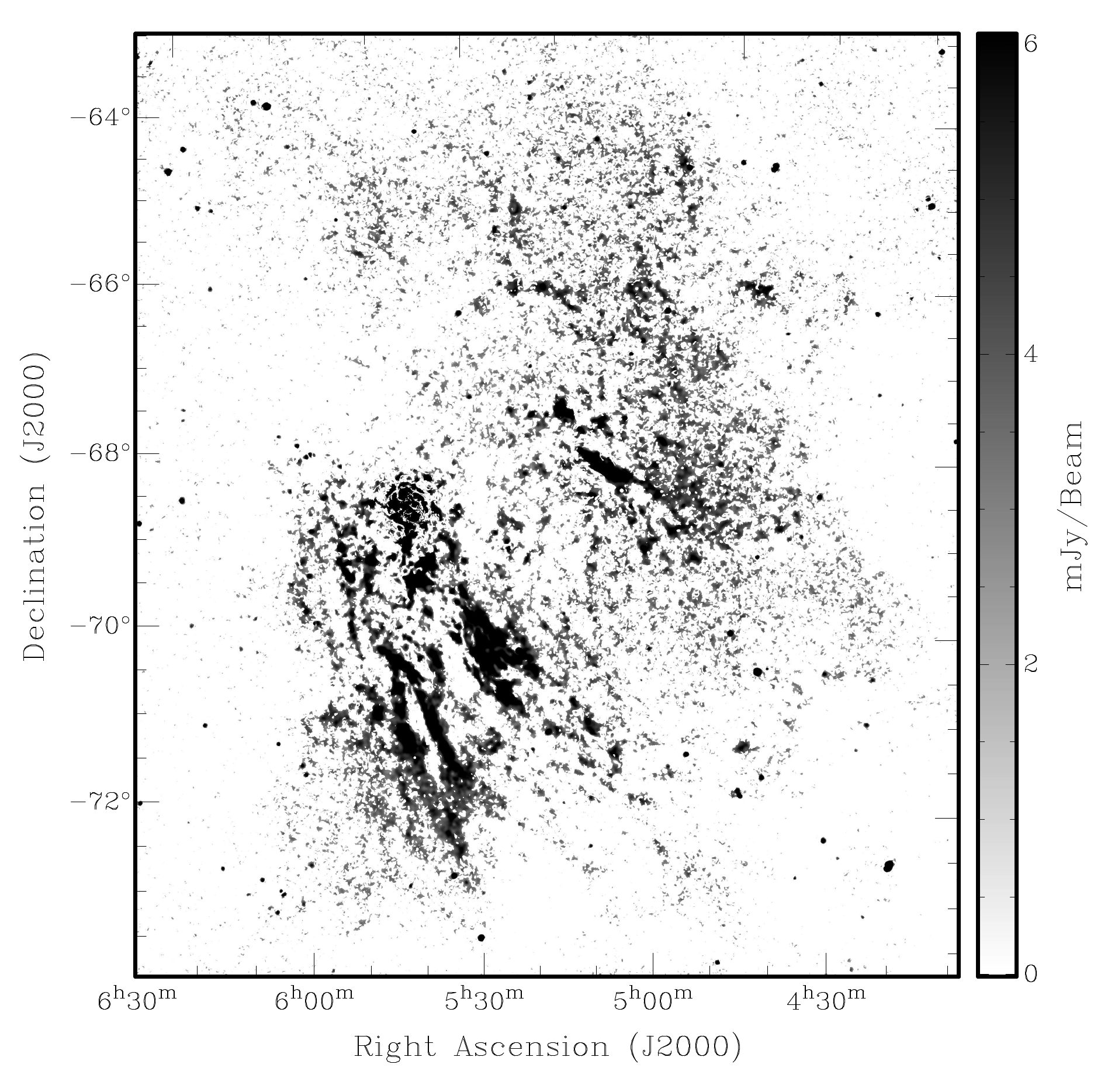}
\caption{De-biased linearly polarized intensity map of the LMC using the combined Parkes  and ATCA data at 1.4 GHz, smoothed to a resolution of 3'.}
\label{fig:combined_pi_3am}

\end{figure}

To maximize our sensitivity to diffuse emission, we have smoothed the combined ATCA and Parkes Stokes I map of the LMC to an angular resolution of 3'. Figure~\ref{fig:sb_on_I} clearly demonstrates that the LMC's total intensity at 1.4 GHz is dominated by the star-forming 30 Doradus region near its eastern edge, with decreasing intensity away from it. Bright HII regions with strong thermal emission are also visible in radio continuum \citep[see also][]{hughes2007}.

As was pointed out by \cite{klein1993}, the total intensity image of the LMC at 1.4 GHz resembles its HI column density distribution. Supergiant HI shells LMC SGS 4 and SGS 6 identified by \cite{kim1999} appear as low emission regions in the 1.4 GHz radio continuum map (see Figure~\ref{fig:sb_on_I}). The association of radio continuum holes with neutral hydrogen holes is rare in external galaxies and has only been observed in the face-on grand-design spiral NGC 6946 \citep{beck2007,braun2007}. The lack of total synchrotron intensity interior to the two supergiant shells SGS 4 and SGS 6 could reflect low cosmic ray electron densities and/or magnetic field strengths. ``Arm S" and ``Arm B", HI arms in the southern part of the LMC's main body \cite[as defined by][]{staveleysmith2003} are also faintly visible in radio continuum emission (Figure ~\ref{fig:sb_on_I}) with little coinciding H$\alpha$ emission. Hence, most of the 1.4 GHz continuum emission from Arm S and B is likely to be non-thermal. 

\subsection{Polarized emission from the LMC at 1.4 GHz} 
\label{subsection:large_scale_pi}

\begin{figure}
\centering
\plotone{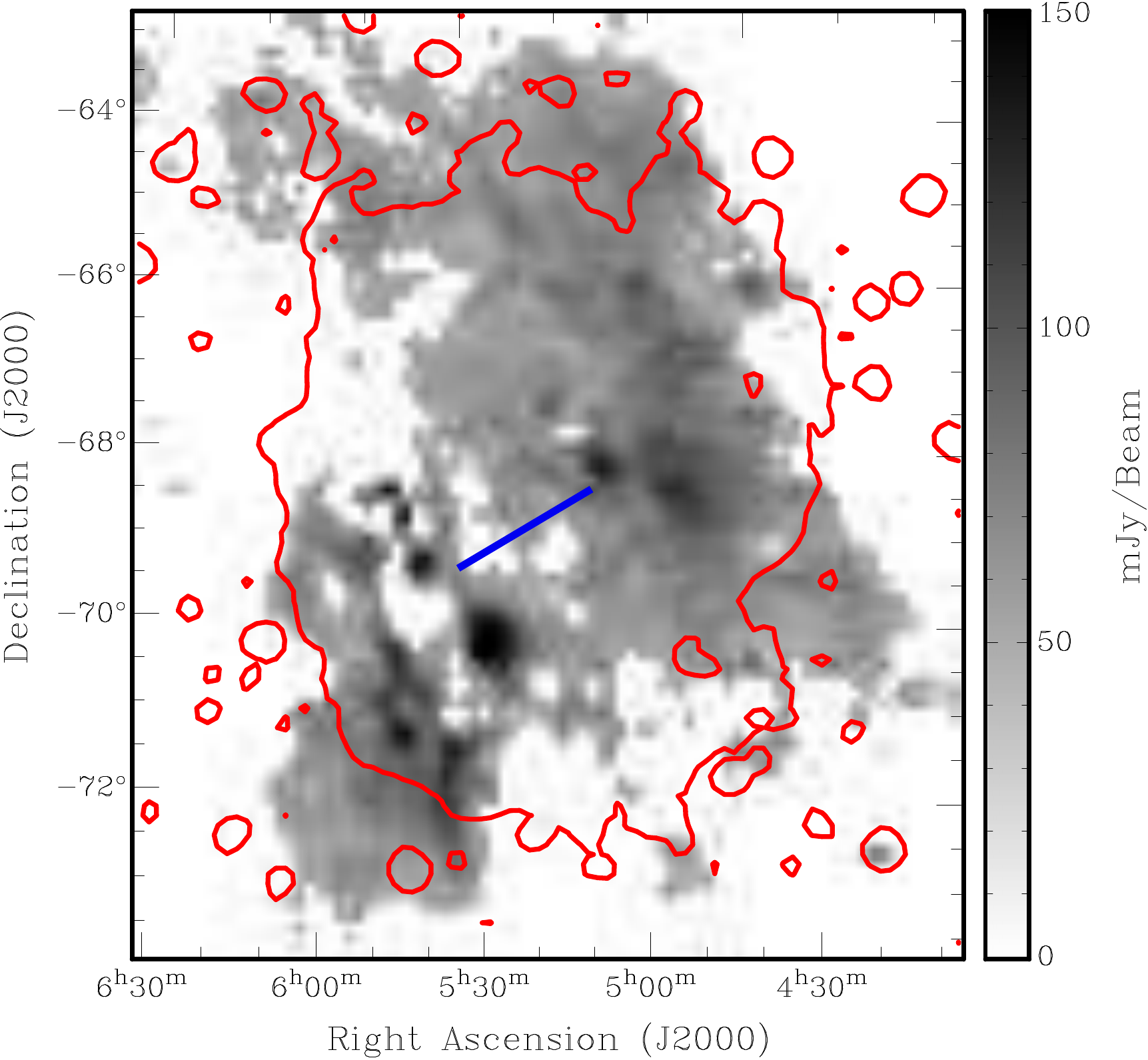}
\caption{Parkes-only linearly polarized intensity at 1328 MHz with the Parkes-only Stokes I contour at 0.15 Jy/beam level overlaid in red.  Pixels with polarized intensity lower than 0.045 Jy/beam are masked. The rough location of the LMC's optical bar is indicated by the blue line segment. }
\label{fig:I_on_PI_contour}
\end{figure}

Unlike total intensity, diffuse polarized emission in our Parkes and ATCA combined map does not resemble the galaxy's HI column density distribution. At 1.4 GHz, the diffuse polarized emission from the LMC is dominated by two filaments south of 30 Doradus region, near the eastern edge of the optical bar (see Figure~\ref{fig:combined_pi_3am}, \ref{fig:I_on_PI_contour}). The longer eastern filament extends from (5$^h$43$^m$, $-$70$^d$30$^m$) to (5$^h$34$^m$, $-$73$^d$15$^m$) while the shorter western filament extends from (5$^h$31$^m$, $-$70$^d$20$^m$) to (5$^h$23$^m$, $-$71$^d$21$^m$). The filaments were identified in earlier surveys conducted by \cite{haynes1991}, \cite{klein1993} and \cite{gaensler2005}. We will discuss the polarization properties, magnetic field structure and the possible origins of the polarized filaments in  Section~\ref{section:origin_filaments}. We also observe a brightly polarized region near the western edge of the optical bar (see Figure~\ref{fig:I_on_PI_contour}). More generally, diffuse emission is observed across most of the LMC's main body, particularly in the western half of the galaxy. 

\begin{figure}
\centering
\plotone{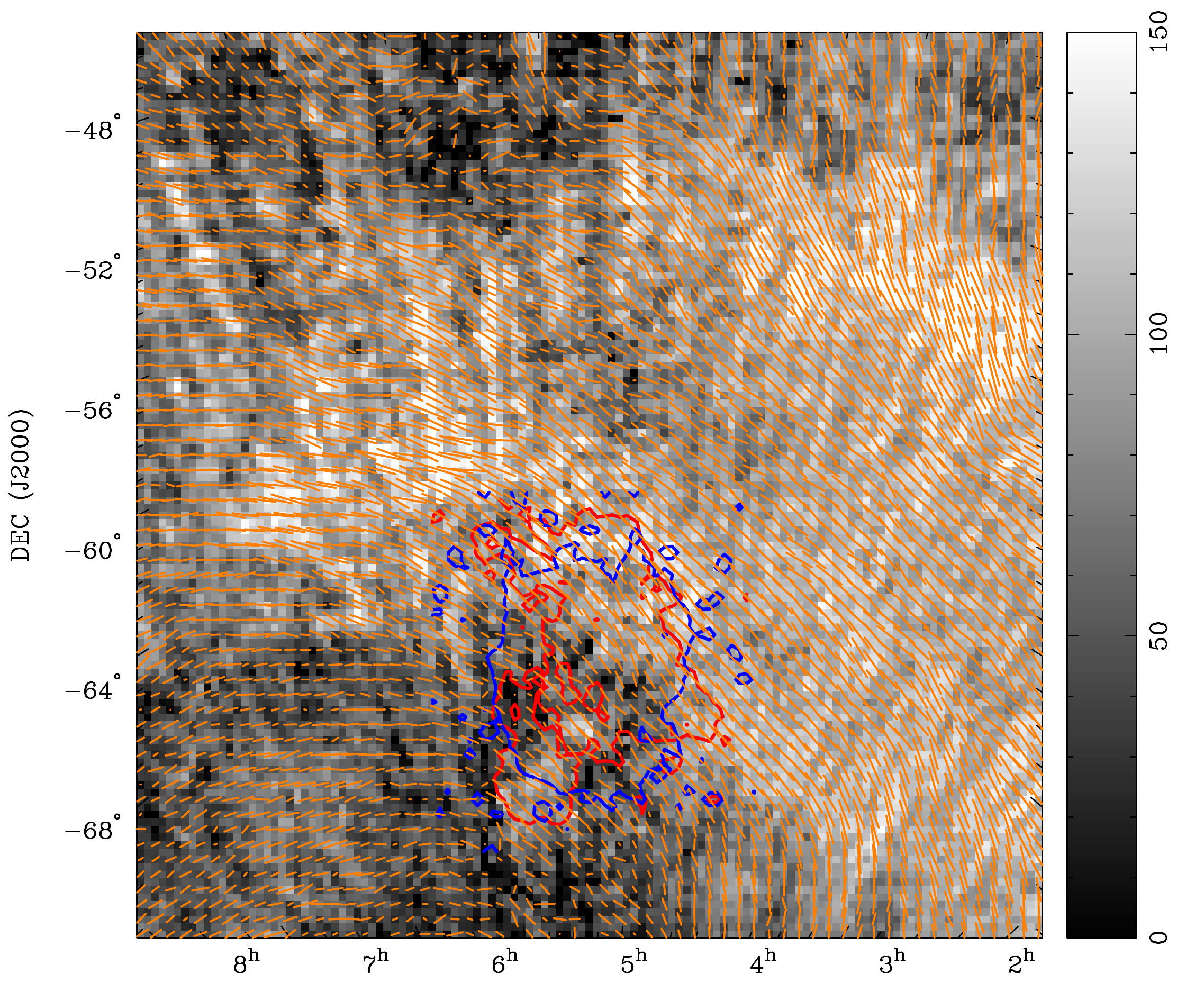}
\caption{The \cite{testori2008} polarized intensity map at 1.4 GHz (in units of brightness temperature mK) of a 30$^\circ$$\times$30$^\circ$ region around the LMC at $36'$ resolution. The Stokes I contour at 0.15 Jy/Parkes beam (blue) and polarized intensity contour at 0.045 Jy/Parkes beam level (red) of our Parkes observation at 1.4 GHz is overlaid. Polarization position angles of E-vectors recorded in the \cite{testori2008} survey are overplotted as red line segments. The intensity grey scale of this figure is reversed compared to earlier figures.}
\label{fig:testori_on_pi}
\end{figure}

We also detect significant polarized emission exterior to the main body of the LMC. This is demonstrated in Figure~\ref{fig:I_on_PI_contour}, where we have overlaid a Stokes I contour on a polarized intensity image of the LMC at 14' resolution. Substantial polarized emission exists north of the LMC. The same structure was evident in the 2.3 GHz polarized intensity map presented by \cite{klein1993}, but the authors did not comment on it. In Figure ~\ref{fig:testori_on_pi}, we have plotted the Parkes total and polarized intensity contours on a 30$^\circ$$\times$30$^\circ$ map of the absolutely calibrated 1.4 GHz polarization survey of the entire Southern sky of \cite{testori2008} at 36' resolution. The bright polarized feature north of the LMC seen in Figure~\ref{fig:I_on_PI_contour} appears to be part of a large polarized structure, extending from the southwest to the northeast, obstructing the northwestern half of the LMC. The same loop is evident in the total intensity map from the Continuum HI Parkes All Sky Survey (CHIPASS), an absolutely calibrated all sky survey at 1.4 GHz (Calabretta, priv. comm.), as well as in the preliminary S-band Polarized All Sky Survey (S-PASS) map at 2.3 GHz (Haverkorn, unpublished).  This suggests a foreground Milky Way polarized synchrotron emitting layer in the direction of the LMC. This Galactic polarized emission towards the LMC could be correlated with dust: starlight polarization measurements toward Milky Way foreground stars in the direction of the LMC indicate a similar magnetic field orientation as that inferred from radio polarization \citep{schmidt1976}. To properly remove the Milky Way foreground emission from our Parkes multi-wavelength data requires subtracting the foreground Stokes Q and U values from our Parkes channel maps. Unfortunately, the \cite{testori2008} survey consists of only a single frequency. Thus it is not possible to remove the Milky Way foreground emission from our Parkes data.

\section {Rotation Measure Determination}
\label{section:rm_determination}

Faraday rotation is a birefringence effect that occurs when linearly polarized light travels through a magnetized medium. The plane of polarization rotates through an angle $\Delta$$\psi$ (in radians) given by
\begin{equation}
\label{eq:angle_lambda}
\Delta\psi= {\rm RM}  \lambda^{2}
\end{equation}
where $\lambda$ is the wavelength of the radiation measured in meters and RM is the rotation measure, defined as
\begin{equation}
{\rm RM} =0.812 \int ^{observer}_{source} {n_{e}(l){B_{\parallel} (l)}} dl~~~\rm{rad~m^{-2}}
\label{eq:rmdef}
\end{equation}
\citep[e.g.,][]{rybicki1986}. In the above equation, $n_e(l)$ (in cm$^{-3}$) is the thermal electron density, $B_{\parallel}(l)$ (in $\mu$G) is the line of sight magnetic field strength and d$\it{l}$ (in pc) is a line element along the line of sight. The sign of the RM provides the direction of the average line of sight  magnetic field: a positive (negative) RM represents a field that is directed towards (away) from us.

We have computed rotation measures using RM Synthesis  \citep{brentjens2005} and RMCLEAN \citep{heald2009}, following the algorithm described by \cite{mao2010}. Recently, \cite{farnsworth2011} demonstrated that using RM synthesis alone might not be sufficient to determine the underlying Faraday structure, even in the simplest case of two components with different intrinsic polarization angles. This is because RM Synthesis does not have an equivalent of a reduced $\chi^2$ (as in the  least square fit of  polarization position angle as a function of $\lambda^2$) to measure the goodness of fit. As a result, the solution can  converge to the incorrect RM value. To ensure the reliability of RM found using the RM Synthesis technique, we have computed the reduced $\chi^2$ of the polarization angle against  $\lambda^2$ relation (Equation~\ref{eq:angle_lambda}). We only accept fits with a reduced $\chi^2$ $\le$ 2. In addition, we have examined the behavior of Stokes Q, U and the polarized intensity as a function of  $\lambda^2$ to ensure that the result from RM synthesis provides a reasonable fit. We have computed the RM from both the extended polarized emission of the LMC itself and from the extragalactic point sources in the surveyed field.

\subsection{RMs derived from the diffuse polarized emission }
\label{subsection:diff_rm_determination}

\begin{figure}
\centering{
\epsscale{0.6}
\plotone{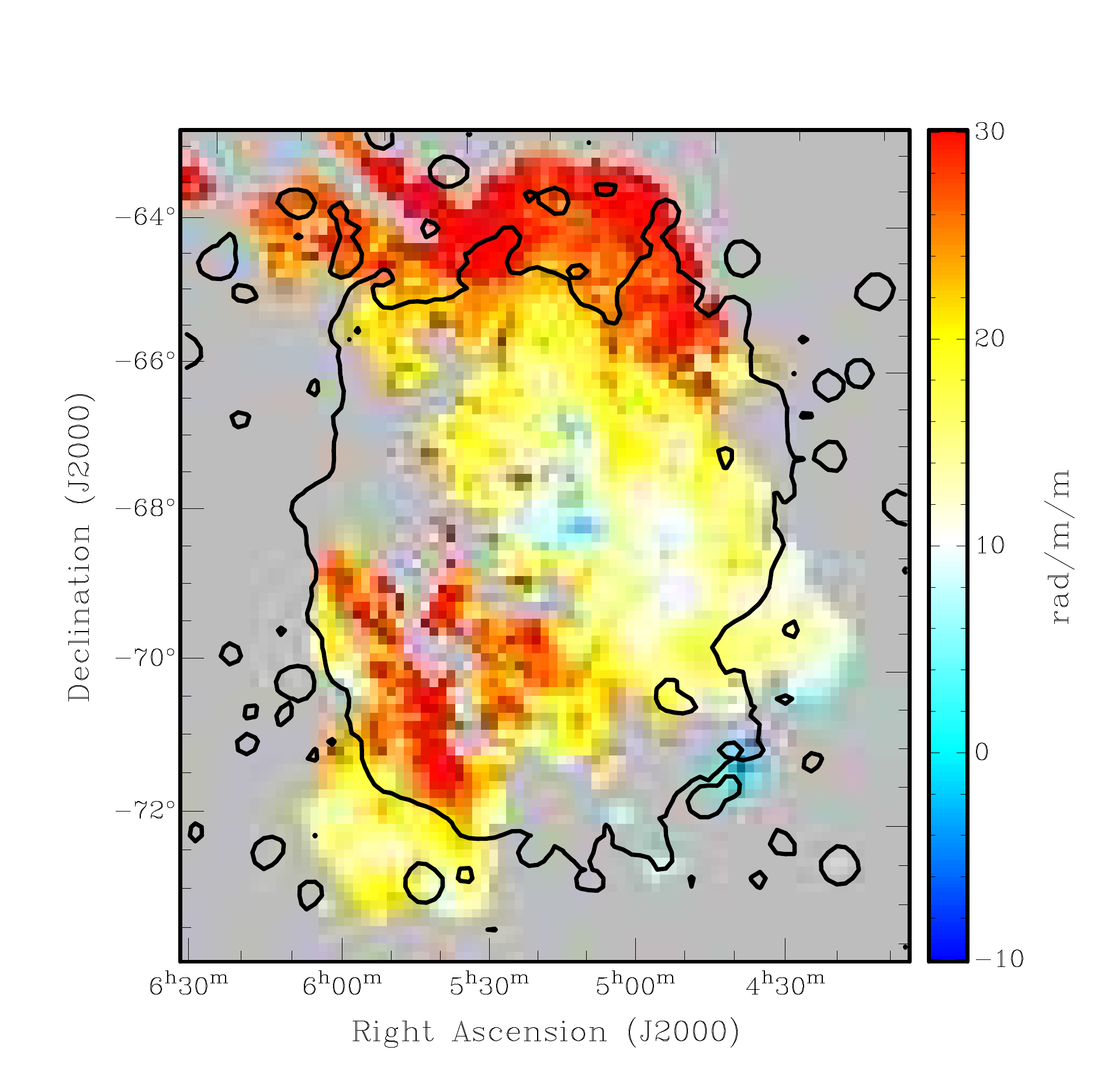}
\vspace{.1in}
\epsscale{0.6}
\plotone{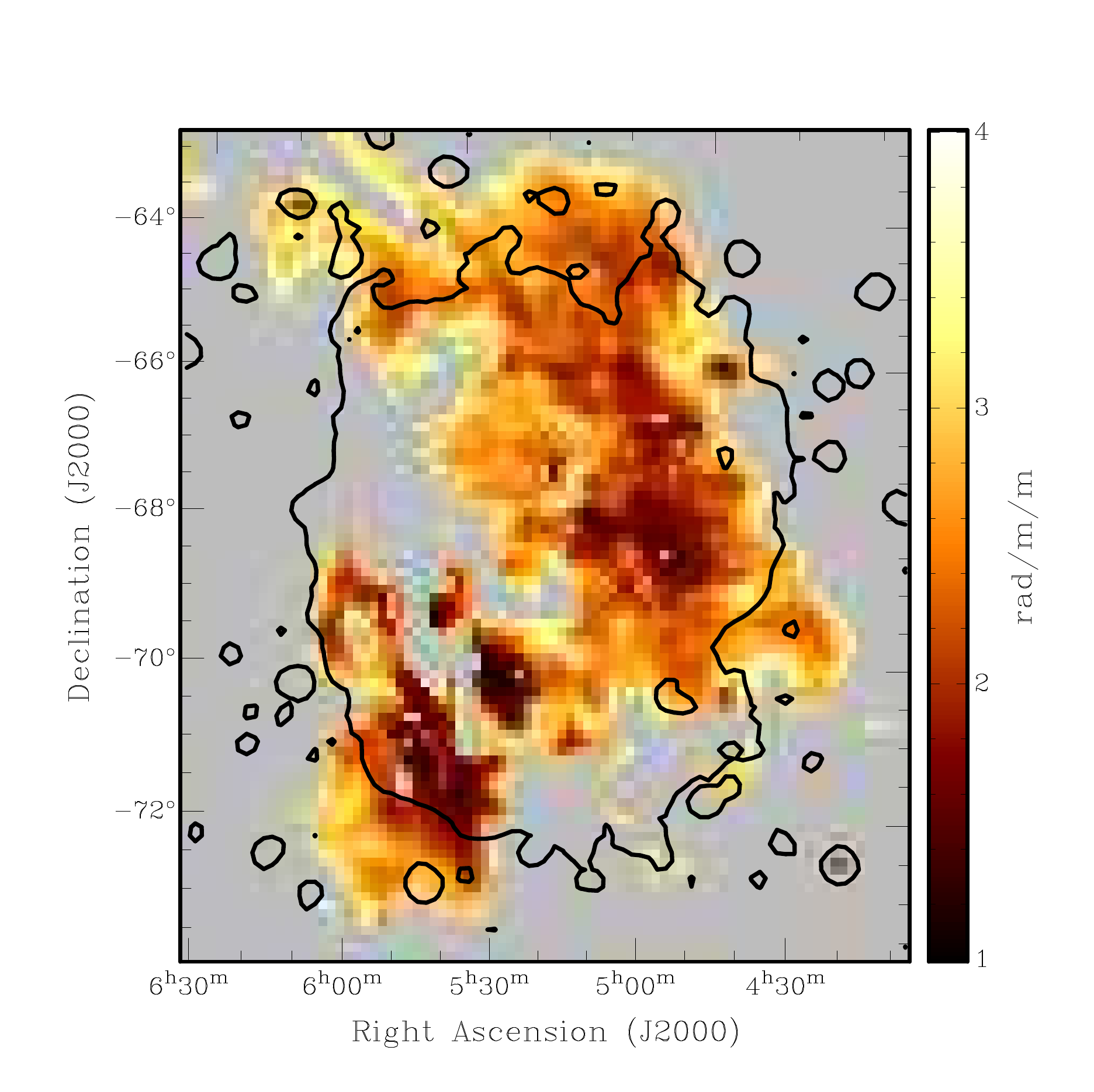}
}
\caption{Top panel: Faraday rotation measures derived from diffuse polarized emission at 1.4 GHz using the Parkes-only data. Pixels with a polarized intensity lower than 0.045 Jy/beam and a reduced $\chi^2$ greater than 2 have been blanked. Bottom panel: The error associated with the Faraday rotation map in the top panel. The black contour indicates the Parkes-only total intensity level at 0.15 Jy/beam.}
\label{fig:rm_parkes}

\end{figure}

\begin{figure}
\centering
\plotone{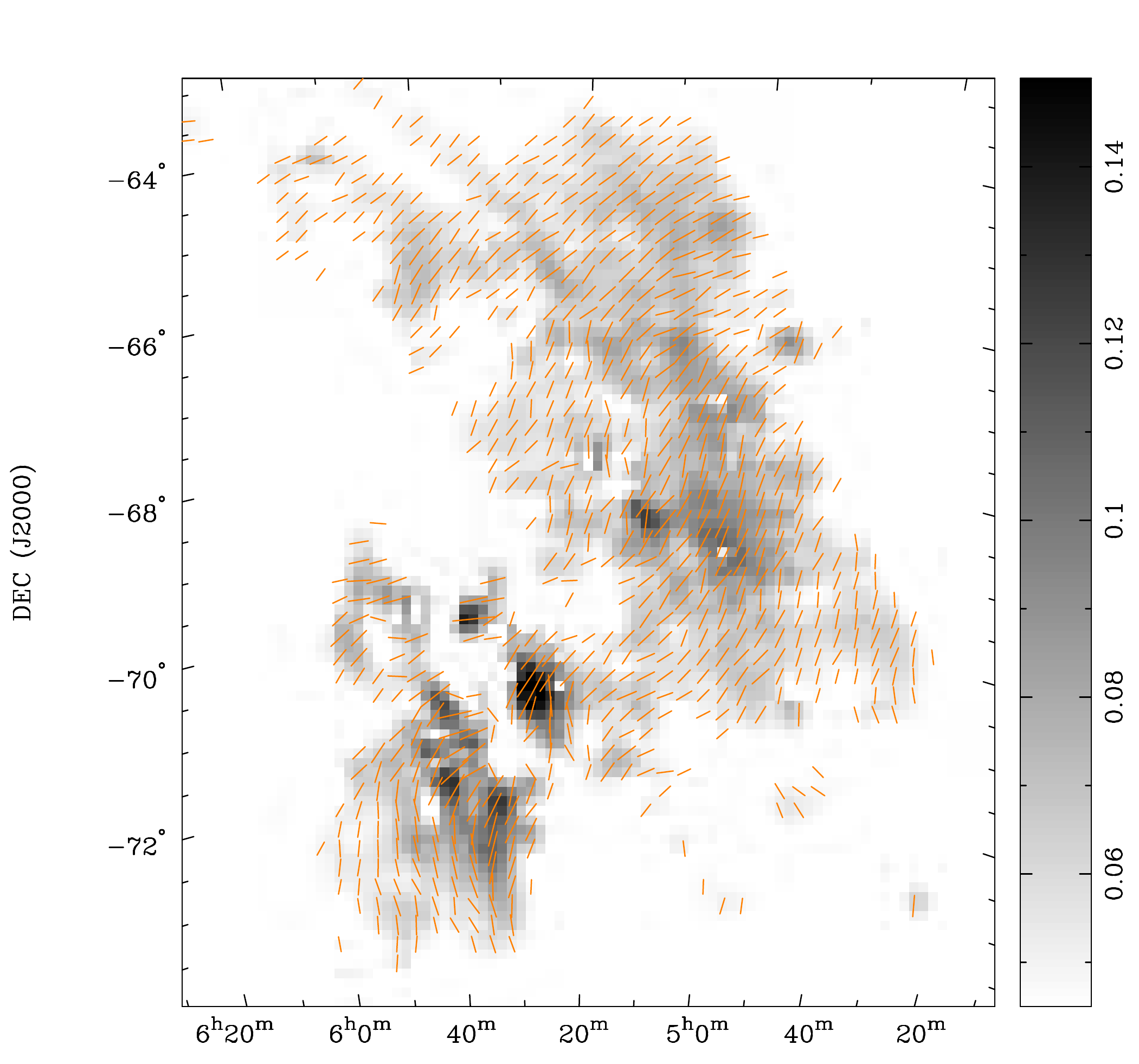}
\caption{Intrinsic polarization position angles of the LMC overlaid on the Parkes polarized intensity map at 1.4 GHz . Pixels with a polarized intensity lower than 0.045 Jy/beam and a reduced $\chi^2$ greater than 2 have been blanked. The color scale is in units of Jy/beam.}
\label{fig:intrinsic_angle}
\end{figure}

RMs of the diffuse polarized emission were determined using Parkes-only Stokes Q and U channel maps. We choose to use the Parkes-only channel maps instead of the Parkes and ATCA combined maps to compute RMs since the former have lower noise levels and a larger total bandwidth that leads to higher RM accuracy.  In order to determine RMs accurately and to maximize the sensitivity to extended RM structure, we have used the  total available bandwidth of the Parkes observation(8-MHz wide channels, covering the frequency range from 1264 MHz to 1504 MHz) for RM computation. The full width half maximum (FWHM) of the rotation measure transfer function (RMTF) is 227 rad m$^{-2}$. The largest extended structure in Faraday depth to which our data are sensitive is  76 rad m$^{-2}$. We compute the RM of a pixel if the measured polarized intensity exceeds 0.045 Jy/Parkes beam. Pixels with a reduced $\chi^2$ greater than 2 are discarded and blanked in the final map. The peak RM and the corresponding RM error maps are shown in Figure ~\ref{fig:rm_parkes} and the distribution of intrinsic polarization position angles are shown in Figure~\ref{fig:intrinsic_angle}. 

\begin{figure}
\centering{
\epsscale{0.6}
\plotone{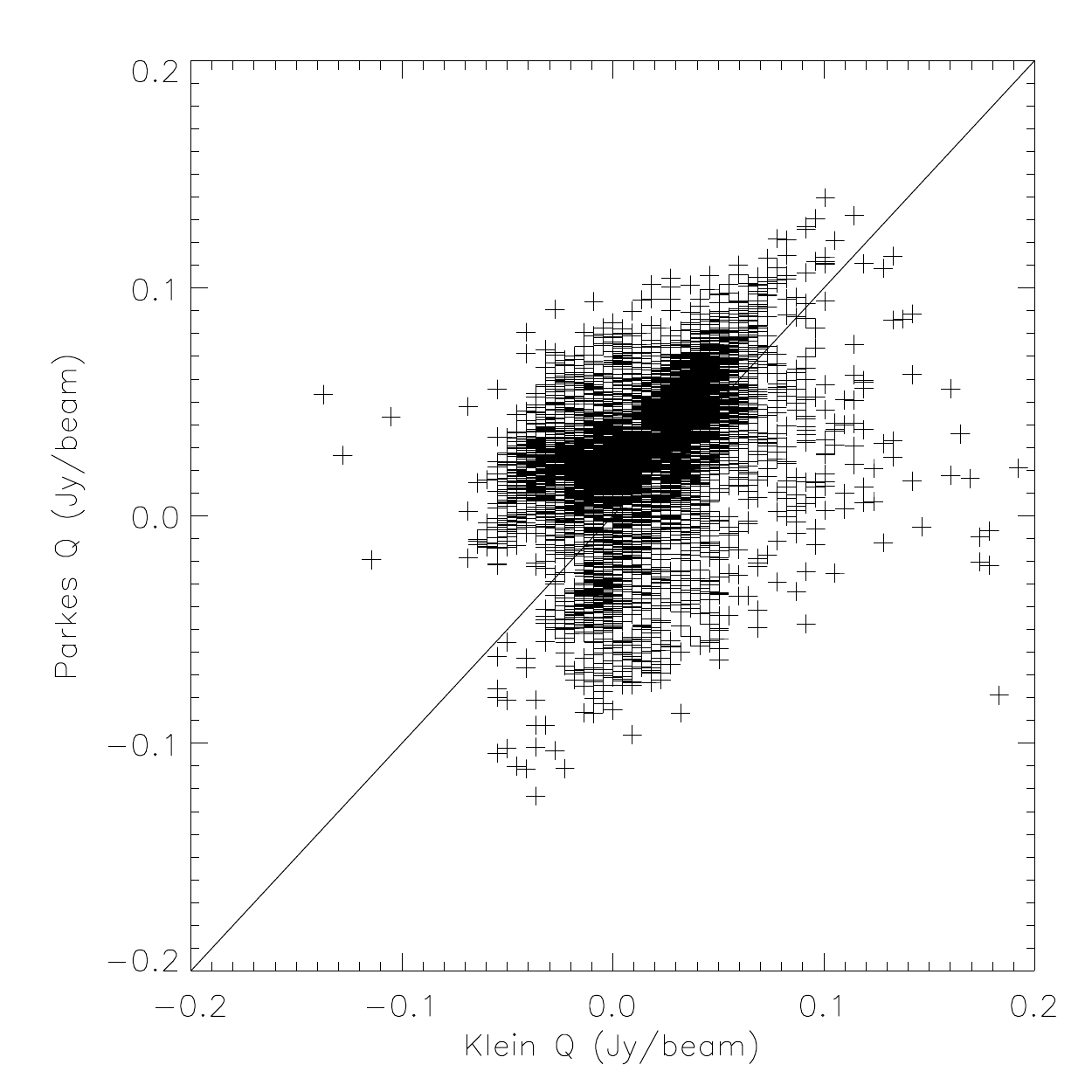}
\vspace{.1in}
\epsscale{0.6}
\plotone{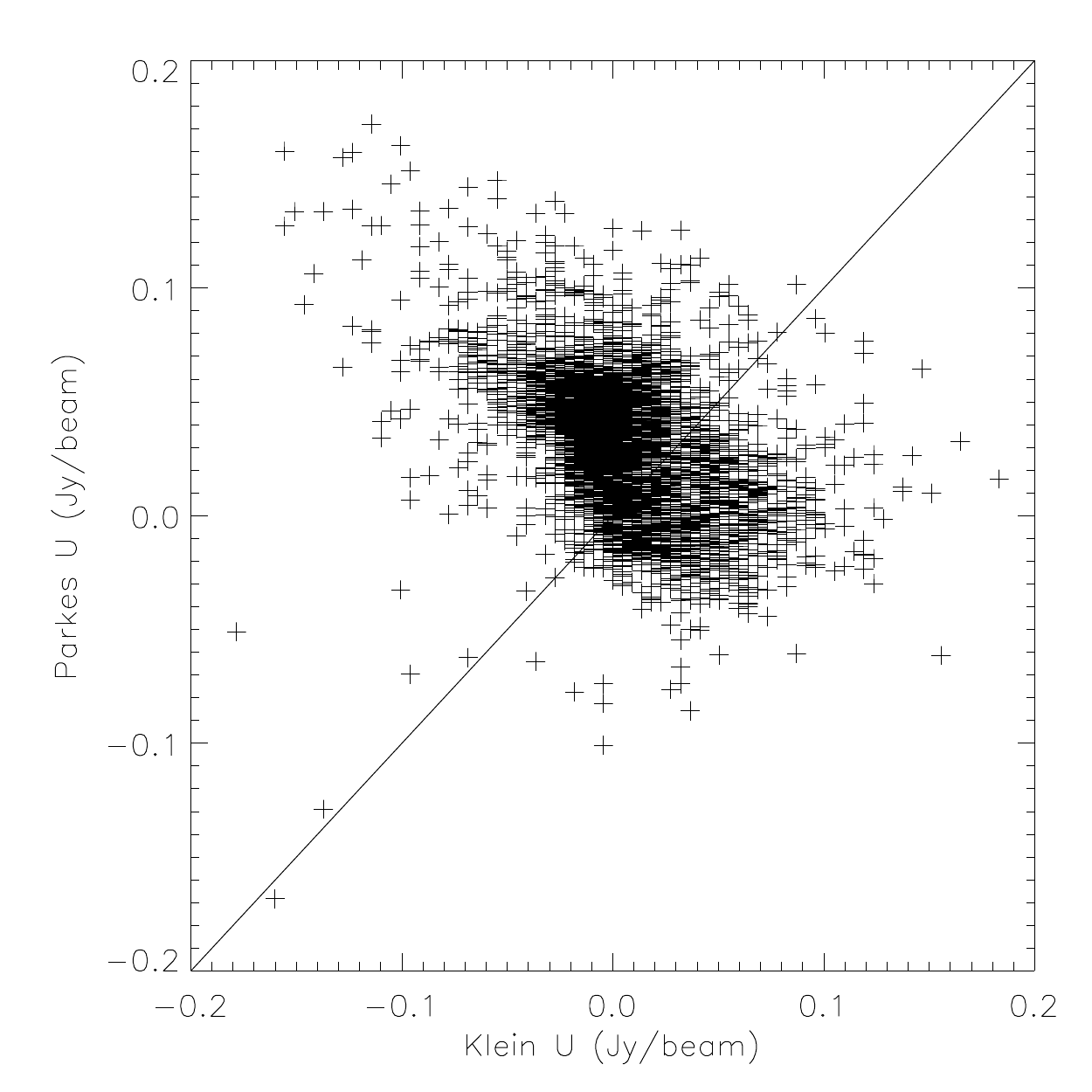}
}

\caption{Comparison of the Stokes Q and U measurements of our Parkes LMC data at 1.4 GHz with the \cite{klein1993} 1.4 GHz polarimetric observations. The pixel to pixel Q, U fluxes are adjusted by assuming that the absolute flux calibrations of the two data sets are equivalent. Top panel: Stokes Q of our Parkes observation plotted against that of  \cite{klein1993}. Bottom panel: Stokes U of our Parkes observation plotted against that of \cite{klein1993}. The dashed line of  slope 1 indicates where data points should lie if the Stokes parameters of the \cite{klein1993} data and ours agree.}
\label{fig:klein_comparison}

\end{figure}

Although \cite{klein1993} computed the rotation measure of diffuse polarized emission with Parkes observations in the eastern half of the LMC using data at 1.4 GHz and 2.45 GHz (see their Figure 3), their data suffer from several shortcomings. First, according to  \cite{klein1993}, their polarization coverage of the LMC was incomplete due to a faulty polarimeter, hence not all flux on scales of the galaxy was mapped. Second, since the diffuse synchrotron emission is subject to wavelength dependent depolarization effects, a linear relationship between position angle and  $\lambda^2$ is not expected. Therefore, RMs computed using a least-square linear fit of the polarization angle against  $\lambda^2$ with data at only two widely separated frequency bands might not be physically meaningful. Moreover, the \cite{klein1993} RMs are subject to a possible n-$\pi$ ambiguity when unwrapping polarization angles since polarization information at only two wavelengths were used. Finally, there is a sign error associated with the 1.4 GHz Stokes U map by \cite{klein1993}. This is illustrated in Figure~\ref{fig:klein_comparison}, where our Parkes Stokes Q and U data are plotted against those of \cite{klein1993}. While Stokes Q measurements of the two data sets agree with each other, Stokes U measurements have opposite signs. Therefore, the RM results presented by \cite{klein1993} and their subsequent analysis are likely unreliable. 

\subsection{RMs of Extragalactic Background Sources}
\label{subsection:egs_rm_determination}

Using the ATCA channel maps made without the shortest baselines, we identified the brightest polarized pixel of each extragalactic source found using the MIRIAD task SFIND \citep{hopkins2002}. We then extracted its Stokes $Q$ and $U$ values across the frequency band for RM determination. We discarded sources with signal-to-noise of polarization detection below 7 and sources with unusually high polarized fraction ($>$ 30 $\%$). A total of 305 reliable EGS RMs were thus computed in the field surveyed by the ATCA. The source coordinates, RMs, RM uncertainties and their flux information are listed in Table~\ref{table:ptsrc_rm}. We define sources to be within LMC's main body if the EGS is less than 3.5$^\circ$ from RA$_{LMC}$=5$^h$16$^m$, DEC$_{LMC}$=-68$^d$41$^m$ (the center of LMC's H$\alpha$ distribution, \citealt{gaensler2005}).  These sources are marked by ``lmc" in the 6th column of Table~\ref{table:ptsrc_rm}. Assuming that the LMC disk is inclined at an angle 35$^\circ$, with position angle of line of nodes at 123$^\circ$ \citep{vandermarel2006} located at a distance of 50 kpc \citep{kallivayalil2006a}, we define sources to be outside the LMC body if the de-projected radial distance from the LMC center to the EGS is larger than 5 kpc. These sources are marked by ``fg" in the 6th column of Table~\ref{table:ptsrc_rm}.

As RM is a line-of-sight integral, to extract the RM produced by the LMC's ISM requires one to subtract the RM arising in the Milky Way foreground. We can use RMs of EGSs whose sightline do not pass through the LMC main body in the field of view to estimate the smooth component of the Milky Way RM foreground.  Since the surveyed field is relatively large in size, a constant foreground RM is insufficient to describe the possible spatial variation of the foreground RM. Therefore, we have fitted a plane to 97 EGSs outside the main body of the LMC in the field of view. The fitted plane has the following form
\begin{align}
\label{eq:fgrm}
RM_{fg,MW} &=  28.7+0.68\times(RA-RA_{LMC})-0.39\times(DEC-DEC_{LMC}) 
\end{align}

\begin{figure}
\epsscale{0.8}
\plotone{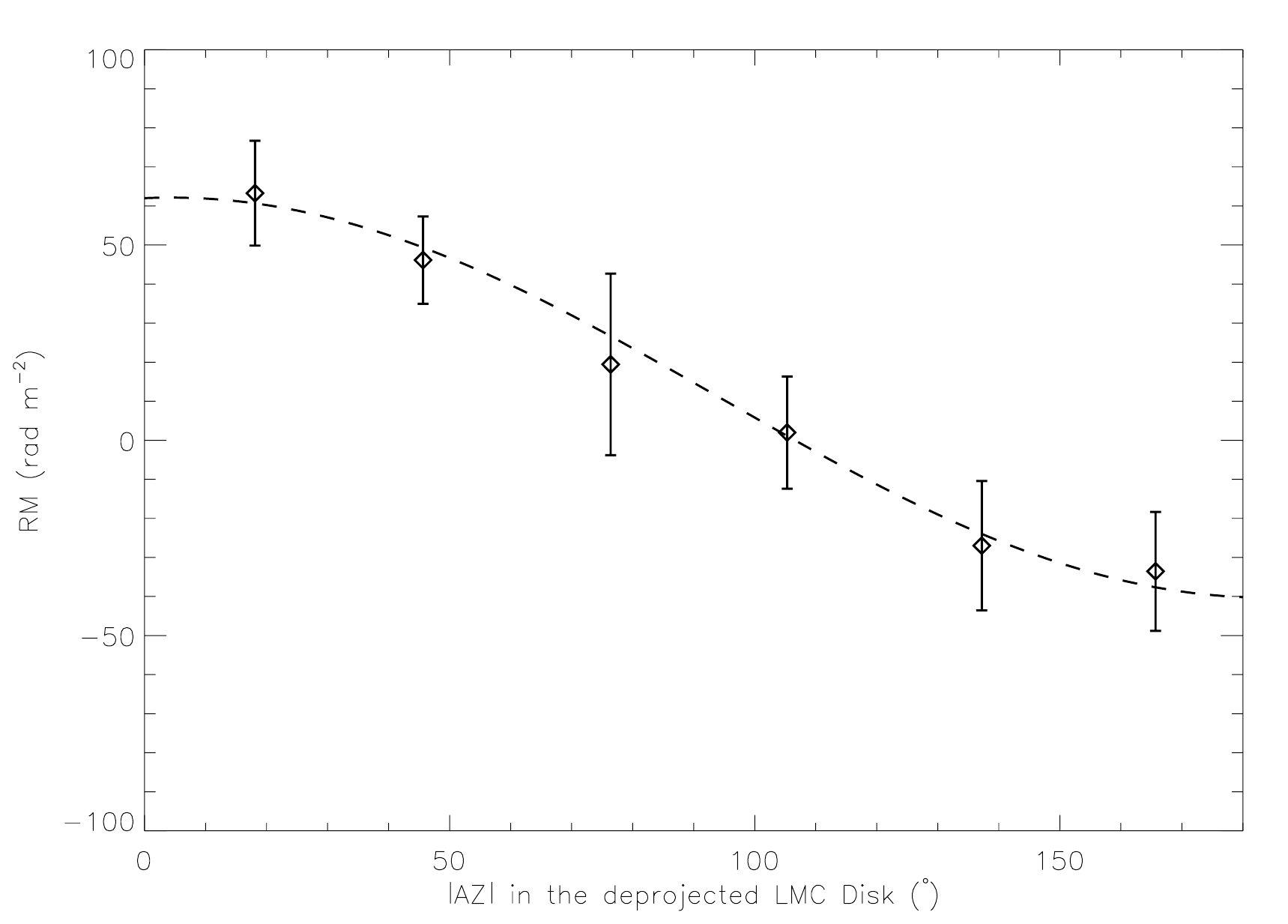}
\caption{The behavior of RMs of EGSs that lie directly behind the LMC as a function of the absolute value of azimuth angle in the deprojected LMC disk. Data points are the average foreground subtracted EGS RMs in 30$^\circ$ azimuthal angle bins. The error bars represent the standard error of the mean within each bin. The data have been folded across the semi-major axis of the LMC. The dotted line is the best sinusoidal fit to the data: 
RM$_{\rm~LMC}$ = (51 $\pm$ 11) $\cos$  ($|$AZ$|$+(-4$^\circ$ $\pm$ 10$^\circ$))+ (11$\pm$ 7)~rad m$^{-2}$.}
\label{fig:egs_rm_fit}

\end{figure}

\begin{figure}
\plotone{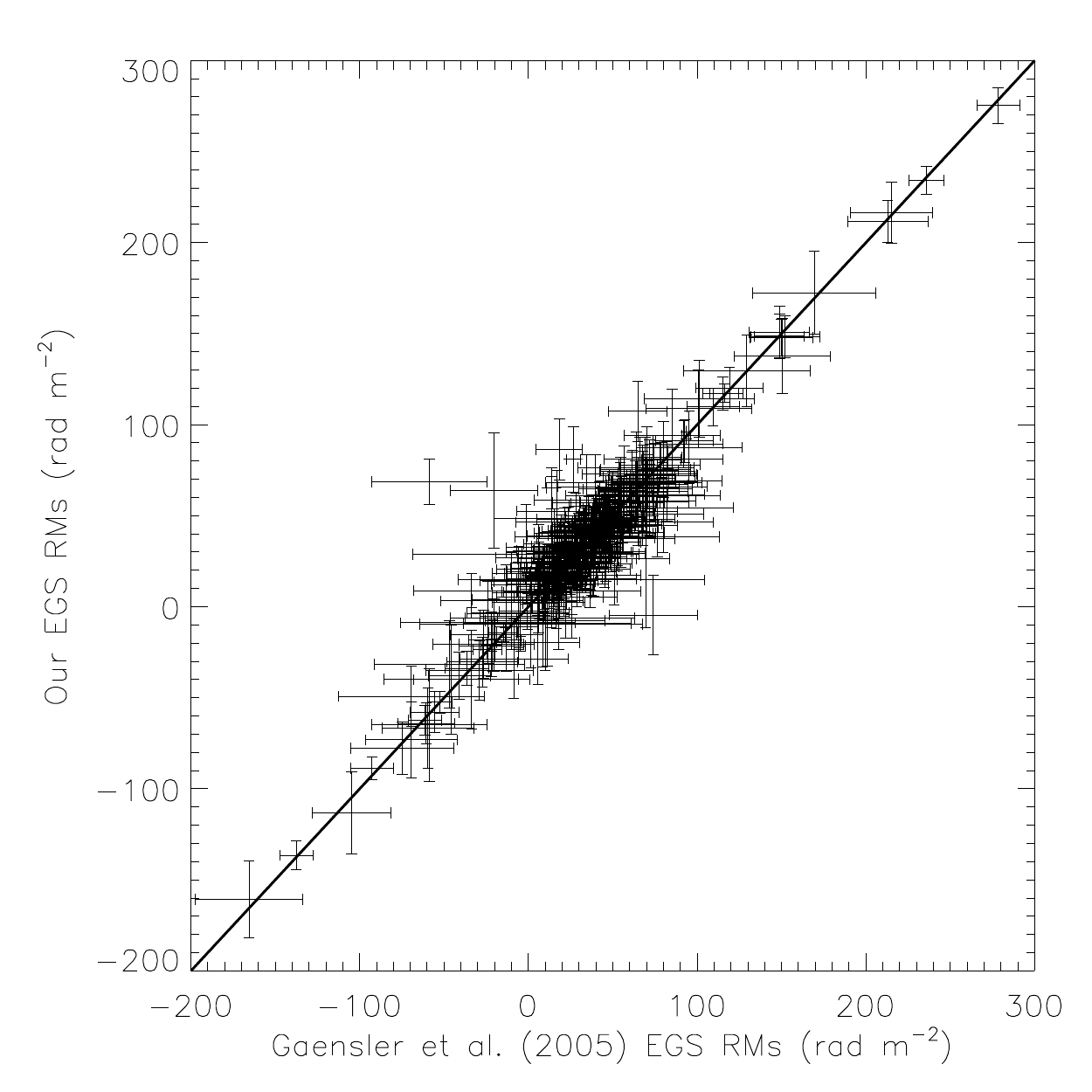}
\caption{Comparison between \cite{gaensler2005} EGS RMs  and the recomputed EGS RMs in this paper. The solid line of slope 1 indicates agreement between the \cite{gaensler2005} RMs and our recomputed RMs.}
\label{fig:compare_bmg_mao}
\end{figure}

After removing the RM arising in the Milky Way foreground, the distribution of EGS RM behind the LMC main body is plotted as a function of the absolute value of the azimuthal angle (AZ) within the LMC disk (measured counter-clockwise from the approaching major axis), as shown in Figure~\ref{fig:egs_rm_fit}. The behavior of RM as a function of $|$AZ$|$ can be fitted by a cosine curve:
\begin{equation}
\label{eq:ptsrc_rm_az_fit}
RM_{LMC} = (51 \pm 11)\cos{(AZ+(-4^\circ \pm 10^\circ))} + (11\pm7)~\rm{ rad~m^{-2} }, 
\end{equation}
with a reduced $\chi^2$ of 41. The large reduced $\chi^2$ reflects the scatter of RM intrinsic to the EGSs and the RM produced by the turbulent ISM in the LMC, both of which are not modeled in this large-scale field fit. The fitted parameters of the RM vs azimuth trend are consistent withe \cite{gaensler2005} results within uncertainties. The observed sinusoidal pattern of EGS RMs can be produced by a purely azimuthal field in the LMC disk of coherent magnetic field strength  $\sim$ 1$\mu$G (estimated using Equation (2) of  \citeauthor{gaensler2005} \citeyear{gaensler2005}). The fit places loose constraint on the phase of the cosine: the pitch angle is consistent with zero. The offset of the sinusoid is also consistent with zero at 1.6 $\sigma$, suggesting a negligible vertical component of the disk field probed. We note that \cite{gaensler2005} reported a significant positive offset of 9 rad m$^{-2}$ for the sinusoidal fit. However, they estimated the Galactic foreground RM towards the LMC by dividing the field of view into 4 sections and obtaining the average RM off the LMC in each section. We suggest that the offset of the sinusoid will no longer exist if  the Milky Way foreground RM is modeled with a higher-order polynomial fit as that in Equation~\ref{eq:fgrm}. As shown in Figure~\ref{fig:compare_bmg_mao}, reprocessed EGS RMs are consistent with those obtained by \cite{gaensler2005}: the linear correlation coefficient between the two data sets is 0.95. We note that EGS RMs are only sensitive to the component of magnetic field that is symmetric with respect to the LMC disk, as any anti-symmetric components cancels.

\section{Global Magnetic Fields in the LMC}
\label{section:global_bfield}

In this section, we demonstrate that the observed diffuse radio polarization at 1.4 GHz likely originates in a layer above the LMC mid-plane on the near side. We then deduce properties of this large scale halo\footnotemark[4]\footnotetext[4]{Throughout this paper, we use ``halo" to represent regions outside the LMC mid-plane} magnetic field.

\subsection{Characterizing the foreground Milky Way Faraday Rotating Medium}
\label{subsection:fg_medium}
To establish the location of the Milky Way foreground emission, we compare RMs of EGS and those derived from diffuse emission along the same sight lines to the north of LMC's main body, where Galactic emission dominates.  

\begin{figure}
\centering
\plotone{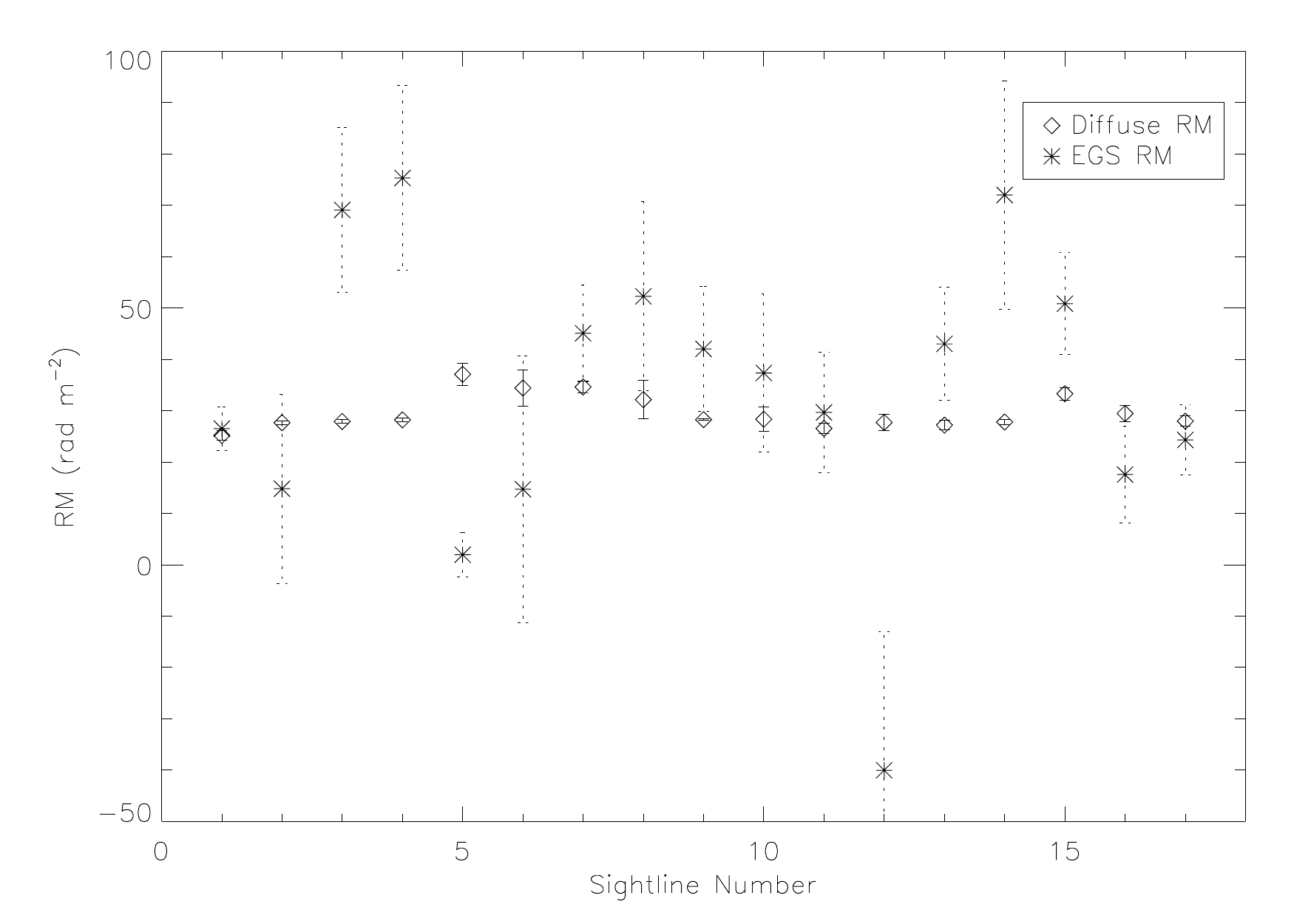}
\caption{Sightline-by-sightline comparison between EGS and diffuse RMs in the region north of the LMC that is contaminated by Milky Way polarized emission. For each sightline, both the diffuse RMs (denoted by diamonds and solid line error bars) and the EGS RMs (denoted by asterisks and dotted line error bars) are plotted.}
\label{fig:compare_spur}

\end{figure}

For each sightline where a reliable EGS RM is derived, we measure the ``off-source" diffuse emission RM. This value is computed by averaging the RMs of  4 pixels located 1 beam away (to the north, south, east and west) from the EGS. A sightline is discarded if any of the pixels used to calculate the diffuse emission coincides with the location of an EGS. Sightline-by-sightline comparison of the EGS RMs against the diffuse RMs, as shown in Figure~\ref{fig:compare_spur}, demonstrates that most EGS RMs and diffuse emission RMs are in agreement with each other within their measurement errors. The weighted mean diffuse emission RM is +28.2 rad m$^{-2}$ with a standard deviation of 1.1 rad  m$^{-2}$, while the weighted mean of the EGS RM is +25 rad m$^{-2}$ with a standard deviation of 19 rad m$^{-2}$. This agreement in RM suggests that the foreground emission component of the observed diffuse polarized emission is Faraday thin.


\subsection{Where is the polarized emitting material within the main body of the LMC at 1.4 GHz?} 
\label{subsection:whereispol}

\begin{figure}
\centering
\plotone{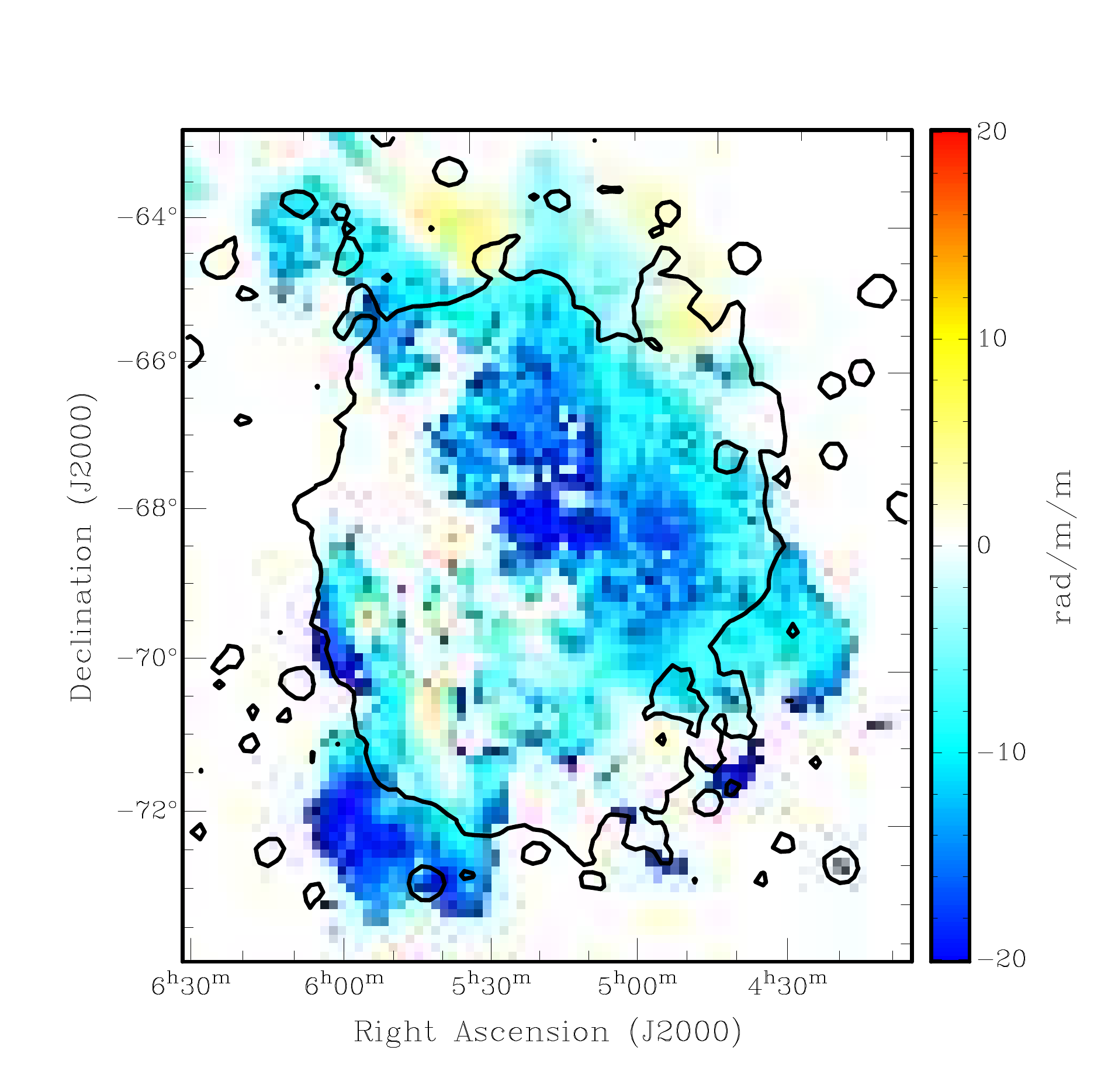}
\caption{Faraday rotation measures derived from diffuse polarized emission at 1.4 GHz using the Parkes-only data, after subtracting foreground Milky Way Faraday rotation. Pixels with  polarized intensities lower than 0.045 Jy/beam have been blanked.}
\label{fig:diff_rm_fgsub}
\end{figure}

In this section, we establish the location of the polarized emitting material within the LMC at 1.4 GHz. Assuming the observed foreground Milky Way diffuse polarized emission (See \ref{subsection:large_scale_pi}) is located behind the Milky Way Faraday rotating medium, we subtracted the Milky Way's RM contribution from the measured diffuse RM using Equation~\ref{eq:fgrm}. The resulting RM map is shown in Figure~\ref{fig:diff_rm_fgsub}. Although the polarized emission from the Milky Way is strong in the northwestern half of the LMC, some of the observed polarized emission must originate in the LMC because diffuse RMs change value sharply at LMC's northwestern edge, as seen in Figure~\ref{fig:rm_parkes}. Moreover, the measured diffuse RM  $\sim$ +17 rad m$^{-2}$ (with a standard error in the mean  of 0.2 rad m$^{-2}$)  does not agree  with the foreground Milky Way value  ($\sim$ +28 rad m$^{-2}$) which is expected if the Milky Way emission dominates. Even though one cannot extract information on the plane-of-the-sky magnetic field from the measured diffuse polarization position angles (Figure~\ref{fig:intrinsic_angle}) when the foreground contamination dominates, information on the LMC's line-of-sight magnetic field in these regions can still be extracted from the RMs derived from the diffuse polarized emission.

\subsubsection{Depolarization due to HII regions}

Because of the high electron densities and fluctuating magnetic fields of H II regions, they are known to depolarize background synchrotron emission due to a combination of beam depolarization and wavelength dependent depolarization effects \citep{landecker2002}. H II regions in our own Galaxy have been used to estimate the polarization horizon, the distance beyond which we no longer detect any polarized emission \citep{gray1999,gaensler2001,kothes2004}. Since H II regions in the LMC have broadly similar H$\alpha$ intensities to those in the Milky Way and are spatially resolved, with angular extents larger than 1 beam, we can estimate the location within the LMC that the polarized emission arise at 1.4 GHz by comparing the polarized intensity and H$\alpha$ maps of the LMC. 

\begin{figure}
\centering
\plotone{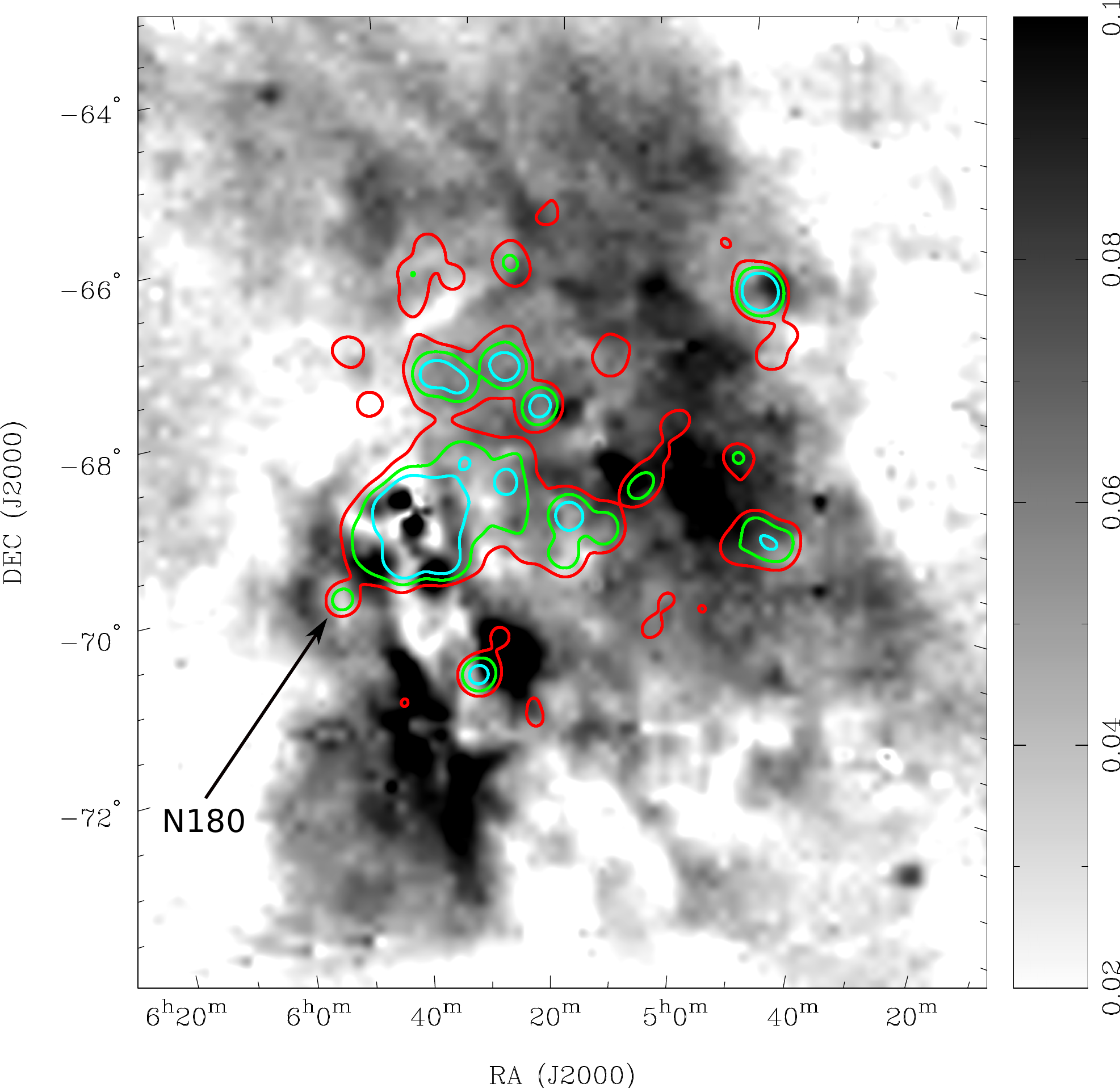}
\caption{ H$\alpha$ intensity contours of the LMC from the SHASSA survey \citep{gaustad2001} overlaid on the de-biased linearly polarized intensity maps of the LMC at 1328 MHz from Parkes single dish observations. The red, green and cyan contours represents H$\alpha$ intensity at 25, 50 and 100 Rayleighs, respectively. The gray scale is in units of Jy/Parkes beam. The arrow indicates the position of the H II region N 180.}
\label{fig:hacontour_on_pi}
\end{figure}

We assume that H II regions reside in the LMC disk just as H II regions in the Milky Way are tightly confined to the galactic disk\footnotemark[5]\footnotetext[5]{The vertical distribution of Galactic H II regions is centered at 7 pc below the Galactic plane with a scatter of 35 pc \citep[see e.g.,][]{fish2003}.}. If there exists little synchrotron emission between the HII regions and us, then we expect these H II regions to depolarize any background polarized emission, producing voids in the polarized intensity map centered around the HII regions. Upon careful visual inspection\footnotemark[6]\footnotetext[6]{We note that because of the foreground contamination in the northwestern part of the galaxy, depolarization caused by HII regions might not result in zero polarized intensity but rather just a reduction in polarized intensity if the foreground polarized emission is smooth.}, we have only found HII complex N180 ( 5$^h$48$^m$38$^s$,$-$70$^d$02$^m$04$^s$, on the eastern edge of the LMC) to correspond to a reduction in the Parkes polarized intensity map (see Figure~\ref{fig:hacontour_on_pi}). This suggests that N180 is extraplanar, on the near side of the LMC.  Since depolarization by other HII regions is not observed, it is likely that they lie behind where the observed diffuse polarized emission originates. Under the assumption that all HII regions (except N180) reside in the LMC disk, we suggest that the polarized emission visible at 1.4 GHz primarily arises in the near-side halo of the LMC.

If polarized emission lie behind the Faraday rotating turbulent medium in the LMC, due to various polarization processes as described earlier in this section, one might expect an anti-correlation between polarized emission and H$\alpha$ intensity. Therefore, we have quantified any possible correlation between H$\alpha$ and polarized intensity by calculating the Pearson linear correlation coefficient between the Parkes polarized intensity map\footnotemark[7]\footnotetext[7]{We did not compare the fractional polarization map against the H$\alpha$ map of the LMC since computing the non-thermal fractional polarization requires accurate separation and subsequent removal of the thermal component, which is beyond the scope of this paper.} produced by RM Synthesis and the H$\alpha$ intensity map. The resulting correlation coefficient of $\sim$ 0.1 gives further evidence that polarized intensity at 1.4 GHz and H$\alpha$ intensity are independent of each other.

\subsubsection{EGS RMs and Diffuse RMs distribution}
\label{subsection:egsrm_diffrm}
\begin{figure}
\centering
\plotone{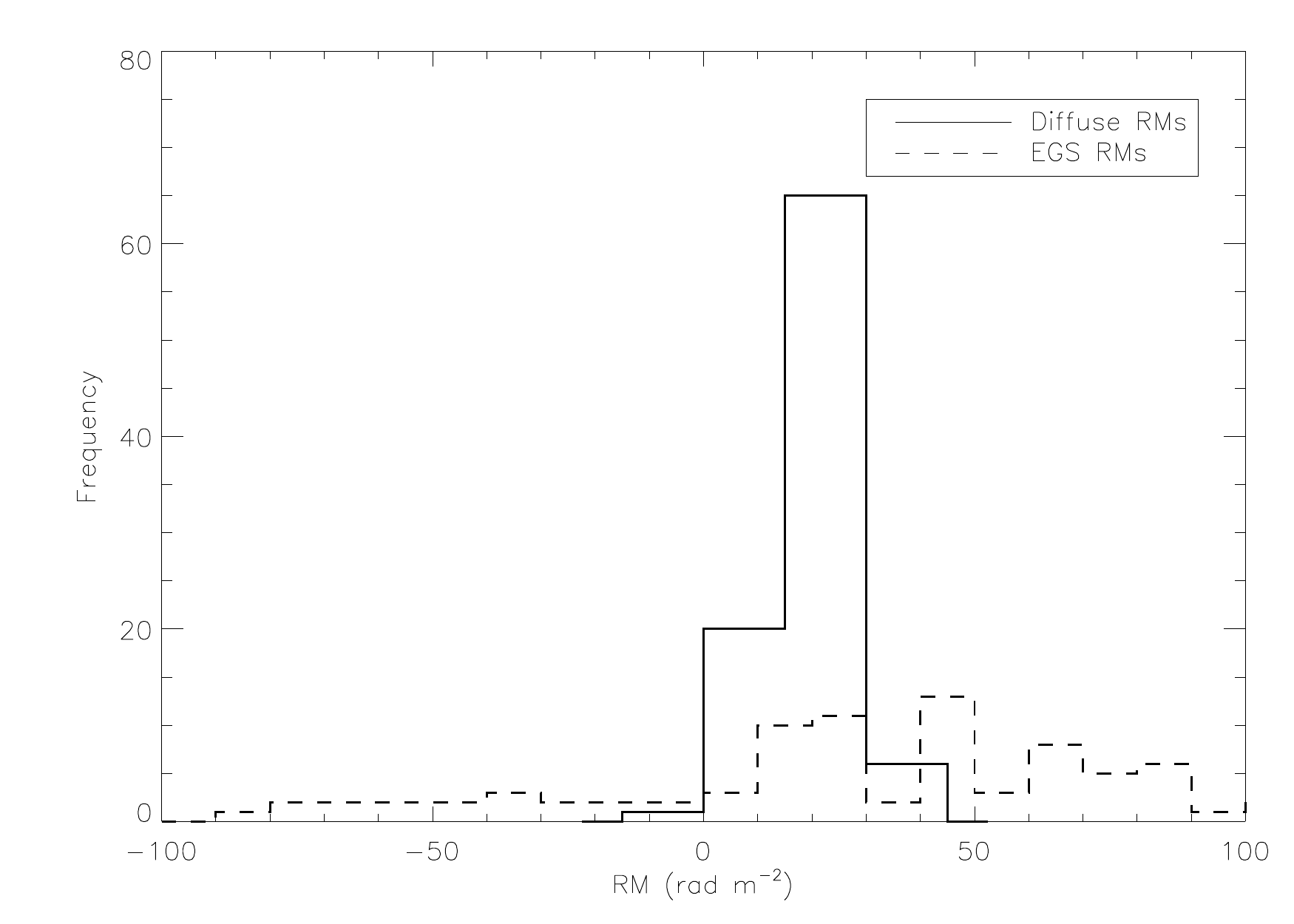}
\caption{The distribution of EGS RMs (dotted line), compared to that of diffuse RMs measured 1 beam away from each EGS sightline (solid line).}
\label{fig:compare_histo}
\end{figure}

Comparing EGS RMs and those derived from the diffuse polarized emission, we can further confirm the location of the LMC polarized emitting medium at 1.4 GHz. We have plotted the distribution of EGS and diffuse RMs in the form of histograms in  Figure~\ref{fig:compare_histo}. It is clear that the scatter in EGS RMs  ($\sim$ 80 rad m$^{-2}$) far exceeds that of the diffuse RMs  (10 rad m$^{-2}$). In our own Milky Way, the standard deviation of EGS RMs at low Galactic latitudes, with long integral path length through the turbulent Galactic disk, exceeds 250 rad m$^{-2}$ \citep{haverkorn2006}. On the other hand, sight lines towards high Galactic latitudes show much smaller scatter ($\sim$ 10 rad m$^{-2}$) \citep{mao2010,schnitzeler2010}. As strong random magnetic fields and large fluctuations in electron density within the turbulent mid-plane would lead to larger RM variance, the disparity between the standard deviation of EGS RMs behind the LMC and that of diffuse RMs indicates that the diffuse RMs probe through less pathlength than the EGS RMs. Specifically, we suggest that the diffuse polarized emission originates above the LMC disk and hence has a much smaller standard deviation than the EGS RMs. EGS RMs, on the other hand, probe through the entire sightline including the turbulent LMC disk, which introduces a large RM scatter.

\subsubsection{The Depolarizing Turbulent LMC Disk} 
\label{subsection:lmc_disk}
In this subsection, we illustrate that the turbulent LMC disk is internally depolarized and the disk is sufficient to depolarize any emission located behind the disk. 

In the azimuth range 300$^\circ$-330$^\circ$, the LMC is neither blocked by the Milky Way polarized emission (Section~\ref{subsection:fg_medium}) nor dominated by the two polarized filaments (Section~\ref{subsection:large_scale_pi}). Therefore diffuse polarized emission from this region must be intrinsic to the LMC. Using our Parkes observations and assuming a non-thermal fraction of 0.63 at 1.4 GHz, we find a non-thermal polarized fraction of the LMC to be $\sim$ 7 \%. The coherent magnetic field in the LMC contributes a rotation measure R$_{tot}$ of 37.5 rad m$^{-2}$ in this azimuthal range (Equation~\ref{eq:ptsrc_rm_az_fit}), with a scatter of $\sigma_{RM}$ $\sim$ 80 rad m$^{-2}$ due to random magnetic fields (\S~\ref{subsection:egsrm_diffrm}). 

We can use Equation (34) of \cite{sokoloff1998} to compute the expected complex polarized fraction from the LMC disk P$_{\rm in}$:
\begin{equation}
\label{eq:intern_depol}
P_{in}=P_0 \frac{1-exp(-2\sigma_{RM}^2 \lambda^4+2i\lambda^2R)}{2\sigma_{RM}^2 \lambda^4+2i\lambda^2R}
\end{equation}
where P$_0$ is the intrinsic degree of polarization of synchrotron radiation which is related to the spectral index $\beta$ by 
\begin{equation}
P_0 = \frac{(3-3\beta)}{(5-3\beta)}
\end{equation}

\begin{figure}
\centering
\plotone{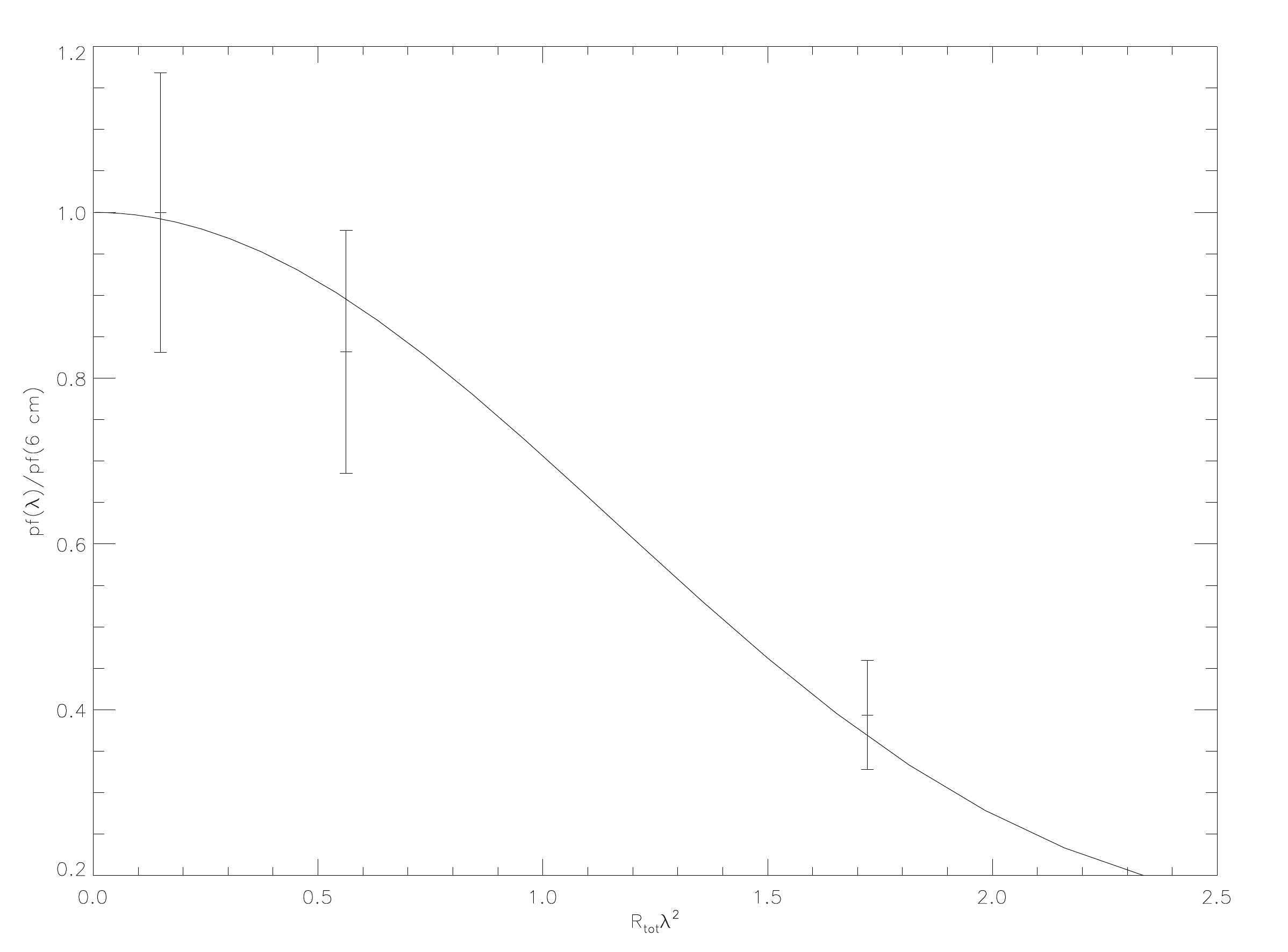}
\caption{The integrated depolarization of the LMC in the azimuthal range 300$^\circ$-330$^\circ$ as a function of R$_{tot}$$\lambda^2$, computed using available polarization data at 4.75 GHz, 2.45 GHz and 1.4 GHz. The ratio of the polarized fraction at 4.75 GHz, 2.45 GHz and 1.4 GHz to that at 4.75 GHz is plotted against the total RM through the disk  (R$_{tot}$) times wavelength squared. The solid line represents the expected depolarization due to a coherent field with R$_{tot}$ = 37.5 rad m$^{-2}$ combined with a random field of RM dispersion 16 rad m$^{-2}$.}
\label{fig:disk_depol}
\end{figure}

Since both R$_{tot}$$\lambda^2$ and $\sigma_{RM}$$\lambda^2$ exceed 1 at centimeter wavelengths, the polarized LMC disk is subjected to differential Faraday depolarization by a coherent field as well as internal Faraday dispersion by a random field. We make use of the LMC polarimetric data of \cite{haynes1991} at 2.45 GHz and 4.75 GHz to study the wavelength dependent depolarization of the LMC disk. We assume that at 4.75 GHz, the LMC disk suffers little depolarization and hence we can use the fractional polarization of the LMC at 4.75 GHz as the intrinsic fractional polarization.  In Figure~\ref{fig:disk_depol}, we have plotted the ratio of the observed polarized fraction to the fractional polarization at 4.75 GHz as a function of R$_{tot}$$\lambda^2$. The solid line plots the expected polarization of the LMC disk using Equation~\ref{eq:intern_depol}. We see that a total regular magnetic field through the disk, represented by $|R_{tot}|$ $\sim$ 37.5 rad m$^{-2}$ and a random field component represented by $\sigma_{RM}$$\sim$ 16 rad m$^{-2}$ can reproduce the observed multi-wavelength polarized emission well. We note that the diffuse polarized emission from the LMC at 4.75 GHz (fractional polarization $\sim$ 17\%) is inconsistent with the random to ordered field ratio of 3.6 in the LMC disk derived by \cite{gaensler2005} from a different approach, as the latter predicts a polarized fraction of $\sim$  7\% at short wavelengths (at 4.75 GHz). The higher observed fractional polarization is likely due to an additional layer of polarized emission above the turbulent disk, as we have advocated in this section.

We note that Equation~\ref{eq:intern_depol} provides the exact complex polarization of diffuse synchrotron emission (in this case, the LMC disk) only if the total number of turbulent cells within the beam cylinder (beam area $\times$ total path length through the synchrotron emitting medium) is large. Given the proximity of the LMC to us and assuming a turbulent cell size similar to that in the Milky Way of $\sim$ 90 pc \citep{gaensler2005}, there are approximately 4 turbulent cells within the beam area and a few 10s of cells along the line of sight. Therefore, Equation~\ref{eq:intern_depol} likely overestimates the amount of depolarization. As a result, a smaller $\sigma_{RM}$ than that implied from the scatter of EGS RMs ($\sim$ 80 rad m$^{-2}$)  is sufficient to describe the observed polarized fraction of the LMC disk. We note that any polarized emission from the LMC that originates in a layer above the turbulent LMC disk also leads to a higher observed polarized intensity than predicted, since the polarized fraction is now the sum of two components.

The LMC's turbulent disk can act as a random external Faraday screen which depolarizes synchrotron radiation emitted behind it, on the far-side halo of the galaxy \citep{sokoloff1998}: 
\begin{equation}
P_{ex} = P_0 exp(-2\sigma_{RM}^2 \lambda^4)
\end{equation}
Even if the intrinsic fractional polarization of the emission from the far side of the LMC is at the theoretical maximum (P$_0$ $\sim$ 72 \%), the emerging polarized fraction P$_{ex}$ would only be 1$\times$10$^{-9}$ at 1.4 GHz for $\sigma_{RM}$ =  80 rad m$^{-2}$. This suggests that any diffuse polarized radiation from behind the LMC disk is completely depolarized. The above arguments demonstrate that the observed fractional polarization at 1.4 GHz cannot be explained by the turbulent LMC disk alone.

\subsubsection{The Azimuthal Variation of Diffuse RMs}
\begin{figure}
\centering
\plotone{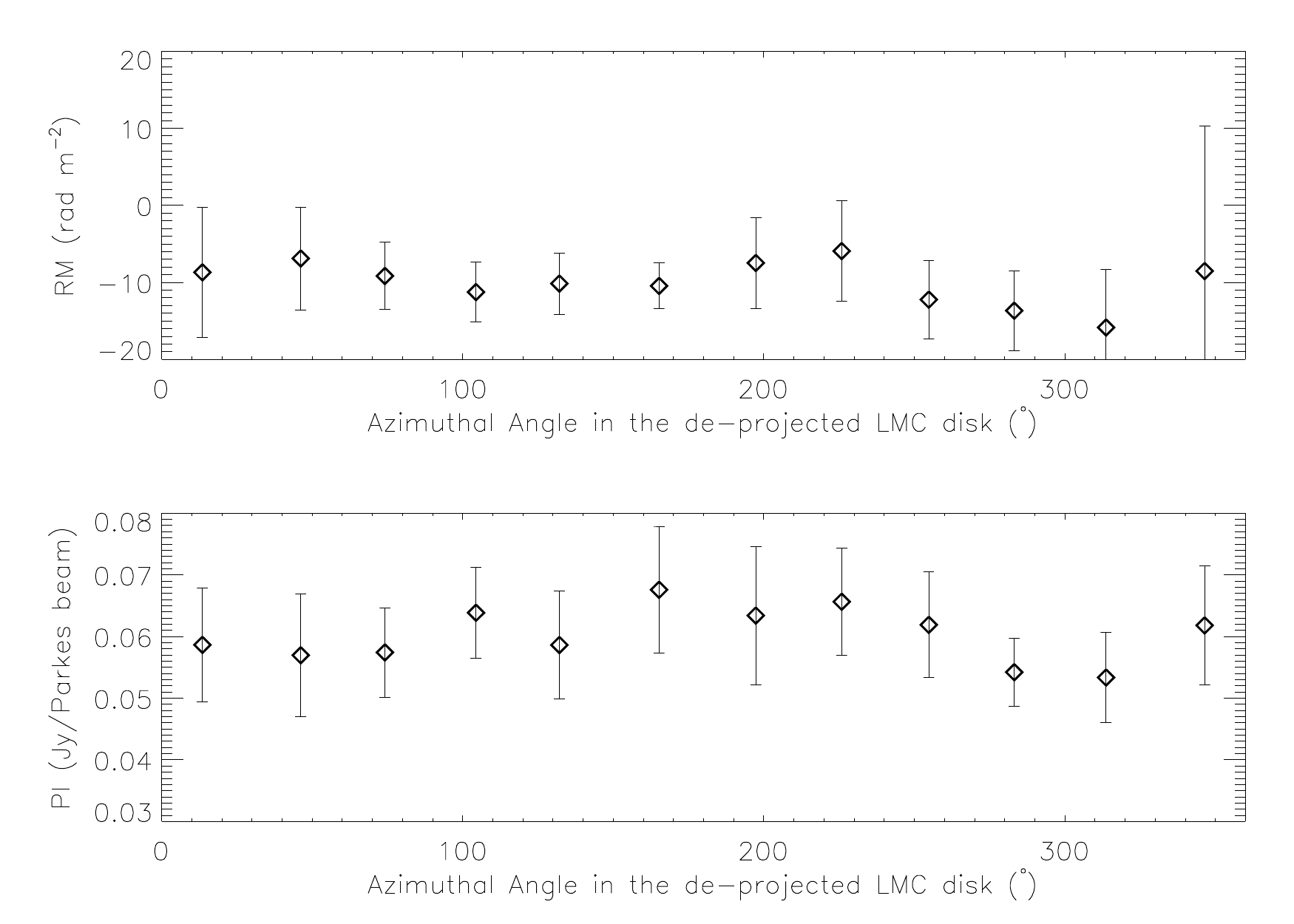}
\caption{The variation of rotation measure derived from diffuse emission (upper panel) and polarized intensity (bottom panel) as a function of azimuth angle in the de-projected LMC disk using our Parkes-only 1.4 GHz observations. Data have been averaged in 20$^\circ$ bins, with error bars indicating the root mean square dispersion within each bin. We note that only pixels with polarized intensity greater than 7 are plotted. We have also masked out positions where local polarized features in the LMC dominate. This includes the two polarized filaments in the south eastern corner of the galaxy as well as the enhanced polarized region located close to the northwestern end of the optical bar.}
\label{fig:rm_pi_az}
\end{figure}

The variation of the foreground rotation measure subtracted diffuse RMs as a function of azimuth in the de-projected LMC disk provides yet another means to locate the LMC polarized emitting material at 1.4 GHz. As shown in the upper panel of Figure ~\ref{fig:rm_pi_az}, diffuse RMs do not exhibit symmetric variation about zero as the variation of EGS RMs  does (Figure~\ref{fig:egs_rm_fit}). Instead, the foreground subtracted diffuse RMs across the main body of the LMC are consistently offset to negative values with a weighted mean of -9.9$\pm$0.2 rad m$^{-2}$. This suggests that  EGS and diffuse RMs do not probe the same magnetic field. As shown in the upper panel of Figure~\ref{fig:rm_pi_az}, the diffuse polarized emission from the LMC has a nearly constant RM as a function of azimuth. 

If the Galactic foreground polarized emission is located behind the Galactic Faraday rotating medium, then the measured diffuse RM is the mean of the Milky Way foreground value and that of the LMC emitting region, weighted by their relative polarized intensity\footnotemark[8]\footnotetext[8]{For example, if the polarized emission from the LMC dominates, the foreground rotation measure removed diffuse RM would be equal to the true total RM through LMC's halo. If the polarized emission from the Milky Way and the LMC are comparable, then the foreground subtracted diffuse RM would be equal to half of the total RM through LMC's halo.}. Hence, the foreground-removed diffuse RMs set a limit on the minimum $|$RM$|$ that originates from the LMC halo of 9.9 rad m$^{-2}$. We adopt this value as the RM produced in the LMC's halo for the remainder of this paper. 

We note that a non-zero $|$RM$|$ of  9.9 rad m$^{-2}$ originating in the halo of the LMC implies that the scale height of its Faraday rotating medium is equal to or greater than its synchrotron emitting medium. If the Faraday rotating layer has a smaller vertical extent than the synchrotron emitting region, then most of the polarized emission will not experience any Faraday rotation. This prediction on the relative scale heights of the rotating and emitting layer is consistent with properties of edge-on galaxies: their typical synchrotron scale height is $\sim$1.8 kpc \citep{krause2009}, while their diffuse ionized gas scale height is in the range of 1-2 kpc \citep{rossa2003,gaensler2008}. An extreme case is M31, for which the thermal electron distribution has a scale height three times thicker than the synchrotron emitting disk \citep{fletcher2004}.

\subsection{The Symmetry of the Large Scale Magnetic Field in the LMC }
\label{subsection:large_scale_sym}
\begin{figure}
\centering
\plotone{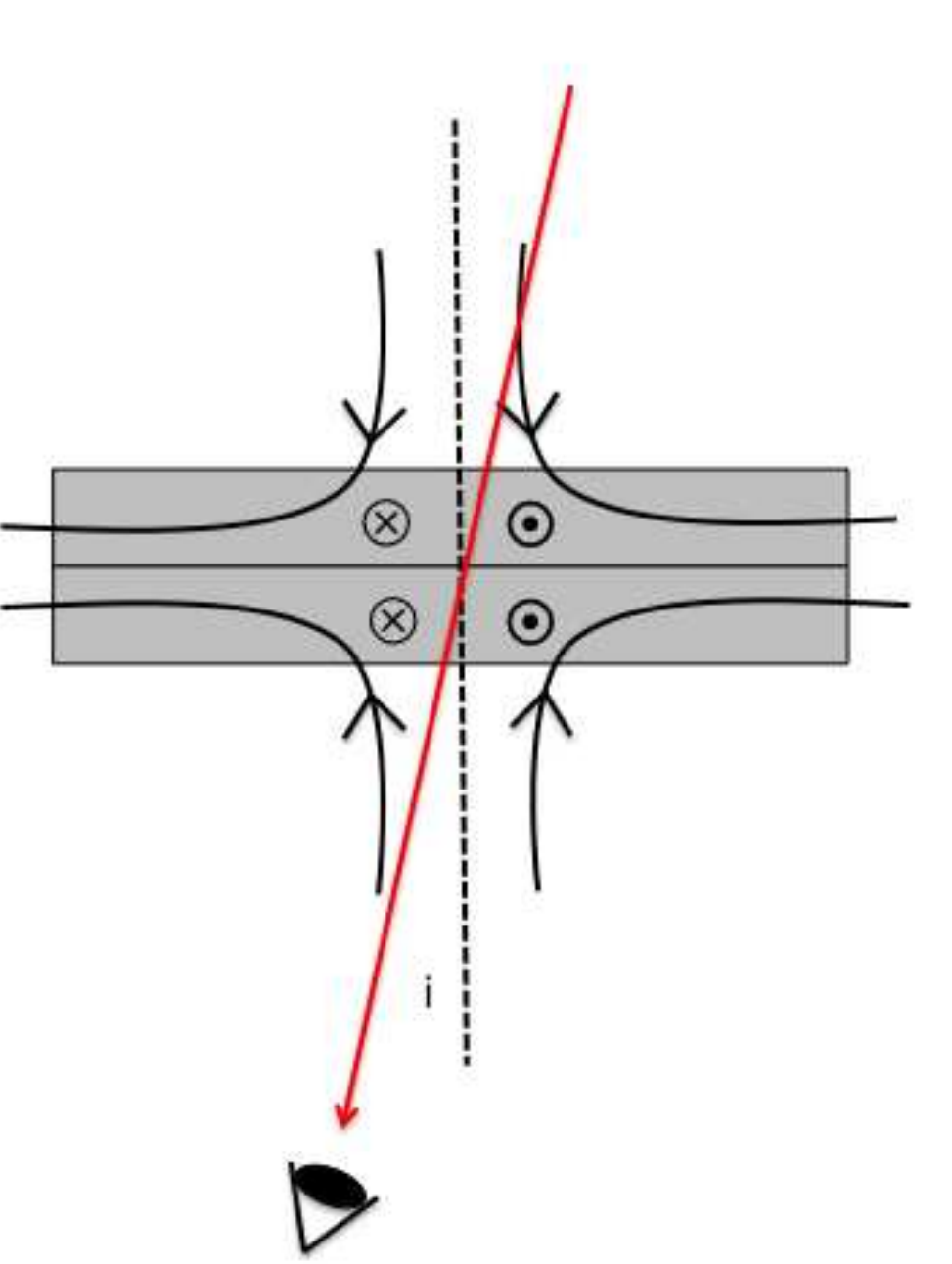}
\caption{Cross-sectional schematic of the proposed quadrupole-type large scale magnetic field in the LMC. The shaded region represents the turbulent mid-plane. The crosses  (dots) represent magnetic fields directed away (towards) the observer at the bottom of the figure. The in-disk azimuthal magnetic field is directed clockwise both above and below the plane. Non-shaded regions represent the LMC ``halo", with vertical magnetic fields reversing direction across the mid-plane. The normal to the LMC disk is denoted by the dashed black line while the line of sight vector at an inclination i$\sim$33$^\circ$ is denoted by the red arrow.}
\label{fig:quad_symmetry}
\end{figure}

We have established in Section~\ref{subsection:whereispol} that the polarized emission visible at 1.4 GHz originates in a layer above the LMC mid-plane on the near side. The consistent offset of $\sim$ -10 rad m$^{-2}$ of the diffuse RM from the foreground value (see Figure~\ref{fig:rm_pi_az}) suggests that the line-of-sight component of the magnetic field in the near side halo is coherently directed away from the observer. We propose that the observed behaviors of EGS and diffuse RMs are consistent with a purely azimuthal disk field that is symmetric with respect to the mid-plane and with a vertical magnetic field in the halo which reverses direction across the mid-plane. Since the EGS RMs probe the entire galaxy, the RM produced on the near and far sides of the halo would cancel out. Therefore, EGS RMs are only sensitive to the magnetic field in the symmetric LMC disk, as evident in the sinusoidal azimuthal variation seen in Figure~\ref{fig:rm_pi_az} and previously obtained by \cite{gaensler2005}. Due to the finite polarization horizon at 1.4 GHz, only the near side of the LMC halo is visible in diffuse polarized emission. The overall geometry of the LMC's large scale magnetic field is likely to be of quadrupolar symmetry and a proposed schematic is illustrated in Figure~\ref{fig:quad_symmetry}.

A recent study of Spitzer Infrared Nearby Galaxies Survey (SINGS) galaxies observed in polarization at 18 cm and 22 cm \citep{braun2010} exhibits some similar RM-vs-azimuth trends as seen for the LMC's diffuse emission: the RMs derived from diffuse polarized emission from these galaxies show asymmetric variation about the foreground RM value as a function of azimuthal angle in the galactic disk. However, unlike our RM trend which stays nearly flat as a function of azimuth, \cite{braun2010} found maximum  $|$RM$|$ near the kinematic receding semi-major axis and maximum polarized intensity near the approaching semi-major axis. This characteristic RM and polarized intensity modulation were successfully fitted by an axisymmetric disk field with a quadrupolar extension into the halo. Unfortunately, due to the presence  of foreground Milky Way polarized emission, we cannot perform meaningful fits of $|$RM$|$ and polarized intensity against azimuth to our LMC diffuse emission data. 

This section suggests that even though the LMC is a dwarf irregular galaxy, its magnetic field symmetry with respect to its mid-plane appears to be similar to other typical spiral galaxies. We will discuss the origin of the LMC's large scale quadrupolar field in Section~\ref{section:origin_large_scale}.

\section{Is magnetic energy density in equipartition with cosmic ray energy density in the LMC?}
\label{section:eq_B_cr} 
Before we can derive the plane-of-the-sky magnetic field strength from the LMC's non-thermal radio continuum emission, we first need to establish whether the energy densities of cosmic rays and magnetic field are in equipartition. This is an assumption often made when no independent information on the cosmic ray density is available, but the validity of equipartition in the LMC was questioned by \cite{chi1993}. With the recent gamma-ray detection from the LMC made by Fermi LAT \citep{abdo2010}, it is now feasible to test whether equipartition is satisfied in the LMC. 

We assume that gamma rays in the LMC are produced predominantly by neutral pion decay resulting from collisions between cosmic-ray nuclei and protons in the LMC ISM \citep{abdo2010}. Given a cosmic ray proton spectrum in the form of a power law $dN_p/dE_p =N_{p,0}E_{p,GeV}^{-2}$ protons GeV$^{-1}$, the $\gamma$-ray flux from $\pi$$^0$ production at the LMC's distance (50 kpc) can be approximated by \citep{gaisser1998}
\begin{equation}
f_{\pi^0}(E_{\gamma,GeV}) = 5.3 \times 10^{-64} N_{p,0} n E_{\gamma,GeV}^{-2}~{\rm photons~GeV^{-1}~s^{-1}~cm^{-2}},
\end{equation}
where $n$$\sim$ 2 cm$^{-3}$ is the density of the LMC's ISM \cite{kim2003}. Equating the above expression to the integrated gamma ray energy spectrum of the LMC measured by LAT: $2.5 \times 10^{-11} E_{\gamma,erg}^{-2}$ photons~erg$^{-1}$~s$^{-1}$~cm$^{-2}$ between 200 MeV and 20 GeV \citep{abdo2010}, we find that this corresponds to a cosmic ray proton energy spectrum $dn_p/dE_p = 1.5 \times 10^{-14} E_{p,erg}^{-2}$ protons~erg$^{-1}$~cm$^{-3}$. The cosmic ray electron energy spectrum is of the form
\begin{equation}
dn_e/dE_{e,erg} = n_{e,0} E_{e,erg}^{\alpha_e}~{\rm electrons~erg^{-1}~cm^{-3}}, 
\end{equation}
where ${\alpha_e}$ is the power law index of cosmic ray electrons. Assuming that cosmic ray electrons do not suffer from radiation losses and that cosmic ray protons outnumber electrons by a factor of 100, then we obtain $n_{e,0}$$\sim$ 1.5 $\times$ 10$^{-16}$ electrons~erg~cm$^{-3}$ and $\alpha_e$ = -2. However, the observed non-thermal spectral index of the LMC as presented in Section~\ref{subsection:thermal_nonthermal}  is $\beta$ = -0.7. This corresponds to a cosmic ray electron energy spectral index $\alpha_e$ of -2.4, steeper than the  proton power law index of -2 implied from gamma ray observations, which suggests that relativistic electrons have aged. If the escape of cosmic ray electrons is negligible, the true normalization of the cosmic ray electron energy power law $n_{e,0}$ must exceed 1.5 $\times$ 10$^{-16}$ electrons~erg$^{-1}$~cm$^{-3}$. 

Using  $n_{e,0}$$=$1.5$\times$10$^{-16}$~electrons~erg$^{-1}$~cm$^{-3}$ as a lower limit, one can derive the upper limit to the mean total magnetic field strength in the LMC from the galaxy's observed gamma ray and radio spectrum using the standard synchrotron spectrum expected from a power law distribution of relativistic electrons \citep{rybicki1986,longair1994}. We use a total integrated synchrotron flux of 461 Jy at 1.4 GHz and a solid angle subtended by the LMC of $\sim$ 0.02 steradian (assuming a 46-Jy contribution from background continuum sources, \cite{hughes2007}). We further assume that the synchrotron emitting layer in the LMC has a similar scale height as edge-on spirals \citep{krause2009} of $\sim$ 2 kpc. We find an upper limit on the LMC's total magnetic field strength of 7 $\mu$G. 

To test if the assumption of equipartition holds, we can separately derive the total equipartition magnetic field strength B$_{eq}$  in $\mu$G by equating the energy density in cosmic rays  ($\epsilon_{CR}$)  to that in the magnetic field  ($\epsilon_{B}$). The total energy density in cosmic rays ($\epsilon_{CR}$) is dominated by protons as they exceed the number of electrons:
\begin{equation}
\epsilon_{CR} = \int ^{E_{max,erg}}_{E_{min,erg}} 1.5 \times 10^{-14} E_{p,erg}^{-2} E_{p,erg} dE ~\rm{erg}~cm^{-3}= \epsilon_{B} = \frac{B_{eq}^2}{8\pi}
\end{equation}
Assuming $E_{max}$=10$^4$ GeV \citep{lagage1983} and $E_{min}$ = 0.94 GeV (the spectral energy break for protons) yields an equipartition magnetic field strength $B_{eq}$ $\sim$ 2 $\mu$G, which is consistent with the upper limit of magnetic field strength (7 $\mu$G) that we found without evoking equipartition. This value is also consistent with a total magnetic field of $\sim$ 4 $\mu$G estimated from EGS RMs \citep{gaensler2005}. Our calculation shows that the equipartition assumption does not appear to be violated in the LMC, as opposed to the claim of \cite{chi1993}. Also, unlike the findings of \cite{pohl1993}, we suggest that using the standard cosmic ray proton to electron ratio of 100 can yield a magnetic field strength consistent with the equipartition assumption in the LMC. These results are in accord with the examination of equipartition assumption performed by \cite{abdo2010} using GALPROP, a code for calculating the propagation of cosmic rays and the diffuse emission produced by them.

\section{Bright Polarized Filaments in the Southeast of the LMC}
\label{section:filaments}
\subsection{The 3D structure of the polarized filaments}
\label{subsection:3d_filaments}

The brightest polarized feature in the diffuse emission map are the two filaments southeast of the LMC, just south of the 30 Doradus region (Figure~\ref{fig:combined_pi_3am}).

As demonstrated in Section~\ref{subsection:lmc_disk}, the turbulent LMC disk suffers severely from internal depolarization at 1.4 GHz. It also acts as an external Faraday screen that depolarizes any synchrotron emission emerging from behind the LMC disk. Since the observed median fractional polarization of the filaments is $\sim$ 8 \% at 1.4 GHz, the polarized filaments must be located on the near side of the LMC above its disk. This is in agreement with the conclusion drawn by \cite{klein1993}, who used the fact that no depolarization is seen towards HII region N206 A (5$^h$31$^m$22$^s$,$-$71$^d$04$^m$10$^s$) and the fact that the $|$RM$|$ they measured towards the filaments (few 10s of rad m$^{-2}$) is too small if they reside in the LMC mid-plane. Hence, \cite{klein1993} concluded that the polarized filaments must be extra-planar to the LMC disk and on the near side. 

\begin{figure}
\centering
\plotone{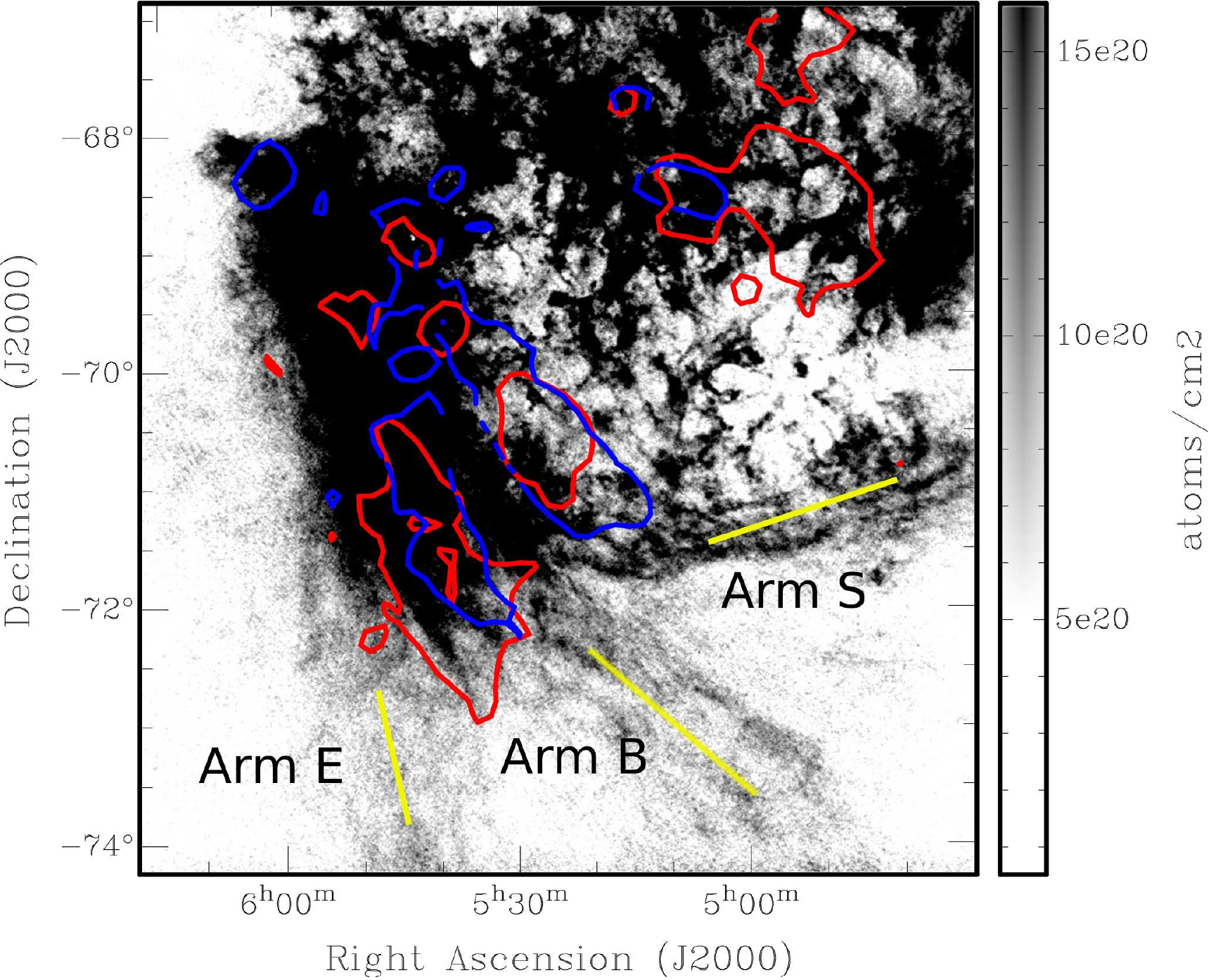}
\caption{The polarized intensity contours of the bright polarized filaments observed by Parkes at 1.4 GHz (red, at 0.06 Jy/ beam) and 2.3 GHz  (blue, at 0.06 Jy/beam) smoothed to 14' resolution overplotted on the HI column density map of the LMC \citep{staveleysmith2003} in the southeastern high neutral hydrogen column density region. Rough positions of Arms E, S and B defined by \cite{staveleysmith2003} are indicated using yellow line segments.}
\label{fig:pi_on_colden_arm}
\end{figure}

\begin{figure}
\centering
\plotone{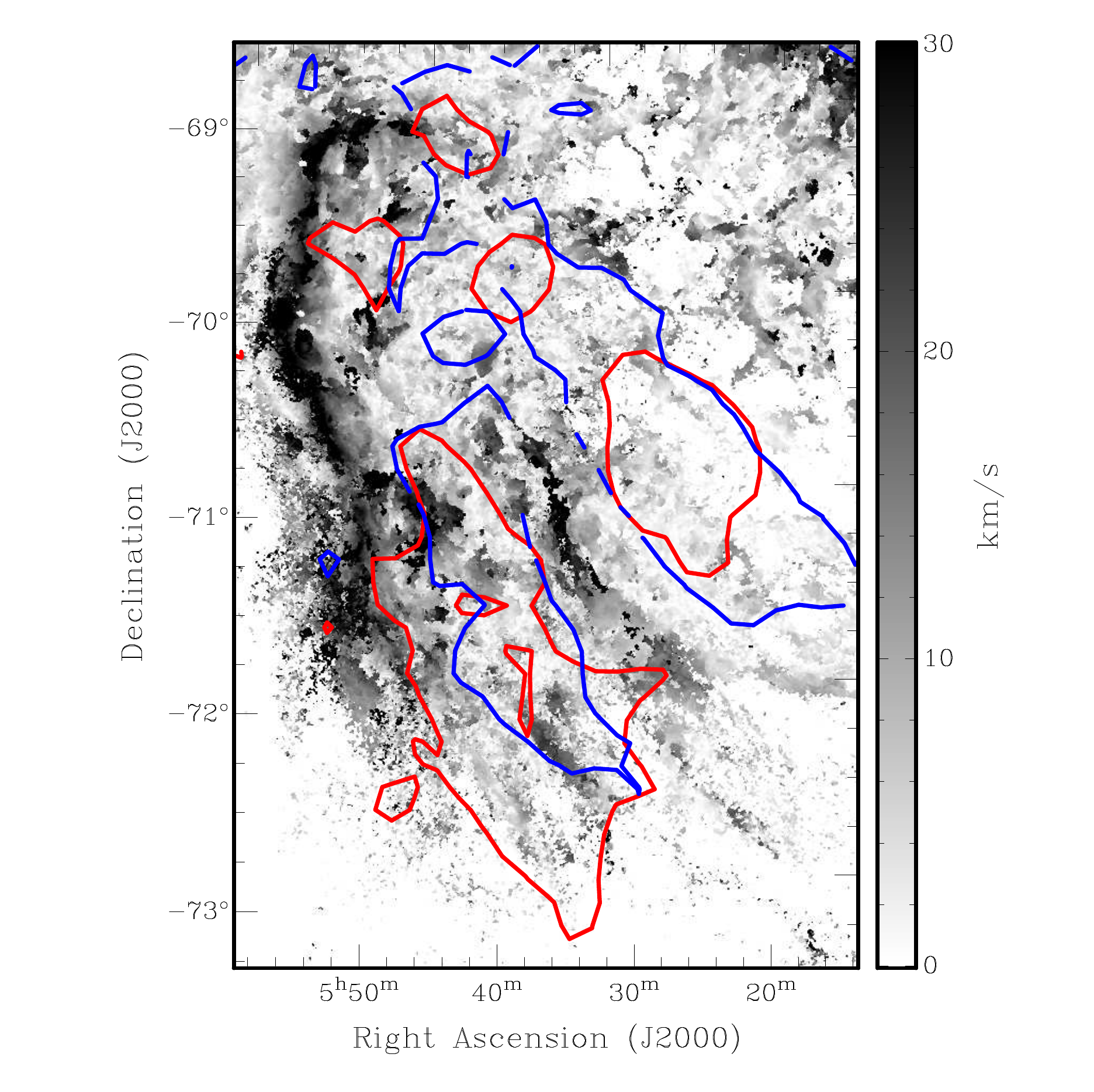}
\caption{The polarized intensity contours of the bright filaments southeast of the LMC at 1.4 GHz (red, at 0.06 Jy/beam level) and 2.3 GHz (blue, at 0.06 Jy/beam level) smoothed to 14' resolution overplotted on the vertical  HI velocity dispersion map \citep{staveleysmith2003} in the southeastern over-density region of the LMC. Dark regions correspond roughly to the location of the  L-component defined by \cite{luks1992}.}
\label{fig:pi_on_l_component}
\end{figure}

In Figure~\ref{fig:pi_on_colden_arm}, we plot the polarized intensity contours of the filaments at 2.45 GHz and 1.4 GHz on the HI column density map of the LMC \citep{staveleysmith2003} just south of the 30 Doradus region. The polarized filaments appear to be associated with extra-planar gas -- the eastern filament is situated between neutral hydrogen Arms ``E" and ``B" as defined by \cite{staveleysmith2003}, whereas the western filament appears to be co-located with Arm ``S". \cite{klein1993} suggested that the filaments are physically associated with the ``L-component" gas (i.e. at a lower radial velocity compared to the disk component) identified by \cite{luks1992}. In Figure~\ref{fig:pi_on_l_component}, we have plotted the vertical velocity dispersion of HI in the LMC (high values corresponds to the L-component position). The brightest parts of the polarized filaments do not directly correlate with the L-component; rather, they appear to avoid regions with high vertical velocity dispersion. This could be similar to the magnetic arm phenomenon seen in NGC 6946 \citep{beck2007}.
Despite the apparent anti-correlation between the L-component and the polarized intensity of the filaments, their similarities in morphologies could imply a physical relationship. There is evidence, based on a 21 cm absorption line measurement towards the continuum source MDM56 located in the LMC disk, that the L-component is located on the near side of the LMC halo \citep{marxzimmer2000}. If the polarized filaments and the L-component are physically related, then this suggests that the polarized filaments are located on the near side of the LMC halo (the same as the polarized emission from the rest of LMC's main body), yielding a consistent geometric picture when applying the depolarization arguments presented in Section~\ref{subsection:whereispol} specifically to the filaments.

\subsection{The Integrated Polarization Properties of the Filaments}
\label{subsubection:filament_pi_model}

We define the filament region to be between azimuthal angle  0$^\circ$ and 72$^\circ$ in the LMC's deprojected disk. The measured maximum brightness of the Milky Way foreground polarized emission is $\sim$ 0.06 Jy/beam at 1.4 GHz, therefore we assume that a pixel belongs to the filament if its emission exceeds 0.06 Jy/beam in this azimuth range. Using this boundary, we find 13 reliable EGS RMs behind the filaments and 17 off the filaments. The total RM through the filaments can be obtained by comparing the on/off-filament EGS RMs, assuming both sets of sightlines pass through the LMC's disk, as well as through its near and far-side halos. The median on-filament EGS RM after foreground removal is +19$\pm$9 rad m$^{-2}$, while the median off-filament EGS RM after foreground removal is +13$\pm$7 rad m$^{-2}$. Since the net RM through the filaments is consistent with zero ( $+$6 $\pm$ 12 rad m$^{-2}$ ), there seems to be no measurable coherent line-of-sight magnetic field through the filaments. A much denser RM grid of background polarized EGSs is needed to study the detailed line-of-sight magnetic field structure through the filaments. 

\begin{figure}
\centering
\plotone{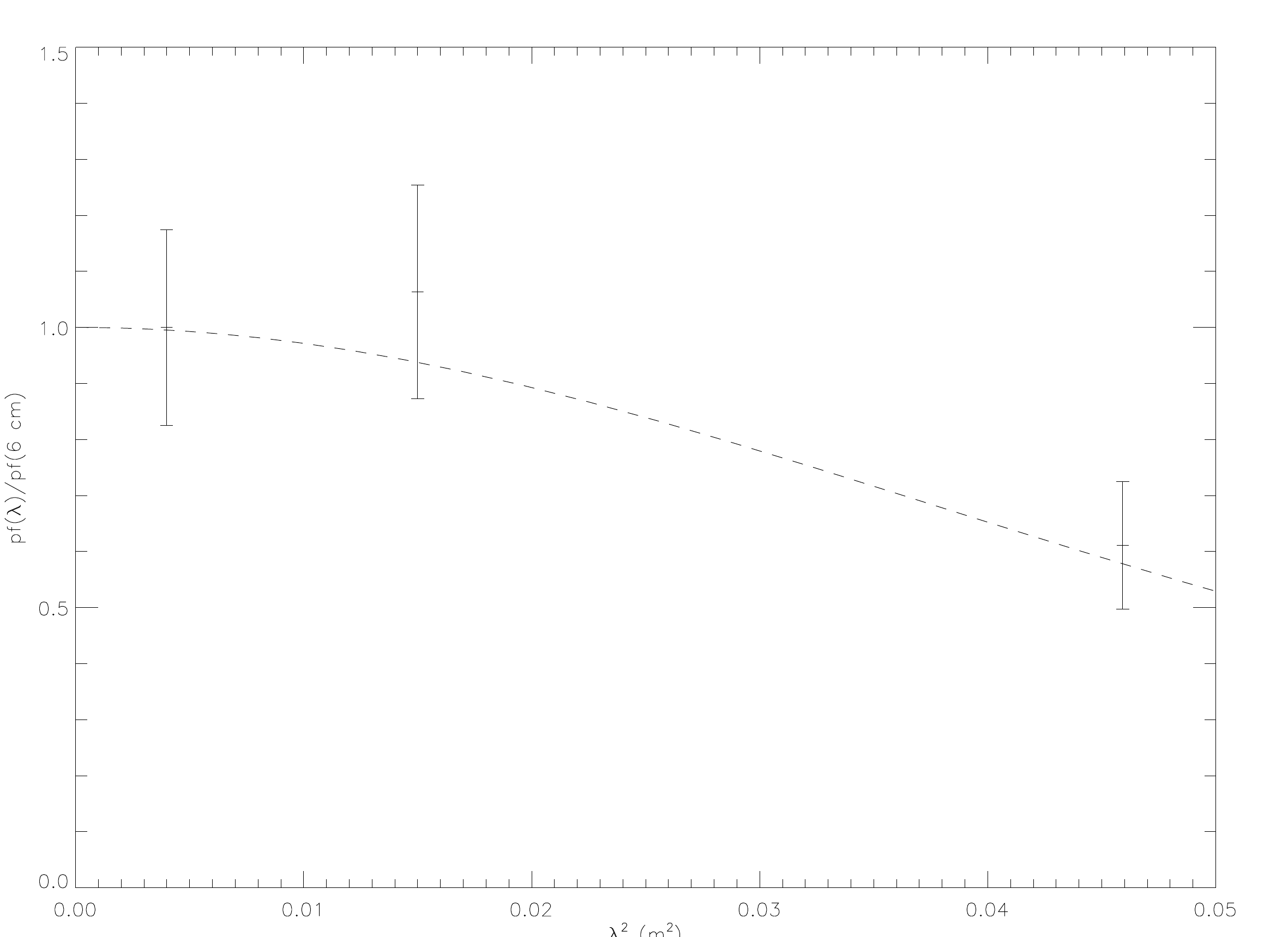}
\caption{The integrated depolarization of the filaments southeast of the LMC as a function of wavelength squared. The ratio of polarized fraction at 4.75 GHz, 2.45 GHz and 1.4 GHz to that at 4.75 GHz is plotted against wavelength squared. The depolarization trend can be well fitted using a model of internal Faraday dispersion with a random field characterized by an RM dispersion of 17 rad m$^{-2}$, as denoted by the dashed line.}
\label{fig:depol_filament}
\end{figure}

In Figure ~\ref{fig:depol_filament}, we have plotted the ratio of median fractional polarization from the filaments at 1.4 GHz, 2.45 GHz and 4.75 GHz to that at 4.75 GHz as a function of $\lambda^2$. The 2.45 and 4.75 GHz data are from  \cite{haynes1991}.  The median fractional polarization is computed from polarized intensity maps smoothed to a common resolution of 14'  within the azimuth range 0$^\circ$ and 72$^\circ$. Assuming that non-thermal fractions in the filaments are the same as the global LMC values (as listed in Section~\ref{subsection:thermal_nonthermal}), we find that the median fractional polarization decreases with increasing wavelength: from $\sim$ 32 \% at 4.75 GHz and 2.45 GHz to $\sim$ 19 \% at 1.4 GHz. This suggests that the filaments suffer from wavelength dependent depolarization. In this analysis, we have assumed that the fraction of total continuum emission that originates from the filaments remains the same at all wavelengths. The integrated polarization properties of the filaments can be well fitted by a model of internal Faraday dispersion with a random field represented by an RM scatter of 17 rad m$^{-2}$. The RM analysis on/off the filaments suggests that the filaments are unlikely to possess a genuine large scale field. The strong polarization observed in this region probably reflects an anisotropic random field.

\subsection{Plane-of-the-sky Magnetic Field Strength in the Filaments}
\label{section:pos_b_field}
Before estimating the plane-of-the-sky ordered magnetic field in the filaments, it is important to understand whether the enhancement in polarized intensity is due to an increase in cosmic ray electron density, in the total magnetic field strength or in the degree of ordering of magnetic fields.

\begin{figure}
\centering
\plotone{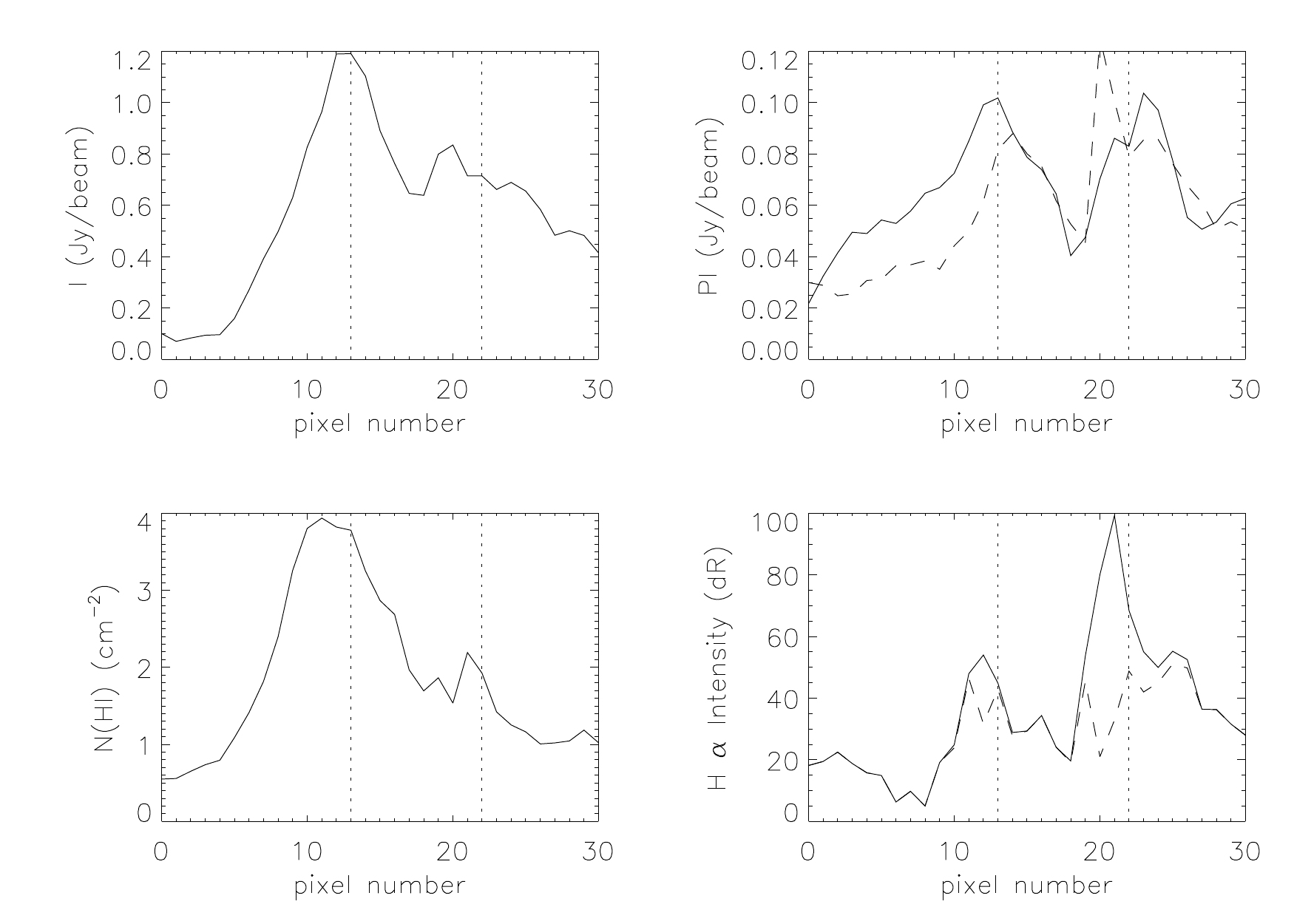}
\caption{Profiles across the polarized filaments southeast of the LMC. We have plotted the trend in total intensity (upper left panel), polarized intensity (upper right panel), HI column density (lower left panel) and H$\alpha$ intensity (lower right panel) across the filaments. The pixel locations of the filaments are indicated by vertical dotted lines in each plot. The solid and dashed lines in the polarized intensity plot denote the polarized intensities at 1.4 GHz and 2.45 GHz, respectively. In the H$\alpha$ intensity plot, the dashed line represent the trend in H$\alpha$ map after bright H II regions have been masked out.}
\label{fig:slice_through_fil}
\end{figure}

From the spatial distribution of gamma-ray emissivity of the LMC derived from {\em Fermi} LAT observations (Figure 10 of \citealt{abdo2010}), the polarized filaments do not correspond to areas of enhanced gamma-ray emissivity. In contrary, the filaments appear to be located in regions close to / below the mean $\gamma$-ray emissivity. If $\gamma$-ray emissivity distribution reflects the cosmic ray proton distribution, this indicates that the increase in polarized intensity of the filaments is not due to an increase in cosmic ray electron density. In the upper left panel of Figure~\ref{fig:slice_through_fil}, we have plotted the total intensity profile in the radio wavelengths across the filaments. The total intensity at 1.4 GHz peaks at the locations of the filaments. This is because of thermal emission from the coincidence of HII regions near the filaments, evident in the H$\alpha$ profile across the filaments in the lower right panel of Figure~\ref{fig:slice_through_fil}. However, since the higher H$\alpha$ intensity of the eastern filament does not correspond to a higher total intensity, this suggests there may be an increase in the non-thermal emission in the filaments at 1.4 GHz, assuming extinction is negligible. Therefore, the polarization enhancement in the filaments is due to an increase in the degree of ordering of the magnetic field and potentially an increase in the total magnetic field strength.

We can estimate the total and ordered magnetic field strengths using the integrated total intensity and polarization properties  of the filaments. We assume that the filaments are related to the neutral hydrogen L-component. Since  $\sim$ 60 $\%$ of the total HI column density toward the filaments belongs to the L-component, we assume that  $\sim$ 60\% of the observed total synchrotron intensity south east of the LMC is from the filaments. In addition, we assume that the non-thermal fraction in the filaments is the same as the global value in the LMC (63 \% at 1.4 GHz) and that the filament emission dominates within the Parkes polarized intensity boundary at 0.06 Jy/beam. We estimate a total synchrotron flux from the filaments of 55 Jy at 1.4 GHz within $\sim$ 2 $\times$10$^{-3}$ steradian (430 7' pixels). Using the standard synchrotron emission formula \citep[e.g.,][]{pacholczyk1970} and  $n_{e,0}$$=$1.5$\times$10$^{-16}$~electrons~erg$^{-1}$~cm$^{-3}$ (Section~\ref{section:eq_B_cr}), we obtain an upper limit on the total plane-of-the-sky magnetic field in the filaments of 11 $\mu$G for a path length of 800 pc, assuming the filaments have similar depth along the line-of-sight as their width in the plane-of-the-sky. 
The ordered component of the magnetic field can be estimated using $q$, the ratio of random to ordered field strength. We derive this ratio using the polarized fraction of the filaments at 4.75 GHz (at $\sim$14' resolution)
\begin{equation}
P_{6cm}/P_0 = \frac{1}{1+\frac{2}{3}q^2}
\end{equation}
\citep{sokoloff1998}. The non-thermal polarized fraction of the filaments at 4.75 GHz is 0.55, while the intrinsic polarized fraction of synchrotron radiation P$_0$ is $\sim$ 72 \% for a spectral index of $-$0.7. This suggests that q $\sim$ 0.76. Hence, the ordered plane-of-the-sky magnetic field strength in the filaments is $\le$ 9.0 $\mu$G.  The derived magnetic field strengths are upper limits not only because the normalization $n_{e,0}$ of the power-law distribution is a lower limit (see Section~\ref{section:eq_B_cr}), but also because the enhancement in polarization might be  due to an anisotropic field (magnetic fields with same plane-of-the-sky orientations but with frequent field reversals), rather than a  large-scale coherent field without reversals.

The orientation of the ordered magnetic field can be inferred from measured polarization position angles (corrected for Faraday rotation effects as measured by the RM synthesis analysis in Section~\ref{section:rm_determination}). We obtain a median polarization position angle in the filaments of +149$^\circ$, measured north through east, from our 1.4 GHz Parkes observations. The magnetic field orientation is therefore at a position angle of +59$^\circ$ $\pm$ 1$^\circ$ \footnotemark[9]\footnotetext[9]{This is in rough agreement with the implied magnetic field orientations at 4.75 GHz of +50$^\circ$ $\pm$ 1$^\circ$ and at 2.45 GHz of +59$^\circ$ $\pm$ 1$^\circ$.}.

As the estimated plane-of-the-sky magnetic field strength is 11 $\mu$G while the line-of-sight component is consistent with zero, we assume that the magnetic fields in the filaments lie mostly in the sky plane. One can now compare the orientation of the filaments' magnetic field with various characteristic vectors in the Magellanic system: The proper motion of the LMC is directed at a position angle of +78$^\circ$ \citep{kallivayalil2006a}, measured north through east. Meanwhile, the Small Magellanic Cloud is situated at a position angle of -129$^\circ$, which is parallel to a position angle of +51$^\circ$, measured north through east. By comparing the LMC's proper motion direction and the SMC's location with the position angle of the magnetic field orientation at +59$^\circ$, we suggest that it is more likely that the filaments' magnetic field is aligned with the direction towards the SMC than along the LMC's proper motion direction. We discuss the implication of this potential alignment in Section~\ref{section:origin_filaments}.

\section{Origin of the LMC's Large Scale Magnetic Field}
\label{section:origin_large_scale}

In this section, we consider the mechanism responsible for producing the observed large-scale quadrupolar magnetic field in the LMC (Section~\ref{subsection:large_scale_sym}). Galactic-scale coherent magnetic fields are thought to be generated by the dynamo mechanism. The classic $\alpha$-$\omega$ dynamo requires turbulence to rise above or below the galactic disk to transform an azimuthal field into a poloidal one. The poloidal field is then transformed back into an azimuthal component by differential rotation of the disk \citep{beck1996}. This process is thought to be at work in spiral galaxies.  We can test if the  $\alpha$-$\omega$ dynamo could work in the LMC by computing the dynamo number, which characterizes the growth rate of magnetic fields,
\begin{equation}
\label{eq:dynamonumber}
D = \frac{ 9\Omega sh_0^2}  {u_{0}^2}  \frac {\partial \Omega}{\partial s} ~.
\end{equation} 
where $s$ is the radial distance from the center of the galaxy,  $h_0$$\sim$ 600 pc \citep{kim1998} is the scale height of the neutral gas disk,  $u_0$$\sim$16 km s$^{-1}$ \citep{kim1998} is the velocity of turbulent motion of gas in the LMC and $\Omega$ is the angular velocity of the rotating disk as inferred from the HI rotation curve measured by \cite{kim1998}. Under the condition that shearing of the disk dominates, we compute $D$ in the range 0.5 kpc$\le$s$\le$3 kpc. We find a mean dynamo number of 10, which is close to the critical value ( $|$$D_{\rm critical}$$|$ $\sim$ 8 $-$ 10) \citep{shukurov2004}. This suggests that magnetic fields in the LMC can grow exponentially via the $\alpha$-$\omega$ dynamo.
The predicted pitch angle and the strength of the magnetic field from a dynamo mechanism can be calculated using expressions given by \cite{shukurov2004}. The predicted pitch angle is -10$^\circ$, while the observed value is consistent with zero. This could be attributed to the uncertainty associated with the position angle of the line of node of the LMC disk. A finer background RM grid is needed in order to refine the pitch angle measurement. The predicted magnetic field strength from a dynamo is 1.6 $\mu$G which is consistent with the observed $\mu$G field (see Section~\ref{subsection:egs_rm_determination}). The observed quadrupolar symmetry of the LMC's large scale magnetic field is consistent with the prediction from the dynamo mechanism  as well -- it is the most readily excited dynamo mode in disk-like systems due to its large scale height and longer decay time than a dipolar-type field \citep{shukurov2004}. Therefore, we conclude that LMC's large scale coherent field could potentially be produced by the mean field dynamo.

However, the growth time scale, $\Gamma$, of coherent magnetic fields for the mean field dynamo is estimated to be $\sim$ 2.7 Gyrs using \citep{shukurov2004}:
\begin{equation}
\Gamma = \frac{3 h_0^2}{l_0 u_0 (\sqrt{|D|}-\sqrt{|D_{critical}|})}, 
\end{equation}
(where l$_0$ is the turbulent scale) much longer than time scale of star-formation triggered by the interaction between the Magellanic Clouds \citep[see for example][]{harris2009}. 
Large levels of energy injection into the ISM by bursts of star formation within the past Gyr should prevent the build up of a coherent large-scale field by the classical mean field dynamo mechanism \citep{chyzy2004,gaensler2005}. 

\cite{gaensler2005} suggested that the cosmic ray driven dynamo \citep{hanasz2009}, in which vigorous star formation results in cosmic rays whose pressure inflates magnetic loops into the LMC halo, effectively enhancing the $\alpha$ effect, is at work in the LMC because of its fast ($\sim$ 700 Myr) amplification time scale. Recently, \cite{siejkowski2010} have implemented the cosmic ray driven dynamo in irregular galaxies using typical rotation velocities, shear and star formation rates of irregulars  to test whether the dynamo can effectively amplify magnetic fields. We find their model `SF10R.03Q.5' to be most similar to the properties of the LMC. This model allows a build up of disk magnetic field strength to $\sim$ 1 $\mu$G with an e-folding time of $\sim$ 400 Myrs. Since the most recent star forming episodes in the LMC (12 Myrs and 100 Myrs ago) were concentrated at the 30 Doradus region and the LMC's bar, the disruption of the global field might be limited to these local regions and hence the large-scale magnetic field could be preserved.

The consistent negative RM of diffuse polarized emission at 1.4 GHz across the face of the LMC suggests that the near-side halo magnetic field is coherently directed away from the observer. The Parker loops (from the cosmic ray driven dynamo) will produce large fluctuations in RM of opposite signs \citep{haverkorn2011}. If the loops have small vertical extent and they remain embedded in the LMC disk, the magnetic fields of the loops will leave imprints on the EGS RMs, but not the diffuse RM from the near-side LMC halo. Also, these loops would have a relatively low volume filling factor, which leads to weak diffuse emission. Magnetic fields of quadrupolar symmetry follow naturally from dynamo theory, for which even parity is favored as it has a twice larger scale in the vertical direction \citep{shukurov2004}.  Unfortunately, no prediction on the symmetry of magnetic fields perpendicular to the galactic disk was produced by \cite{hanasz2009} in their cosmic-ray driven dynamo simulations. Therefore, we cannot directly compare the observations with simulation predictions for the magnetic field structure in the halo. None-the-less, in this section, we have shown that the magnetic field properties of the LMC are consistent with the mean field dynamo but its amplification time scale is too long to explain the coherent fields observed in the LMC, since there were many bursts of star formation in the past $\sim$ 2 Gyrs. Instead, we suggest that the cosmic-ray driven dynamo could be at work in the LMC to produce the observed signatures in EGS and diffuse RMs.

\section{Magnetic Fields Originating from Tidal Filaments }

\label{section:origin_filaments}
In this section, we consider several possible mechanisms for illuminating the filaments in polarization southeast of the LMC. 

\subsection{Magnetic draping of the Milky Way halo field?}
Recently, \cite{pfrommer2010}  proposed that the asymmetric polarization properties of galaxies in the Virgo cluster can be explained by the draped intracluster medium (ICM) magnetic field around member galaxies as they orbit the cluster center. The fields are then illuminated by cosmic rays accelerated in supernova remnants within the draped layer, emitting polarized synchrotron radiation. This is analogous to the Large Magellanic Cloud's motion through the Milky Way halo. Therefore, we consider the possibility that draped Galactic halo magnetic fields around the LMC produce the polarized filaments.

If draping is the dominant effect, one does not expect density enhancements to coincide with increases in polarized intensity as draping is not a compression effect but rather a boundary effect. This is contrary to what is seen in the southeastern filaments:  the bottom left panel of Figure~\ref{fig:slice_through_fil} demonstrates that the positions of the filaments correspond to peaks in HI column density (more so for the western filament than the eastern one). Moreover, if draping is responsible for producing the polarized emission in the filaments, to supply enough relativistic electrons, one expects the optical / IR emission from the stellar population to overlap with the magnetic drape and to lead the polarized emission, which is not in agreement with our observations either. Therefore, we conclude that even though draping of the Milky Way halo magnetic field around the LMC is likely to be occurring, it is not the main mechanism responsible for producing the observed polarized intensity and rotation measure of the filaments.

\subsection{Ram pressure effects on the ISM in the LMC }

Ram pressure of the Galactic halo hot gas can act on the ISM in the LMC as the LMC plows through the Milky Way halo. As the lowest density constituent of the ISM, the relativistic ISM (cosmic rays and magnetic fields) can be redistributed easily by ram pressure because the ram pressure experienced by the LMC ($\sim$2 $\times$10$^{-13}$ dynes cm$^{-2}$) is comparable to the total pressure in LMC's relativistic ISM assuming equipartition ($\sim$3 $\times$10$^{-13}$ dynes cm$^{-2}$). 
The effects of ram pressure on non-relatistivic components of the LMC's ISM  have been modeled in detail by \cite{mastropietro2009}. These simulations showed that ram pressure acting on the LMC when it was moving face-on ($>$ 30 Myrs ago) with respect to its proper motion vector produced the patchy and filamentary distribution of neutral hydrogen and H$\alpha$ observed across the LMC disk at present. On the other hand, the recent (30 Myrs ago) edge-on motion of the LMC with respect to its proper motion could have reproduced the high density HI filaments and high concentration of H$\alpha$ emission on the eastern side of the LMC disk. 
The ram pressure to which the LMC's relativistic ISM is subjected is analogous to that produced by the motion of cluster galaxy members' in the ICM. \cite{ov2003} (OV2003 hereon) modeled the evolution of magnetic fields of a galaxy moving nearly edge-on through the ICM at different stages of its orbit using 3D magnetohydrodynamic (MHD) simulations. The fact that the current  LMC proper motion vector lies mainly in the sky plane makes comparison between their simulation and our observations feasible. If the LMC is  on its first passage into the Milky Way \citep{besla2007}, its space velocity and the density of the ambient medium are both increasing. In this case, the ``pre-ram pressure peak" phase in the OV2003 simulations would be most similar to  the LMC's current motion in the Milky Way halo. The polarized intensity map of the simulated galaxy during this phase shows a brightly polarized ridge near the leading edge of the galaxy's motion, shifted slightly towards the direction of galactic rotation. The polarized emission is nearly symmetric about the leading edge and has the form of half-ring around the galaxy's center. This prediction is at odds with our observations -- the polarized filaments we observe are not symmetric about the current proper motion direction of the LMC. Instead, the brightest polarized emission is offset from the leading edge in the direction opposite the LMC's rotation. More importantly, the ram pressure experienced by the LMC is $\sim$ 300 times smaller than the simulated galaxy in OV2003, so it is unclear if the ram pressure acting on LMC's ISM is strong enough for similar signatures to be present in polarized emission. Therefore,  we conclude that ram pressure effects alone are unlikely to have been able to produced the polarized filaments.

\subsection{Tidal interactions originated Magnetic Fields}

The Large Magellanic Cloud is being subjected to tidal forces from the Small Magellanic Cloud as the two are likely to have been in a binary orbit for the last 2 Gyrs \citep[][e.g.,]{besla2010,diaz2011}. As pointed out in Section~\ref{subsection:3d_filaments}, both polarized filaments appear to associate with gaseous arms in the LMC: the eastern filament appears to be in the inter-arm region between Arm B and E while the western filament appear to be part of Arm S. Arm E leads to the Magellanic leading arm clouds, while Arm B connects to the Magellanic Bridge (which joins the Magellanic Clouds). Arm S appears to be the southern boundary of the main body of the LMC but its origin is unclear. The possible alignment of  the magnetic field orientation in the filaments at +59$^\circ$ with the direction of the SMC (along position angle -129$^\circ$) suggests that magnetic field in the filaments could have been shaped by tidal interactions between the Magellanic Clouds. 

As suggested by \cite{staveleysmith2003}, the HI arms in the LMC are likely to be extra-planar. Therefore, we propose that when these HI arms formed, frozen-in magnetic fields were displaced along with the gas. As the gas was stretched along the direction of interaction, the ordered component of the frozen-in magnetic field parallel to the direction of the SMC become preferentially enhanced, resulting in a high fractional polarization parallel to the position angle of the SMC. Polarized emission of such nature has been observed in the merging spirals NGC 4038/4039 (The Antennae Galaxies)-- a highly polarized synchrotron ridge (with apparent magnetic field vectors running parallel to the ridge) appears to emerge at the base of the southern tidal tail of the system \citep{chyzy2004}. The polarized ridge reveals a coherent large scale field likely due to gas stretching motions. According to the latest numerical simulations of the NGC 4038/4039 system, the interaction that formed the tidal tail occurred roughly 600 Myrs ago \citep{karl2010} and the polarized emission remains bright until the present day. A recent MHD simulation of this galaxy pair has modeled the magnetic field evolution during the interaction and has successfully reproduced the synchrotron ridge near the base of the southern tidal tail \citep{kotarba2010}.

If the magnetic fields in the polarized filaments of the LMC are similarly of tidal origin, this fits well within the ``pan-Magellanic" magnetic field hypothesis framework. 
\cite{mathewson1970} first proposed, using starlight polarization measurements towards  Magellanic Clouds stars, that the plane-of-the-sky magnetic field orientation of the filaments appears to be preferentially aligned with the Magellanic Bridge, suggesting a large scale magnetic field encompassing the Magellanic System. The possible alignment of the filaments' magnetic field with the position angle of the SMC is suggestive that the filaments may indeed trace part of the ``pan-Magellanic" magnetic field. Since the eastern filament is likely to be associated with Arm E which is connected to the leading arm feature \citep{staveleysmith2003}, the magnetic field found in the filaments gives support to the recent discovery of  a 6$\mu$G coherent magnetic field in the Leading Arm high velocity cloud (HVC)  297.5+22.5+240 \citep{mg2010}, as this is indicative that this HVC was magnetized when it was first stripped off.

\section{Conclusions} 
We have mapped the diffuse synchrotron polarization from the Large Magellanic Cloud  with the Parkes radio telescope at 1.4 GHz. Combined with higher resolution interferometric data from the ATCA, we have produced a map of polarized emission sensitive to all scales. Analyzing the RMs of extragalactic polarized sources in conjunction with RMs derived from diffuse polarized emission, we demonstrate that diffuse polarized emission at 1.4 GHz likely originates from a region above the LMC's mid-plane. The consistent negative sign of the foreground RM subtracted diffuse RM suggests a magnetic field directed coherently away from us on the LMC's near-side halo. We propose that the LMC possesses a quadrupolar  large-scale magnetic field consistent with a dynamo: an azimuthal in-disk field that is symmetric across the mid-plane and a vertical field in the halo that reverses direction across the mid-plane. 
We find two brightly polarized filaments south of the 30 Doradus region. We suggest that the alignment of magnetic fields in the filaments with the position angle of the Small Magellanic Cloud in the sky plane could mean that the magnetic fields in the filaments are part of a pan-Magellanic magnetic field which is of tidal origin.

Future observations of a dense RM grid of EGSs behind the LMC with the Australian Square Kilometer Array Pathfinder and the eventual Square Kilometer Array will provide a much more detailed view of LMC's global field as well as the magnetic fields in the polarized filaments. All-sky absolutely calibrated multi-frequency polarization surveys will be able to clarify the nature of the Milky Way polarized foreground in the vicinity of the LMC and allow proper foreground Stokes Q and U subtraction so that the magnetic fields in the LMC in the sky plane can be extracted.  More sensitive multi-frequency observations of the LMC's diffuse polarized emission at short wavelengths will also be essential for developing  a complete model of LMC's magnetism.

\textbf{Acknowledgements} 
The authors thank Ettore Carretti for kindly providing the IDL mapmaking software to process our Parkes data. We thank Dominic Schnitzeler and Tim Robishaw the discussions and coding of RM synthesis and deconvolution in IDL. The authors thank Pat Slane for helpful discussions on computing the equipartition magnetic field strength in the LMC. We also thank Tom Landecker, Mark Calbertta, Annie Hughes, Doug Finkbeiner, Troy Porter, John Raymond, You-Hua Chu, Chris Pfrommer, Robert Gruendl, Steve Snowden, Sean Points, John Dickel, Anvar Shukurov, Amanda Kepley, Greg Madsen and Michal Hanaz for useful discussions and email correspondence. 
The Australia Telescope Compact Array and the Parkes radio telescope are part of the Australia Telescope National Facility which is funded by the Commonwealth of Australia for operation as a National Facility managed by CSIRO. 
B. M. G. acknowledges the support in part by an Australian Research Council Federation Fellowship (FF0561298).  M. H. acknowledges funding from the European Union's Seventh Framework Programme (FP7/2007-2013) under grant agreement number 239490 and research programme 639.042.915, which is (partly) financed by the Netherlands Organisation for Scientific Research (NWO). 

 {\it Facilities:} ATCA Parkes


\clearpage
\clearpage

\clearpage

\clearpage

\clearpage

\clearpage

\clearpage

\clearpage

\clearpage

\clearpage

\clearpage

\clearpage

\clearpage

\clearpage

\clearpage

\clearpage

\clearpage

\clearpage

\clearpage

\clearpage

\clearpage

\clearpage

\begin{deluxetable}{lrrr}
\centering
\tablecolumns{3} 
\tablewidth{0pc} 
\tablecaption{Calibration Scale Factor used to combine ATCA and Parkes data} 
\tablehead{   
\colhead {Frequency (GHz)} & \colhead{Scale Factor} & \colhead{Parkes Beam Size (arcsec$^2$)}
}

\startdata 
1.424 &   1.114 &   903$\times$876  \\ 
1.416 &   1.136 &  910$\times$884  \\  
1.400 &  1.083 &  912$\times$888   \\
1.392 &   1.058 &  915$\times$897 \\
1.384 &  1.062 &  918$\times$897 \\ 
1.376 &  1.064 &  918$\times$898 \\ 
1.368 &    1.055 &  924$\times$903\\  
1.360 &  1.061 &  927$\times$903 \\  
1.352 &   1.029 &  928$\times$909 \\ 
1.344 &  1.033 &  933$\times$914   \\
1.336 &   1.030 &  936$\ltimes$919  \\
1.328 &   1.018 &  940$\times$919\\

\enddata 
\label{table:cal_factor}

\end{deluxetable} 
\clearpage

\clearpage
\LongTables

\begin{deluxetable}{llrrrc}
\tablecolumns{8} 
\tabletypesize{\footnotesize}
\tablewidth{0pc} 
\tablecaption{Rotation Measures of Extragalactic Sources towards the LMC from ATCA observations} 
\vspace{0.01in}
\tablehead{   
\colhead{RA(J2000)(hms)} &     \colhead{DEC(J2000)(dms)}   &    \colhead{RM(rad m$^{-2}$)}   & \colhead{PI (mJy)\tablenotemark{a}} & \colhead{I (mJy)\tablenotemark{a}} & \colhead{Used in fit?\tablenotemark{b}}}
\startdata 
 6:17:21.53 & -68:30:27.36 & $+$ 15$\pm$16 &    7.70 &    45.6 &   - \\
 6:28:43.06 & -72:56:23.28 & $+$ 45$\pm$10 &    8.00 &     319 &  fg \\
 6:16:56.59 & -69:15:43.92 & $+$ 25$\pm$14 &    4.40 &     213 &   - \\
 6:23:55.82 & -71:51:28.44 & $-$  4$\pm$28 &    2.30 &    40.3 &  fg \\
 6:06:26.54 & -64:12:57.96 & $+$ 27$\pm$18 &    3.80 &    63.4 &  fg \\
 6:09:52.06 & -66:32:53.16 & $+$ 28$\pm$ 6 &    8.70 &    62.8 &  fg \\
 6:09:30.31 & -66:21:10.08 & $+$ 32$\pm$ 9 &    6.10 &    55.5 &  fg \\
 6:04:05.26 & -63:14:12.84 & $+$ 48$\pm$14 &    4.00 &    33.5 &  fg \\
 6:13:48.34 & -68:48:45.72 & $+$ 23$\pm$ 6 &    8.60 &    39.4 &   - \\
 6:22:33.86 & -72:02:00.60 & $+$  8$\pm$14 &    3.70 &    43.1 &  fg \\
 6:22:18.70 & -72:01:15.96 & $+$ 65$\pm$10 &    5.00 &    45.9 &  fg \\
 6:13:01.87 & -68:48:02.52 & $+$ 30$\pm$23 &    2.00 &    16.2 &   - \\
 6:02:42.89 & -63:21:16.20 & $+$ 48$\pm$12 &    3.90 &    76.2 &  fg \\
 6:12:27.10 & -68:54:49.32 & $-$  5$\pm$17 &    2.60 &    25.9 &   - \\
 6:01:54.31 & -63:14:10.68 & $+$ 30$\pm$19 &    2.40 &    34.9 &  fg \\
 6:08:29.81 & -67:23:34.44 & $+$ 29$\pm$22 &    2.10 &    21.1 &   - \\
 6:05:17.76 & -66:25:13.08 & $+$ 23$\pm$16 &    2.80 &    77.6 &   - \\
 6:00:30.38 & -63:13:15.60 & $+$ 30$\pm$17 &    2.60 &    28.6 &  fg \\
 6:14:46.58 & -71:13:44.76 & $+$ 57$\pm$20 &    2.50 &    72.3 &   - \\
 6:12:31.39 & -70:44:14.28 & $+$ 27$\pm$27 &    1.80 &    35.0 &   - \\
 6:00:53.71 & -64:27:53.64 & $+$ 41$\pm$17 &    2.70 &    44.1 &  fg \\
 6:07:46.92 & -68:39:47.52 & $+$ 28$\pm$ 6 &    6.80 &    82.0 &   - \\
 6:05:50.50 & -67:43:05.52 & $+$ 16$\pm$11 &    3.90 &    23.6 &   - \\
 6:00:42.22 & -64:28:22.44 & $+$ 44$\pm$ 3 &    13.2 &     173 &  fg \\
 6:04:17.93 & -66:51:29.16 & $+$ 48$\pm$14 &    3.10 &    38.0 &   - \\
 6:05:32.52 & -67:36:07.56 & $+$ 22$\pm$12 &    3.70 &    51.7 &   - \\
 6:15:48.50 & -72:23:02.04 & $+$ 24$\pm$14 &    3.50 &    38.3 &  fg \\
 6:09:35.62 & -70:08:19.68 & $+$ 49$\pm$11 &    4.10 &    82.4 &   - \\
 6:19:54.60 & -73:53:06.36 & $+$ 41$\pm$21 &    2.20 &    14.2 &  fg \\
 6:18:49.18 & -73:40:45.12 & $+$ 41$\pm$14 &    3.40 &     129 &  fg \\
 6:07:22.37 & -69:26:33.00 & $+$ 63$\pm$22 &    2.00 &    21.2 &   - \\
 6:00:08.50 & -65:12:42.84 & $+$ 41$\pm$11 &    4.40 &    40.8 &  fg \\
 6:00:04.66 & -65:13:35.76 & $+$ 44$\pm$15 &    3.20 &    43.8 &  fg \\
 6:17:40.27 & -73:33:16.20 & $-$ 10$\pm$24 &    1.90 &    39.0 &  fg \\
 5:59:57.70 & -65:32:30.48 & $+$ 68$\pm$20 &    2.30 &    35.1 &  fg \\
 6:10:24.82 & -71:19:17.76 & $+$ 22$\pm$ 6 &    8.30 &     297 &   - \\
 6:01:18.48 & -66:54:29.88 & $+$ 17$\pm$18 &    2.50 &    23.8 &   - \\
 6:00:03.72 & -66:12:53.64 & $+$ 14$\pm$ 9 &    5.00 &     148 &   - \\
 6:13:06.94 & -72:41:07.44 & $+$ 42$\pm$15 &    3.10 &     108 &  fg \\
 6:11:57.98 & -73:00:12.24 & $+$ 45$\pm$ 7 &    6.70 &     148 &  fg \\
 6:02:22.37 & -68:44:25.44 & $+$ 21$\pm$26 &    1.70 &    46.0 &   - \\
 6:05:31.08 & -71:12:31.32 & $+$ 39$\pm$16 &    2.90 &    67.1 &   - \\
 6:08:13.97 & -72:42:04.32 & $+$ 46$\pm$20 &    2.40 &    66.6 &  fg \\
 6:03:20.47 & -70:31:33.96 & $+$ 38$\pm$18 &    2.50 &    42.8 &   - \\
 6:01:34.94 & -69:55:36.48 & $-$  9$\pm$23 &    1.90 &     170 &   - \\
 5:51:58.51 & -62:53:29.40 & $+$  5$\pm$29 &    2.50 &    25.5 &  fg \\
 6:08:41.26 & -73:21:34.20 & $+$ 28$\pm$15 &    3.10 &    46.9 &  fg \\
 5:51:47.42 & -62:51:44.28 & $+$ 17$\pm$18 &    4.40 &    18.0 &  fg \\
 5:59:42.94 & -69:19:24.96 & $+$ 12$\pm$15 &    3.00 &    17.6 &   - \\
 5:51:24.55 & -62:49:47.28 & $+$ 32$\pm$10 &    9.40 &     208 &  fg \\
 5:52:25.46 & -64:01:52.68 & $+$ 27$\pm$ 4 &    10.9 &     199 &  fg \\
 6:00:35.74 & -70:12:16.56 & $+$ 44$\pm$15 &    3.00 &    93.5 &   - \\
 6:01:11.30 & -70:36:10.44 & $+$ 30$\pm$15 &    3.10 &     229 &   - \\
 6:05:58.03 & -73:13:18.84 & $+$ 19$\pm$10 &    4.60 &     111 &  fg \\
 6:00:05.09 & -70:38:34.80 & $+$ 13$\pm$15 &    3.10 &     460 &   - \\
 6:05:33.91 & -73:34:21.00 & $-$ 16$\pm$20 &    2.40 &     122 &  fg \\
 5:54:18.58 & -68:06:01.08 & $-$ 34$\pm$ 9 &    4.80 &    36.9 &   - \\
 6:03:50.66 & -73:25:19.56 & $+$ 64$\pm$32 &    1.50 &    40.9 &  fg \\
 6:00:03.46 & -71:42:09.72 & $+$ 41$\pm$ 6 &    8.10 &     153 &   - \\
 6:01:18.53 & -72:20:32.28 & $+$ 57$\pm$26 &    1.80 &    53.4 &   - \\
 6:01:41.45 & -72:38:29.40 & $+$ 36$\pm$14 &    3.20 &    69.2 &   - \\
 6:01:26.38 & -72:44:30.48 & $+$ 37$\pm$23 &    2.00 &    7.40 &  fg \\
 6:00:55.92 & -72:46:22.08 & $+$ 48$\pm$24 &    1.90 &    91.1 &  fg \\
 5:49:19.39 & -64:37:42.60 & $+$ 15$\pm$18 &    2.50 &    14.3 &  fg \\
 5:55:16.94 & -69:45:20.52 & $+$ 69$\pm$22 &    2.00 &    23.7 &   - \\
 5:53:39.96 & -68:46:33.96 & $+$ 45$\pm$13 &    3.50 &    41.1 & lmc \\
 5:55:02.74 & -69:51:49.32 & $+$ 94$\pm$25 &    1.80 &    49.7 &   - \\
 5:52:05.93 & -68:14:38.76 & $-$ 14$\pm$ 5 &    8.50 &     186 & lmc \\
 5:52:43.42 & -68:51:07.56 & $+$ 57$\pm$22 &    2.00 &    38.7 & lmc \\
 5:55:16.78 & -70:37:37.56 & $+$ 25$\pm$19 &    2.40 &    29.9 &   - \\
 5:56:58.34 & -71:59:53.16 & $+$ 54$\pm$11 &    4.40 &    46.3 &   - \\
 5:59:51.67 & -73:35:17.16 & $+$ 19$\pm$14 &    3.30 &    35.6 &  fg \\
 5:51:39.94 & -68:43:12.72 & $+$ 61$\pm$16 &    2.80 &    75.9 & lmc \\
 5:48:46.99 & -66:15:21.60 & $-$  3$\pm$10 &    4.80 &    44.3 &   - \\
 5:56:34.58 & -72:06:42.48 & $+$ 64$\pm$ 8 &    6.10 &    46.4 &   - \\
 5:47:25.92 & -65:04:05.88 & $+$ 19$\pm$11 &    4.20 &    78.8 &   - \\
 5:52:30.07 & -69:47:57.12 & $+$151$\pm$14 &    3.20 &    75.0 & lmc \\
 5:47:05.35 & -64:47:49.56 & $+$ 69$\pm$16 &    2.80 &    28.1 &   - \\
 5:47:15.12 & -65:05:36.96 & $+$ 21$\pm$12 &    3.90 &    51.3 &   - \\
 5:51:21.96 & -69:04:54.84 & $+$ 73$\pm$20 &    2.20 &    31.1 & lmc \\
 5:51:30.36 & -69:16:30.72 & $+$117$\pm$ 5 &    8.70 &    96.2 & lmc \\
 5:57:42.12 & -73:09:45.36 & $+$ 66$\pm$20 &    2.30 &    31.5 &  fg \\
 5:56:13.87 & -72:26:18.24 & $+$ 22$\pm$17 &    2.60 &    16.6 &   - \\
 5:52:59.23 & -70:39:49.68 & $-$ 20$\pm$23 &    2.00 &    28.1 &   - \\
 5:52:28.63 & -70:21:40.32 & $-$ 14$\pm$ 3 &    16.0 &     191 &   - \\
 5:47:01.20 & -65:35:02.76 & $+$ 35$\pm$26 &    1.70 &    38.1 &   - \\
 5:52:18.62 & -70:21:52.56 & $-$ 21$\pm$ 3 &    13.3 &     167 &   - \\
 5:52:10.49 & -70:31:25.32 & $+$  5$\pm$21 &    2.10 &    53.9 &   - \\
 5:49:42.86 & -69:00:21.96 & $+$138$\pm$21 &    2.20 &    31.8 & lmc \\
 5:47:45.62 & -67:45:06.84 & $+$ 45$\pm$12 &    3.70 &    86.0 & lmc \\
 5:51:05.09 & -71:06:11.52 & $+$ 65$\pm$15 &    3.00 &    47.6 &   - \\
 5:56:23.26 & -74:24:02.88 & $+$ 44$\pm$25 &    2.00 &    24.2 &  fg \\
 5:44:47.62 & -65:44:25.80 & $+$ 24$\pm$22 &    2.00 &    26.6 &   - \\
 5:44:38.02 & -65:34:53.04 & $+$ 28$\pm$ 6 &    7.40 &    94.2 &   - \\
 5:46:53.16 & -68:32:47.76 & $+$ 27$\pm$11 &    4.10 &    22.4 & lmc \\
 5:48:57.31 & -70:39:27.36 & $+$ 42$\pm$10 &    4.20 &    35.3 & lmc \\
 5:45:19.08 & -67:56:41.28 & $-$ 21$\pm$19 &    2.30 &    21.9 & lmc \\
 5:45:51.36 & -68:46:05.52 & $+$172$\pm$23 &    1.90 &    80.4 & lmc \\
 5:40:37.75 & -63:14:24.36 & $+$ 31$\pm$ 4 &    11.9 &     117 &  fg \\
 5:40:31.51 & -63:15:11.16 & $+$ 26$\pm$ 4 &    12.1 &     107 &  fg \\
 5:41:10.18 & -64:23:39.84 & $+$ 75$\pm$18 &    2.50 &    46.5 &  fg \\
 5:40:58.94 & -65:12:51.48 & $+$ 15$\pm$22 &    2.00 &    50.7 &   - \\
 5:46:08.76 & -71:27:42.84 & $+$ 88$\pm$20 &    2.20 &    36.7 &   - \\
 5:49:08.23 & -73:57:04.32 & $+$ 27$\pm$16 &    2.80 &    18.7 &  fg \\
 5:41:58.92 & -68:15:44.28 & $+$ 33$\pm$19 &    2.20 &    39.7 & lmc \\
 5:41:27.46 & -67:39:50.76 & $+$216$\pm$17 &    2.50 &    53.4 & lmc \\
 5:43:28.87 & -70:46:58.08 & $+$ 30$\pm$16 &    2.80 &    30.6 & lmc \\
 5:40:11.04 & -67:18:14.40 & $+$  8$\pm$26 &    1.60 &    77.6 & lmc \\
 5:40:20.54 & -71:24:31.32 & $+$ 21$\pm$25 &    1.80 &    74.0 & lmc \\
 5:42:20.52 & -73:31:35.76 & $+$ 78$\pm$13 &    3.30 &    75.1 &  fg \\
 5:41:51.12 & -73:32:14.64 & $+$ 21$\pm$ 6 &    7.00 &    1240 &  fg \\
 5:35:13.27 & -64:34:31.44 & $+$  2$\pm$ 4 &    10.2 &    61.7 &   - \\
 5:35:29.95 & -67:16:44.76 & $+$109$\pm$21 &    2.00 &    47.0 & lmc \\
 5:34:48.41 & -67:55:57.72 & $-$ 73$\pm$21 &    2.00 &    58.6 & lmc \\
 5:32:51.05 & -63:48:13.32 & $+$ 39$\pm$ 9 &    5.10 &     559 &  fg \\
 5:31:19.39 & -63:47:48.12 & $+$ 58$\pm$25 &    1.80 &    21.1 &  fg \\
 5:31:22.70 & -65:16:34.32 & $-$  9$\pm$23 &    1.90 &    54.1 &   - \\
 5:33:51.10 & -71:47:24.36 & $+$114$\pm$21 &    2.10 &    37.5 & lmc \\
 5:34:10.97 & -72:45:38.52 & $+$ 25$\pm$11 &    3.80 &    38.9 &   - \\
 5:34:25.49 & -73:27:29.52 & $+$ 49$\pm$ 5 &    8.60 &     156 &  fg \\
 5:33:45.05 & -72:16:23.88 & $+$ 89$\pm$13 &    3.40 &     146 &   - \\
 5:31:03.12 & -65:43:32.52 & $+$ 22$\pm$19 &    2.30 &    21.3 & lmc \\
 5:30:08.86 & -62:52:36.48 & $+$ 81$\pm$18 &    7.00 &    87.6 &  fg \\
 5:30:07.56 & -63:08:23.28 & $+$  8$\pm$15 &    3.90 &    55.6 &  fg \\
 5:30:57.46 & -66:48:38.88 & $-$ 11$\pm$ 7 &    6.00 &     120 & lmc \\
 5:29:58.25 & -63:07:49.08 & $+$ 28$\pm$ 4 &    15.9 &     117 &  fg \\
 5:30:01.10 & -63:19:01.56 & $+$ 51$\pm$23 &    2.00 &    48.8 &  fg \\
 5:29:57.38 & -63:58:23.88 & $+$ 62$\pm$20 &    2.20 &    13.2 &   - \\
 5:29:34.90 & -64:09:24.48 & $+$ 15$\pm$26 &    1.70 &    15.1 &   - \\
 5:32:27.98 & -73:32:27.60 & $+$ 65$\pm$15 &    2.80 &    31.1 &  fg \\
 5:29:51.58 & -67:49:31.44 & $+$ 75$\pm$ 6 &    6.80 &     204 & lmc \\
 5:30:15.22 & -74:11:17.52 & $+$ 37$\pm$ 3 &    15.5 &     130 &  fg \\
 5:30:09.82 & -74:10:57.36 & $+$ 32$\pm$ 3 &    15.6 &     163 &  fg \\
 5:29:37.25 & -73:39:59.04 & $+$ 17$\pm$ 9 &    4.80 &    65.0 &  fg \\
 5:27:16.82 & -64:52:22.08 & $+$ 45$\pm$ 9 &    4.70 &    19.2 &   - \\
 5:27:47.83 & -70:36:37.08 & $-$ 49$\pm$16 &    2.60 &    83.0 & lmc \\
 5:26:31.61 & -65:49:08.04 & $-$ 30$\pm$14 &    3.30 &     173 & lmc \\
 5:26:30.02 & -65:47:54.24 & $-$ 67$\pm$22 &    2.00 &    24.3 & lmc \\
 5:26:26.95 & -65:56:06.72 & $+$234$\pm$ 8 &    5.90 &     153 & lmc \\
 5:26:35.28 & -67:49:06.24 & $-$ 78$\pm$14 &    2.90 &    91.3 & lmc \\
 5:26:34.34 & -69:28:14.88 & $+$ 72$\pm$20 &    2.10 &    17.1 & lmc \\
 5:25:55.06 & -66:27:45.00 & $+$130$\pm$20 &    2.30 &    38.2 & lmc \\
 5:25:14.06 & -63:51:33.48 & $+$ 52$\pm$18 &    2.40 &    33.9 &   - \\
 5:23:40.99 & -70:51:12.24 & $+$ 94$\pm$ 2 &    22.0 &     336 & lmc \\
 5:23:09.31 & -73:14:09.96 & $+$ 14$\pm$15 &    2.90 &    50.0 &   - \\
 5:22:59.14 & -68:44:27.60 & $-$ 32$\pm$24 &    1.80 &    28.1 & lmc \\
 5:22:29.33 & -70:37:53.04 & $+$148$\pm$10 &    4.10 &     131 & lmc \\
 5:22:32.26 & -64:11:03.84 & $+$ 42$\pm$12 &    3.60 &    40.9 &   - \\
 5:22:29.69 & -64:00:10.80 & $+$ 37$\pm$15 &    2.90 &    32.4 &   - \\
 5:22:17.09 & -67:27:38.52 & $+$275$\pm$10 &    4.30 &    55.2 & lmc \\
 5:22:24.26 & -64:12:44.64 & $+$ 45$\pm$17 &    2.70 &    41.3 &   - \\
 5:21:28.92 & -71:59:56.40 & $+$120$\pm$11 &    3.90 &    21.0 & lmc \\
 5:18:54.67 & -74:14:48.12 & $-$ 21$\pm$14 &    3.10 &    55.3 &  fg \\
 5:20:07.70 & -68:02:22.56 & $+$ 44$\pm$ 9 &    4.90 &    76.0 & lmc \\
 5:19:08.14 & -71:47:48.48 & $+$ 82$\pm$ 6 &    7.10 &    74.6 & lmc \\
 5:19:25.87 & -67:47:00.24 & $+$ 81$\pm$ 9 &    4.60 &    30.3 & lmc \\
 5:18:13.37 & -71:38:07.44 & $+$ 40$\pm$18 &    2.50 &    47.8 & lmc \\
 5:19:29.23 & -65:27:19.44 & $-$ 21$\pm$18 &    2.60 &    42.6 & lmc \\
 5:18:32.71 & -69:35:22.20 & $+$148$\pm$12 &    3.90 &     340 & lmc \\
 5:19:14.33 & -64:56:00.60 & $+$ 30$\pm$12 &    3.90 &    53.7 &   - \\
 5:18:59.64 & -64:02:42.00 & $-$ 40$\pm$27 &    1.70 &    45.1 &   - \\
 5:16:37.39 & -72:37:08.76 & $+$ 35$\pm$11 &    4.10 &     208 &   - \\
 5:18:01.99 & -67:55:39.36 & $-$137$\pm$ 8 &    5.60 &     169 & lmc \\
 5:18:29.42 & -65:31:33.96 & $+$ 16$\pm$16 &    2.90 &    16.8 & lmc \\
 5:17:15.58 & -70:23:56.76 & $+$110$\pm$10 &    4.40 &    85.4 & lmc \\
 5:16:42.29 & -71:49:06.24 & $+$ 55$\pm$18 &    2.40 &     283 & lmc \\
 5:17:40.90 & -67:12:45.72 & $+$ 63$\pm$12 &    3.90 &    44.7 & lmc \\
 5:14:48.67 & -74:41:26.16 & $+$  8$\pm$11 &    7.20 &     111 &  fg \\
 5:17:15.19 & -62:51:13.68 & $+$  9$\pm$ 9 &    13.2 &    72.2 &  fg \\
 5:16:09.96 & -66:18:13.32 & $-$ 65$\pm$31 &    1.50 &    9.70 & lmc \\
 5:16:01.82 & -65:00:39.96 & $-$  5$\pm$22 &    2.10 &    19.5 &   - \\
 5:15:56.38 & -65:13:30.36 & $+$  8$\pm$13 &    3.60 &    33.1 & lmc \\
 5:15:24.19 & -65:58:38.28 & $+$ 69$\pm$13 &    3.70 &     271 & lmc \\
 5:11:58.54 & -73:06:34.92 & $+$ 61$\pm$11 &    4.10 &    46.7 &   - \\
 5:11:32.52 & -73:33:09.36 & $-$ 40$\pm$30 &    1.50 &    23.7 &  fg \\
 5:12:14.14 & -72:32:44.52 & $+$ 42$\pm$17 &    2.70 &     119 &   - \\
 5:13:39.22 & -69:27:45.36 & $-$113$\pm$23 &    2.00 &    64.3 & lmc \\
 5:14:57.67 & -64:40:34.32 & $+$ 43$\pm$11 &    4.20 &    27.5 &   - \\
 5:14:43.44 & -64:51:36.72 & $+$ 72$\pm$22 &    2.10 &    13.1 &   - \\
 5:14:46.08 & -62:55:13.80 & $+$ 26$\pm$25 &    4.80 &     102 &  fg \\
 5:09:07.20 & -73:33:03.96 & $+$ 15$\pm$15 &    3.00 &     137 &  fg \\
 5:10:27.79 & -69:32:06.72 & $+$149$\pm$12 &    3.70 &    44.9 & lmc \\
 5:11:02.40 & -66:48:03.24 & $+$ 31$\pm$14 &    3.30 &    23.8 & lmc \\
 5:07:36.94 & -71:52:22.80 & $+$ 31$\pm$20 &    2.30 &    51.4 & lmc \\
 5:10:11.28 & -68:02:50.64 & $-$ 52$\pm$ 6 &    8.00 &    71.1 & lmc \\
 5:11:38.64 & -64:02:22.20 & $+$ 51$\pm$10 &    4.60 &     135 &   - \\
 5:10:57.62 & -65:29:28.68 & $+$ 45$\pm$22 &    2.10 &    40.7 & lmc \\
 5:09:51.48 & -67:16:22.08 & $-$ 89$\pm$ 6 &    7.30 &    81.9 & lmc \\
 5:05:39.36 & -71:07:40.44 & $+$212$\pm$12 &    3.90 &    84.3 & lmc \\
 5:08:31.32 & -67:06:18.72 & $+$117$\pm$ 9 &    5.00 &    92.9 & lmc \\
 5:04:02.47 & -72:03:44.64 & $+$ 51$\pm$11 &    4.20 &    88.1 &   - \\
 5:05:36.12 & -70:05:15.72 & $-$ 38$\pm$13 &    3.50 &    84.9 & lmc \\
 5:04:56.14 & -70:45:07.20 & $+$ 44$\pm$15 &    3.00 &    24.3 & lmc \\
 5:05:47.06 & -66:43:28.20 & $+$ 16$\pm$ 9 &    4.90 &    44.4 & lmc \\
 5:00:37.18 & -72:06:25.20 & $+$ 45$\pm$11 &    4.20 &    79.4 &   - \\
 5:05:37.73 & -66:43:00.84 & $+$ 40$\pm$12 &    3.80 &    22.4 & lmc \\
 5:04:54.38 & -67:36:10.44 & $+$ 47$\pm$15 &    3.10 &    40.3 & lmc \\
 5:05:01.06 & -66:45:20.16 & $+$ 81$\pm$21 &    2.20 &     156 & lmc \\
 4:57:35.38 & -73:28:34.68 & $+$ 32$\pm$13 &    3.60 &    47.1 &  fg \\
 5:02:02.90 & -69:32:01.32 & $+$ 86$\pm$17 &    2.70 &     362 & lmc \\
 5:05:06.79 & -64:18:35.28 & $+$ 18$\pm$ 9 &    4.80 &     263 &   - \\
 5:04:28.66 & -64:57:22.32 & $+$ 24$\pm$ 7 &    6.70 &    84.8 &   - \\
 4:59:39.84 & -69:55:04.08 & $+$ 21$\pm$11 &    4.20 &     122 & lmc \\
 4:53:02.57 & -73:41:53.52 & $-$  6$\pm$27 &    1.70 &    9.60 &  fg \\
 5:02:27.62 & -65:57:14.76 & $-$161$\pm$21 &    2.20 &    45.9 & lmc \\
 4:50:46.51 & -74:35:35.88 & $+$ 11$\pm$23 &    4.20 &    92.9 &  fg \\
 4:51:34.37 & -73:57:59.40 & $+$ 17$\pm$14 &    3.20 &    26.0 &  fg \\
 4:49:00.02 & -74:35:31.92 & $+$ 52$\pm$19 &    5.40 &    85.0 &  fg \\
 4:48:48.65 & -74:17:30.48 & $+$  9$\pm$ 5 &    10.3 &     199 &  fg \\
 4:57:26.66 & -68:29:10.68 & $+$ 49$\pm$26 &    1.70 &    10.9 & lmc \\
 5:01:26.90 & -64:16:45.12 & $+$ 14$\pm$21 &    2.10 &    12.7 &   - \\
 4:58:46.99 & -67:05:37.68 & $+$ 49$\pm$14 &    3.20 &    59.3 & lmc \\
 4:48:44.95 & -73:51:46.44 & $-$ 35$\pm$16 &    2.80 &    43.5 &  fg \\
 4:58:44.21 & -66:57:42.12 & $+$  5$\pm$22 &    2.00 &    68.1 & lmc \\
 4:51:21.19 & -72:21:06.48 & $+$ 32$\pm$20 &    2.30 &    61.7 &   - \\
 4:56:18.70 & -68:49:46.20 & $-$ 58$\pm$11 &    4.00 &    47.5 & lmc \\
 5:01:16.15 & -63:51:22.32 & $+$ 39$\pm$18 &    2.50 &    26.2 &   - \\
 4:53:38.28 & -70:28:10.92 & $+$ 43$\pm$ 4 &    11.7 &     159 & lmc \\
 4:59:47.23 & -64:15:36.36 & $+$ 33$\pm$ 9 &    5.00 &    89.1 &   - \\
 4:49:31.85 & -72:16:57.36 & $+$ 31$\pm$10 &    4.70 &    24.2 &   - \\
 4:49:03.89 & -72:20:19.68 & $+$ 29$\pm$ 6 &    7.60 &    33.9 &   - \\
 4:58:35.28 & -64:49:47.64 & $+$ 41$\pm$ 5 &    8.60 &    74.8 &   - \\
 4:49:03.19 & -70:52:11.28 & $+$  7$\pm$ 1 &    33.3 &     263 & lmc \\
 4:51:14.14 & -69:31:47.28 & $-$ 62$\pm$ 8 &    5.50 &     128 & lmc \\
 4:57:59.57 & -63:41:27.60 & $+$ 14$\pm$21 &    2.20 &    27.7 &   - \\
 4:46:10.30 & -72:05:10.32 & $+$ 12$\pm$ 7 &    5.90 &     169 &   - \\
 4:52:21.24 & -68:23:03.48 & $+$  2$\pm$16 &    2.70 &     110 & lmc \\
 4:40:35.23 & -74:21:14.04 & $+$ 29$\pm$27 &    2.00 &    90.4 &  fg \\
 4:47:46.87 & -70:48:44.28 & $+$ 17$\pm$ 5 &    8.40 &    81.1 & lmc \\
 4:50:43.13 & -68:49:27.12 & $+$ 18$\pm$27 &    1.70 &    13.1 & lmc \\
 4:46:21.98 & -71:02:12.48 & $+$  3$\pm$27 &    1.60 &    16.8 & lmc \\
 4:44:37.03 & -71:15:12.96 & $+$ 38$\pm$18 &    2.30 &    23.1 &   - \\
 4:41:40.42 & -72:16:05.88 & $+$ 32$\pm$22 &    2.00 &    9.40 &  fg \\
 4:55:37.66 & -62:54:02.16 & $+$ 61$\pm$28 &    2.60 &    33.8 &  fg \\
 4:51:07.06 & -66:49:33.60 & $-$ 64$\pm$11 &    3.90 &    60.7 & lmc \\
 4:47:24.86 & -69:15:47.88 & $+$107$\pm$17 &    2.70 &    31.3 & lmc \\
 4:45:17.21 & -70:14:29.76 & $+$ 14$\pm$16 &    2.80 &    37.3 & lmc \\
 4:45:41.21 & -69:12:26.28 & $+$ 70$\pm$12 &    3.70 &    24.4 & lmc \\
 4:49:35.35 & -66:41:34.80 & $-$ 29$\pm$22 &    2.10 &    21.4 & lmc \\
 4:53:26.93 & -63:28:59.16 & $+$ 68$\pm$11 &    4.40 &    38.0 &  fg \\
 4:47:11.21 & -67:50:50.64 & $+$ 50$\pm$15 &    3.00 &    37.9 & lmc \\
 4:41:45.86 & -70:25:53.40 & $+$ 67$\pm$13 &    3.20 &    40.9 & lmc \\
 4:51:40.22 & -63:54:36.00 & $+$ 91$\pm$12 &    4.00 &    33.8 &   - \\
 4:47:38.59 & -66:56:14.28 & $+$ 26$\pm$21 &    2.20 &    10.0 & lmc \\
 4:51:29.26 & -63:41:04.56 & $+$ 74$\pm$25 &    1.90 &     156 &   - \\
 4:44:16.03 & -68:42:10.80 & $-$ 24$\pm$ 5 &    8.50 &    96.2 & lmc \\
 4:49:38.33 & -65:05:02.76 & $+$ 24$\pm$16 &    2.90 &     123 &   - \\
 4:51:12.12 & -63:44:19.32 & $+$ 72$\pm$10 &    4.80 &    29.4 &   - \\
 4:34:55.22 & -72:39:57.24 & $+$  7$\pm$ 5 &    8.50 &     202 &  fg \\
 4:39:33.00 & -70:45:08.64 & $+$ 17$\pm$10 &    4.10 &    98.0 &   - \\
 4:50:47.90 & -63:24:17.28 & $-$  8$\pm$28 &    1.70 &    8.00 &  fg \\
 4:43:24.65 & -67:56:22.92 & $-$ 18$\pm$12 &    3.60 &    73.0 & lmc \\
 4:32:08.23 & -72:49:39.72 & $+$ 15$\pm$ 9 &    4.90 &    90.1 &  fg \\
 4:28:22.22 & -73:49:48.36 & $-$  3$\pm$21 &    2.00 &    15.7 &  fg \\
 4:35:13.06 & -71:26:47.04 & $+$  4$\pm$27 &    1.50 &    13.4 &   - \\
 4:39:18.41 & -69:39:50.04 & $-$  6$\pm$29 &    1.50 &    11.1 & lmc \\
 4:45:12.46 & -65:47:07.80 & $+$ 23$\pm$16 &    3.00 &     108 &   - \\
 4:25:57.84 & -74:01:53.76 & $+$ 19$\pm$19 &    2.40 &    16.9 &  fg \\
 4:42:24.38 & -67:28:04.08 & $+$ 55$\pm$ 9 &    4.80 &    24.0 & lmc \\
 4:42:19.73 & -67:27:15.84 & $+$ 69$\pm$21 &    2.00 &    24.9 & lmc \\
 4:43:17.69 & -66:52:05.16 & $+$ 45$\pm$19 &    2.30 &    73.7 &   - \\
 4:32:51.98 & -71:33:18.00 & $+$ 14$\pm$ 9 &    4.80 &     165 &  fg \\
 4:30:51.24 & -72:16:18.84 & $+$ 30$\pm$20 &    2.10 &    9.60 &  fg \\
 4:38:41.98 & -69:00:30.96 & $+$ 70$\pm$13 &    3.40 &    40.4 & lmc \\
 4:29:53.90 & -72:27:10.08 & $+$ 30$\pm$17 &    2.40 &    20.8 &  fg \\
 4:32:02.26 & -71:13:48.36 & $+$ 34$\pm$10 &    4.30 &    29.5 &  fg \\
 4:43:26.78 & -65:26:40.92 & $+$ 58$\pm$12 &    4.10 &    91.0 &   - \\
 4:44:51.77 & -64:27:17.28 & $+$ 58$\pm$20 &    2.40 &    8.40 &   - \\
 4:32:31.06 & -70:41:57.12 & $+$ 21$\pm$13 &    3.10 &    12.9 &   - \\
 4:46:00.58 & -63:13:49.44 & $+$ 14$\pm$19 &    2.50 &    26.5 &  fg \\
 4:40:56.98 & -66:24:24.12 & $+$ 46$\pm$ 5 &    9.00 &     444 &   - \\
 4:38:56.35 & -67:21:54.36 & $+$ 52$\pm$ 7 &    6.20 &     427 &   - \\
 4:25:01.68 & -72:47:15.00 & $+$ 65$\pm$11 &    3.70 &    76.5 &  fg \\
 4:36:57.26 & -67:47:27.96 & $+$ 34$\pm$21 &    2.10 &    43.8 &   - \\
 4:23:49.39 & -72:44:19.32 & $+$ 13$\pm$ 2 &    18.1 &     409 &  fg \\
 4:44:10.99 & -63:05:43.08 & $+$ 54$\pm$25 &    1.90 &    34.9 &  fg \\
 4:36:01.58 & -67:25:12.36 & $+$ 48$\pm$11 &    4.10 &    60.2 &   - \\
 4:18:19.87 & -73:31:35.40 & $+$ 47$\pm$18 &    2.30 &    59.8 &  fg \\
 4:37:55.10 & -66:04:49.44 & $+$ 58$\pm$18 &    2.60 &    53.3 &   - \\
 4:37:21.46 & -66:07:58.44 & $+$ 40$\pm$11 &    4.40 &     396 &   - \\
 4:32:38.74 & -68:24:58.32 & $+$ 16$\pm$20 &    2.20 &    9.20 &   - \\
 4:38:40.06 & -65:03:23.40 & $+$ 58$\pm$ 9 &    5.20 &     295 &   - \\
 4:33:53.88 & -67:18:17.64 & $+$ 35$\pm$13 &    3.30 &    76.3 &   - \\
 4:37:43.92 & -64:54:21.96 & $+$ 10$\pm$22 &    2.20 &    45.8 &   - \\
 4:37:34.44 & -64:53:50.28 & $+$ 32$\pm$ 9 &    5.50 &    57.3 &   - \\
 4:37:08.09 & -64:59:04.20 & $+$ 49$\pm$ 2 &    20.7 &     741 &   - \\
 4:28:07.97 & -69:06:37.44 & $+$ 76$\pm$20 &    2.20 &    16.3 &   - \\
 4:35:07.68 & -65:34:05.16 & $+$ 67$\pm$10 &    4.90 &    54.4 &   - \\
 4:39:06.67 & -63:04:57.72 & $+$ 52$\pm$ 8 &    6.10 &    59.8 &  fg \\
 4:37:40.68 & -63:44:24.36 & $+$ 42$\pm$19 &    2.50 &    36.5 &  fg \\
 4:16:02.69 & -72:16:36.12 & $-$  6$\pm$10 &    4.20 &    75.3 &  fg \\
 4:28:34.32 & -67:41:51.36 & $+$ 39$\pm$ 5 &    9.40 &    71.9 &   - \\
 4:37:00.24 & -63:15:27.00 & $+$ 51$\pm$16 &    3.10 &    23.6 &  fg \\
 4:35:08.28 & -64:01:09.84 & $+$ 65$\pm$17 &    2.90 &    49.0 &  fg \\
 4:26:50.38 & -67:53:19.68 & $+$ 43$\pm$13 &    3.50 &     262 &   - \\
 4:06:59.69 & -73:34:06.60 & $+$  9$\pm$ 7 &    6.90 &     247 &  fg \\
 4:35:41.38 & -62:58:41.88 & $+$ 43$\pm$ 8 &    7.20 &     362 &  fg \\
 4:34:25.51 & -63:37:44.76 & $+$ 54$\pm$13 &    4.70 &     216 &  fg \\
 4:05:18.82 & -73:31:40.80 & $+$ 19$\pm$15 &    3.70 &    21.5 &  fg \\
 4:22:49.01 & -68:37:29.64 & $+$ 48$\pm$ 3 &    19.2 &     133 &   - \\
 4:23:22.18 & -68:16:52.32 & $+$ 24$\pm$11 &    6.10 &    76.3 &   - \\
 4:23:05.81 & -68:23:27.96 & $+$ 18$\pm$16 &    4.10 &    58.9 &   - \\
 4:09:03.10 & -72:19:15.60 & $+$ 44$\pm$12 &    7.20 &    60.2 &  fg \\
 4:08:36.00 & -72:18:45.36 & $+$ 28$\pm$ 6 &    17.0 &    88.1 &  fg \\
 4:31:28.20 & -64:06:33.48 & $+$ 44$\pm$22 &    5.60 &     240 &  fg \\
 \enddata
 \label{table:ptsrc_rm}
\tablenotetext{a}{The de-biased polarized flux and total intensity is estimated at the location of the peak pixel.}
\tablenotetext{b}{Sources used in the Milky Way foreground rotation measure estimation are denoted by ``fg", while sources whose projection lie directly behind the LMC body are denoted by ``lmc".}

\end{deluxetable} 
\clearpage

\end{document}